\documentclass[a4paper,11pt]{article}

\usepackage{jheppub} 
\usepackage{subfigure}

\preprint{KUNS-2967}

\title{\boldmath 
Krylov complexity and chaos in quantum mechanics

}

\author[a]{Koji Hashimoto,}
\author[b]{Keiju Murata,}
\author[c]{Norihiro Tanahashi,}
\author[a]{Ryota Watanabe}

\affiliation[a]{Department of Physics, Kyoto University, Kyoto 606-8502, Japan}
\affiliation[b]{Department of Physics, Nihon University, Sakurajosui, Tokyo 156-8550, Japan}
\affiliation[c]{Department of Physics, Chuo University, Kasuga, Bunkyo-ku, Tokyo 112-8551, Japan}

\emailAdd{koji@scphys.kyoto-u.ac.jp}
\emailAdd{murata.keiju@nihon-u.ac.jp}
\emailAdd{tanahashi@phys.chuo-u.ac.jp}
\emailAdd{watanabe@gauge.scphys.kyoto-u.ac.jp}

\abstract{
Recently, Krylov complexity was proposed as a measure of complexity and chaoticity of quantum systems.
We consider the stadium billiard 
as a typical example of the quantum mechanical system obtained by quantizing a classically chaotic system,
and numerically evaluate Krylov complexity for operators and states. Despite no exponential growth of the Krylov complexity, we find a clear correlation between variances of Lanczos coefficients and classical Lyapunov exponents, and also a correlation with the statistical distribution of adjacent spacings of the quantum energy levels. 
This shows that the variances of Lanczos coefficients can be a measure of quantum chaos. 
The universality of the result is supported by our similar analysis of Sinai billiards.
Our work provides a firm bridge between Krylov complexity and classical/quantum chaos.  
}

\begin{document}

\maketitle
\flushbottom

\section{Introduction}
\label{sec:intro}
Since quantum mechanics is the basis of the description of nature, studying chaos in the quantum realm, or quantum chaos is important both theoretically and experimentally. In recent years, quantum chaos has attracted interest in a variety of studies, including thermalization processes in thermodynamic systems and black hole dynamics. However, a clear standard for the definition of quantum chaos is still overdue. Historically, there have been several proposals to define quantum chaos. The traditional characterization is by the statistical distribution of adjacent energy level spacing in quantum systems \cite{Bohigas:1983er}. Turning to the dynamics of quantum systems, it is expected that local operators in chaotic systems will become more complex with time evolution. To measure this operator growth, the out-of-time-order correlator (OTOC) \cite{Larkin} has attracted much attention as a possible indicator of quantum chaos.

The exponential growth rate of the OTOC, called the quantum Lyapunov exponent, is regarded as a naive quantum counterpart of the Lyapunov exponent in classical systems. Remarkably, the quantum Lyapunov exponent of a finite-temperature quantum many-body system has an upper bound determined by the temperature \cite{Maldacena:2015waa}. This upper bound is expected to be an indicator of the existence of a gravity dual through the AdS/CFT correspondence. The Sachdev-Ye-Kitaev (SYK) model \cite{Sachdev:1992fk,Kitaev-talk}, which saturates the upper bound, has played important roles in the studies of quantum chaos and quantum gravity. Furthermore, the thermodynamic well-definedness of the OTOC suggests an upper bound on the energy dependence of the Lyapunov exponent for very general systems \cite{Hashimoto:2021afd}.

However, it has been shown that even in a classically chaotic system, the OTOC does not necessarily show exponential growth \cite{Hashimoto:2017oit},\footnote{
In quantum mechanical systems with a few degrees of freedom, the condition in \cite{Maldacena:2015waa} that there exists a separation of time scales (between the dissipation time and the scrambling time)
does not hold.
} and conversely, even in a classically non-chaotic system, the OTOC can grow exponentially due to the presence of unstable points in the potential \cite{Xu:2019lhc,Bhattacharyya:2020art,Hashimoto:2020xfr}. Thus, the OTOC alone is unfortunately insufficient to characterize the quantum chaoticity. Also, the relationship between the energy level statistics, which has been previously proposed as a characterization of quantum chaos, and the OTOC is unclear.

Recently, the notion of Krylov complexity for operators, Krylov operator complexity,\footnote{In literature, this is simply called Krylov complexity. In this paper, since we will consider the notion of Krylov complexity for both operators and states, we use the terms ``Krylov operator complexity" and ``Krylov state complexity'' for clarity. See also \cite{Erdmenger:2023shk}.} was proposed as a new indicator to evaluate operator growth more directly and quantitatively \cite{Parker:2018yvk}.\footnote{
The definition of the Krylov complexity depends on the choice of the inner product introduced in the operator space. In this paper, we consider the infinite-temperature inner product. For studies with the other choices of the inner product, see for example \cite{Parker:2018yvk,Dymarsky:2021bjq,Avdoshkin:2022xuw,Guo:2022hui,Kundu:2023hbk}.}
Also, based on the same idea as the Krylov operator complexity, the Krylov complexity for states, Krylov state complexity, has been proposed as a quantitative measure of the complexity of a quantum state in the Schr\"odinger picture \cite{Balasubramanian:2022tpr}.\footnote{The authors of \cite{Balasubramanian:2022tpr} proposed the term ``spread complexity" for the notion of Krylov complexity for states. Its relation to topological phases of matter was recently investigated in \cite{Caputa:2022yju,Caputa:2022eye}.} In the calculation of the Krylov complexity for operators and states, a given quantum system is reduced to a one-dimensional chain model, where the hopping is given by a series of real numbers called Lanczos coefficients. Thus the time evolution of the Krylov operator/state complexity is encoded into the set of Lanczos coefficients.

A natural question is a possible relationship between the Lanczos coefficients and chaos. In this regard, in finite-dimensional quantum systems such as spin systems and the SYK model, novel correlations between certain variances of the Lanczos coefficients, asymptotic values of the Krylov operator complexity and the energy level statistics were reported \cite{Rabinovici:2021qqt,Rabinovici:2022beu}.\footnote{
The authors of \cite{Balasubramanian:2022dnj} analytically studied the statistics of the Lanczos coefficients for a random ensemble of Hamiltonians.
}
Unfortunately, these models have no classical counterparts, and their relation to classical chaoticity is not clear. In fact, if we are to define quantum chaos consistently with classical chaos, the first interest to us is the quantum mechanical system obtained by quantizing the classical chaotic system. In this paper, we consider billiard systems \cite{Sinai:1970,Bunimovich:1974,Bunimovich:1979} which are one of the most-studied classical/quantum system of chaos. We investigate in the billiard systems the relationship between the classical chaoticity, the quantum energy level spacing statistics, and the Lanczos coefficients for operators and quantum states 
and show the existence of a correlation between them by numerical analyses.
This result on the Krylov operator/state complexity and their Lanczos coefficients may also hold for broader general quantum mechanical systems, providing a key step for refining the definitions of quantum chaos.

The novel venue on the bridge connecting complexity and chaos can shed light more on quantum gravity.
Complexity has attracted much attention in black hole studies, and specifically, there are several conjectures in the AdS/CFT correspondence that the complexity of the holographic quantum system probes the interior of the dual black hole \cite{Susskind:2014rva,Stanford:2014jda,Susskind:2014jwa,Susskind:2014moa,Brown:2015bva,Brown:2015lvg}. One of the issues there is the existence of several different proposals on the definition of complexity, allowing ambiguity. In contrast, the Krylov operator/state complexity is defined unambiguously. It is an interesting and important question whether the Krylov operator/state complexity describes the interior of the black hole.\footnote{See \cite{Kar:2021nbm} for studies in this direction. See \cite{Dymarsky:2021bjq,Caputa:2021ori,Avdoshkin:2022xuw,Camargo:2022rnt} for studies on the Krylov operator complexity in QFTs.}

This paper is organized as follows. In Sec.~\ref{sec:2}, we first review the Lanczos coefficients, the Krylov operator/state complexity, and their possible relation with chaos. Then, we illustrate the Krylov complexity by showing some analytically calculable complexity in integrable quantum mechanical systems. In Sec.~\ref{sec:3}, we consider the stadium billiard, which is a typical chaotic system, and numerically investigate the Krylov complexity for the momentum operator. We see that a certain variance of the Lanczos coefficient is correlated with the chaoticity of the system. In Sec.~\ref{sec:4}, we numerically investigate the Krylov state complexity in the stadium billiard. Again, we see that there is a correlation between the variance of the Lanczos coefficients and chaos. In Sec.~\ref{sec:5}, we perform the same analysis for the Sinai billiard as for the stadium billiard. Sec.~\ref{sec:6} is for our summary and discussion. In App.~\ref{app:1}, we discuss the details of numerical calculations.


\section{Review and some analytic examples}
\label{sec:2}

\subsection{Review on Krylov complexity}
\label{sec:2-1}
Here, we review the definitions of Krylov operator/state complexity and their possible relation to chaoticity.\footnote{
See \cite{Muck:2022xfc} for a pedagogical introduction to Krylov operator complexity and some analytical methods. For studies of Euclidean time evolution and Krylov complexity, see \cite{Avdoshkin:2019trj,Dymarsky:2019elm}.
}

\subsubsection{Krylov operator complexity}
Consider a quantum system with Hamiltonian $H$. The Krylov operator complexity of a given operator $\mathcal{O}$ is defined as follows. To begin with, we can rewrite the Heisenberg operator $\mathcal{O}(t) = e^{iHt}\mathcal{O}(0)e^{-iHt}$ as
\begin{equation}
	\mathcal{O}(t) = \sum_{n=0}^\infty \frac{(it)^n}{n!}\mathcal{L}^n\mathcal{O}(0)\,,
\label{eq:BCH}
\end{equation}
where $\mathcal{L}$ is the Liouvillian superoperator, $\mathcal{L}=[H,\,\cdot\,]$. This is a linear combination of the sequence of operators
\begin{equation}
    \mathcal{O},~\mathcal{L}\mathcal{O},~\mathcal{L}^2\mathcal{O},~\cdots\,,
\label{eq:nested commutators}
\end{equation}
where $\mathcal{O}$ stands for $\mathcal{O}(0)$. The operator (sub)space, $\mathcal{H}_\mathcal{O}$, which is spanned by \eqref{eq:nested commutators} is called the Krylov space associated with the operator $\mathcal{O}$. Define $K_\mathcal{O} \equiv \dim\mathcal{H}_\mathcal{O}$. Introducing an inner product between operators $\mathcal{O}_1$ and $\mathcal{O}_2$ as, for example,\footnote{One can choose another definition of the inner product. Although there is often a normalization coefficient in \eqref{eq:inner product} in the literature, we omitted it since such an overall constant does not change the Lanczos coefficients.
}
\begin{equation}
	(\mathcal{O}_1|\mathcal{O}_2) \equiv {\rm Tr}\big[\mathcal{O}_1^\dagger\mathcal{O}_2\big]\,,
\label{eq:inner product}
\end{equation}
we can construct an orthonormal basis for $\mathcal{H}_\mathcal{O}$ by the following procedure, which is known as the {\it Lanczos algorithm}:
\begin{itemize}
\item[1.~] $b_0\equiv0\,, \quad \mathcal{O}_{-1}\equiv0$
\item[2.~] $\mathcal{O}_0\equiv \mathcal{O}/\|\mathcal{O}\|$, where $\|\mathcal{O}\|\equiv\sqrt{(\mathcal{O}|\mathcal{O})}$
\item[3.~] For $n\geq1$: $\mathcal{A}_n=\mathcal{L}\mathcal{O}_{n-1}-b_{n-1}\mathcal{O}_{n-2}$
\item[4.~] Set $b_n=\|\mathcal{A}_n\|$
\item[5.~] If $b_n=0$ stop; otherwise set $\mathcal{O}_n=\mathcal{A}_n/b_n$ and go to step 3.
\end{itemize}
If the dimension $K_\mathcal{O}$ of the Krylov space is finite, the above algorithm ends with $b_{K_\mathcal{O}}=0$. The algorithm produces the orthonormal basis $\{\mathcal{O}_n\}_{n=0}^{K_\mathcal{O}-1}$ called the Krylov basis, and a set of positive numbers $\{b_n\}$ called the Lanczos coefficients.
Now expand the Heisenberg operator $\mathcal{O}(t)$ in terms of the Krylov basis as
\begin{equation}
    \mathcal{O}(t) = \sum_{n=0}^{K_\mathcal{O}-1}i^n\varphi_n(t)\mathcal{O}_n\,.
\label{eq:operator in the Krylov basis}
\end{equation}
Note that $\varphi_n(t)$ obeys the normalization condition,
\begin{equation}
    \sum_{n=0}^{K_\mathcal{O}-1}|\varphi_n(t)|^2=\|\mathcal{O}\|^2\,.
\end{equation}
Substituting \eqref{eq:operator in the Krylov basis} into the Heisenberg equation yields
\begin{equation}
    \dot{\varphi}_n(t) = b_n\varphi_{n-1}(t) - b_{n+1}\varphi_{n+1}(t)\,,
\label{eq:Krylov chain for operator}
\end{equation}
where the dot represents the derivative with respect to time. The initial condition is $\varphi_n(0)=\delta_{n0}\|\mathcal{O}\|$ by definition. It is convenient to normalize the operator in the first place so that the condition simplifies to
$\varphi_n(0)=\delta_{n0}$.
The Krylov complexity for the operator $\mathcal{O}$ is defined as
\begin{equation}
    C_\mathcal{O}(t) \equiv \sum_{n=0}^{K_\mathcal{O}-1}n|\varphi_n(t)|^2\,.
\label{eq:def Krylov operator complexity}
\end{equation}
The operator $\mathcal{O}_n$ contains the nested commutator $\mathcal{L}^n\mathcal{O}$, which is expected to become more complex with $n$. Therefore, the Krylov operator complexity roughly measures the number of the nested commutators in the given Heisenberg operator $\mathcal{O}(t)$.
Equation \eqref{eq:Krylov chain for operator} can be regarded as a hopping model on a one-dimensional chain. The Lanczos coefficients $\{b_n\}$ represent the hopping amplitudes. The Krylov operator complexity can be interpreted as the expectation value of the ``position'' $n$ of the hopping particle on the one-dimensional chain.

Originally, the authors of \cite{Parker:2018yvk} considered many-body systems in the thermodynamic limit. They conjectured that the Lanczos coefficient $b_n$ grows at most linearly with $n$, and when the system is chaotic, $b_n$ asymptotically grows linearly as $b_n\sim\alpha n+\gamma~(n\to\infty)$ with some constants $\alpha$ and $\gamma$. It was also shown that when the Lanczos coefficient behaves as $b_n\sim\alpha n+\gamma$, the Krylov operator complexity grows exponentially as $K_\mathcal{O}(t)\sim e^{2\alpha t}$. Thus, the exponential growth of the Krylov operator complexity may be an indicator of chaos. Note that this argument and conjecture assume the thermodynamic limit, and do not necessarily hold in finite-dimensional systems.\footnote{Recently, the linear growth of Lanczos coefficients has been also observed in some integrable systems \cite{Dymarsky:2021bjq,Bhattacharjee:2022vlt}. Therefore, this property is generally not enough for distinguishing whether a given system is chaotic or not.}

In finite-dimensional systems, the Lanczos algorithm must terminate. Therefore, even if the Lanczos coefficient grows at first, it necessarily turns to decrease at some point \cite{Barbon:2019wsy}. The authors of \cite{Rabinovici:2021qqt} proposed a possible relation between the Lanczos coefficients and the chaoticity. Given an operator $\mathcal{O}(t)$, define the moments $\{\mu_n\}$ of the two-point correlation function $(\mathcal{O}(t)|\mathcal{O}(0))$ by
\begin{equation}
	\mu_n\equiv \left.\frac{d^n}{dt^n}(\mathcal{O}(t)|\mathcal{O}(0))\right|_{t=0}\,.
\label{eq:moments}
\end{equation}
It is known \cite{Parker:2018yvk,Rabinovici:2021qqt,SANCHEZDEHESA1978275} that the Lanczos coefficients $\{b_n\}$ of $\mathcal{O}(t)$ are intimately related to the moments $\{\mu_n\}$ via
\begin{equation}
	b_n^2 = \frac{D_{n-2}D_n}{D_{n-1}}\,, \quad n\geq 1
\label{eq:Lanczos coefficient in terms of moments}
\end{equation}
where $D_n$ is defined as
\begin{equation}
	D_n = 
	\begin{vmatrix}
   		\mu_0 & \mu_1 & \mu_2 & \cdots & \mu_n \\
  		\mu_1 & \mu_2 & \mu_3 & \cdots & \mu_{n+1} \\
		\vdots & \vdots & \vdots & \cdots & \vdots \\
		\mu_n & \mu_{n+1} & \mu_{n+2} & \cdots & \mu_{2n}
	\end{vmatrix}\,.
\label{eq:Hankel determinant}
\end{equation}
The authors of \cite{Rabinovici:2021qqt} argued that when the system is integrable, that is, the distribution of the adjacent energy level spacing obeys the Poisson statistics, there may be more possibility for $D_n$ to become small,
which is in turn expected to cause fluctuation in the Lanczos coefficient $b_n$ through \eqref{eq:Lanczos coefficient in terms of moments}.
Conversely, if the system is chaotic, that is, the level spacing obeys the Wigner-Dyson statistics, the behavior of $b_n$ is expected to become less erratic. The authors of \cite{Rabinovici:2021qqt} then introduced the variance of the Lanczos coefficient as
\begin{equation}
    \sigma^2 \equiv {\rm Var}(x_i)=\langle x^2 \rangle - \langle x \rangle^2\,,\quad x_i\equiv \ln\left(\frac{b_{2i-1}}{b_{2i}}\right)\ ,
\label{eq:variance}
\end{equation}
where $\langle \cdots \rangle$ represents the mean value.\footnote{The part of the series of the Lanczos coefficients for which $x_i$ becomes too large should be excluded from the calculation of the variance since it is numerically less reliable \cite{Rabinovici:2021qqt}.}
This quantity measures the magnitude of the erratic behavior of $b_n$. 
They confirmed the expected behavior in the XXZ spin chain model and in the SYK model. Adding an integrability breaking term in the XXZ model \cite{Rabinovici:2022beu}, they also investigated the correlation between the variance $\sigma^2$ and the average $\langle\tilde{r}\rangle$ of the parameters
\begin{equation}
    \tilde{r}_n=\frac{\min(s_n,s_{n-1})}{\max(s_n,s_{n-1})}\,,\quad s_n=E_{n+1}-E_n,
\label{eq:ratio}
\end{equation}
which is a popular characterization of the level statistics \cite{Oganesyan Huse (2007),Atas Bogomolny Giraud Roux (2013)}. Specifically, $\langle\tilde{r}\rangle$ takes the following values depending on the distribution of adjacent level spacing in the spectrum under consideration \cite{Atas Bogomolny Giraud Roux (2013)}:
\begin{equation}
    \langle\tilde{r}\rangle=
    \begin{cases}
        2\ln2-1\approx 0.38629 & \quad \text{Poisson} \\
        4-2\sqrt{3}\approx 0.53590 & \quad \text{GOE} \\
        2\frac{\sqrt{3}}{\pi}-\frac{1}{2}\approx 0.60266 & \quad \text{GUE} \\
        \frac{32}{15}\frac{\sqrt{3}}{\pi}-\frac{1}{2}\approx 0.67617 & \quad \text{GSE}
    \end{cases}
\label{eq:canonical values of ratio}
\end{equation}
where GOE, GUE and GSE are ensembles of the Hermitian random matrices model (see~\cite{Atas Bogomolny Giraud Roux (2013)} for their definitions).
It is known that the statistical distribution of adjacent energy level spacing of stadium and Sinal billiards obeys that of GOE.  

If the dimension of the Krylov space is finite, the Krylov operator complexity remains at a finite value and often saturates at late time. It is proposed that this late-time saturation value may be concerned with the erratic or non-erratic behavior of the Lanczos coefficients mentioned above \cite{Rabinovici:2021qqt}.
We will discuss this in Sec.~\ref{sec:7}.


\subsubsection{Krylov state complexity}
Similarly to the Krylov operator complexity, we can define Krylov complexity for quantum state \cite{Balasubramanian:2022tpr}. Note that the Schr\"odinger state $|\psi(t)\rangle = e^{-iHt}|\psi(0)\rangle$ is a linear combination of
\begin{equation}
    |\psi\rangle,~ H|\psi\rangle,~ H^2|\psi\rangle,~ \cdots\,,
\label{eq:basis state}
\end{equation}
where we denoted $|\psi(0)\rangle$ as $|\psi\rangle$. The (sub)space, $\mathcal{H}_{|\psi\rangle}$, which is spanned by \eqref{eq:basis state} is also called the Krylov space. Define $K_{|\psi\rangle}\equiv \dim\mathcal{H}_{|\psi\rangle}$. Using the natural inner product, we can orthonormalize \eqref{eq:basis state} by the Lanczos algorithm:
\begin{itemize}
\item[1.~] $b_0\equiv0\,, \quad |K_{-1}\rangle\equiv0$
\item[2.~] $|K_0\rangle \equiv |\psi(0)\rangle\,,\quad a_0=\langle K_0|H|K_0\rangle$
\item[3.~] For $n\geq1$: $|\mathcal{A}_n\rangle=(H-a_{n-1})|K_{n-1}\rangle-b_{n-1}|K_{n-2}\rangle$
\item[4.~] Set $b_n=\sqrt{\langle \mathcal{A}_n|\mathcal{A}_n\rangle}$
\item[5.~] If $b_n=0$ stop; otherwise set $|K_n\rangle=\frac{1}{b_n}|\mathcal{A}_n\rangle\,,~a_n=\langle K_n|H|K_n\rangle$, and go to step 3.
\end{itemize}
If $K_{|\psi\rangle}$ is finite, then this Lanczos algorithm ends with $b_{K_{|\psi\rangle}}=0$. The resulting orthonormal basis $\{|K_n\rangle\}_{n=0}^{K_{|\psi\rangle}-1}$ is called the Krylov basis. Notice that there are two sets of Lanczos coefficients $\{a_n\}$ and $\{b_n\}$ in this case.
Expanding the Schr\"odinger state $|\psi(t)\rangle$ in terms of the Krylov basis as
\begin{equation}
    |\psi(t)\rangle = \sum_{n=0}^{K_{|\psi\rangle}-1}\psi_n(t)|K_n\rangle\,,
\label{eq:state in the Krylov basis}
\end{equation}
and substituting \eqref{eq:state in the Krylov basis} into the Schr\"odinger equation, we have
\begin{equation}
    i\dot{\psi}_n(t) = a_n\psi_n(t)+b_{n+1}\psi_{n+1}(t)+b_n\psi_{n-1}(t)\,.
\label{eq:Krylov chain for state}
\end{equation}
The initial condition is $\psi_n(0)=\delta_{n0}$ by definition. The Krylov state complexity of the state $|\psi\rangle$ is defined as
\begin{equation}
    C_{|\psi\rangle}(t) \equiv \sum_{n=0}^{K_{|\psi\rangle}-1}n|\psi_n(t)|^2\,.
\label{eq:Krylov state complexity}
\end{equation}
In \cite{Balasubramanian:2022tpr}, the Krylov state complexity for the thermofield double (TFD) states in some chaotic systems was considered and found to show the characteristic peak and plateau structure, which \cite{Balasubramanian:2022tpr} presumed was universal for the chaotic systems. Recently, this structure is studied both analytically and numerically in \cite{Erdmenger:2023shk}. We will investigate this feature by considering the Krylov state complexity for states other than the TFD in chaotic quantum mechanics.

Unlike the case of Krylov operator complexity, there are two types of Lanczos coefficients, $a_n$ and $b_n$. 
We introduce variances of Lanczos coefficients $a_n$ and $b_n$ as
\begin{equation}
\begin{split}
    &\sigma^2_a \equiv {\rm Var}(x^{(a)}_i),\quad x^{(a)}_i\equiv \ln\left(\frac{a_{2i-1}}{a_{2i}}\right)\ ,\\
    &\sigma^2_b \equiv {\rm Var}(x^{(b)}_i),\quad x^{(b)}_i\equiv \ln\left(\frac{b_{2i-1}}{b_{2i}}\right)\ .
\end{split}
\label{eq:varianceab}
\end{equation}
It is known that these Lanczos coefficients can be expressed in the same manner as \eqref{eq:Lanczos coefficient in terms of moments}~\cite{SANCHEZDEHESA1978275}. We will later numerically study the variances similar to \eqref{eq:variance} and investigate the possible relation with chaoticity.


\subsection{Analytic examples of quantum mechanical complexity}
\label{sec:2-2}

Here, as an illustration, we present some explicit examples of the analytic calculation of the time evolution of Krylov complexity for operators and states in quantum mechanics. 

\subsubsection{Krylov operator complexity and inverse harmonic oscillator
}

As a typical operator we take a normalized Gaussian operator
\begin{align}
    {\cal O}_0 = \left(\frac{2\alpha}{\pi}\right)^{1/4} 
    \exp\left[-\frac{\hat{x}^2}{\alpha}\right]. 
\label{eq:GO}
\end{align}
Here $\hat{x}$ is the position operator, and $\alpha$ is a real positive constant. The reason why we take this Gaussian operator is that it is normalizable and thus well-defined: $({\cal O}_0|{\cal O}_0) = 1$ where the inner-product is the standard trace in the quantum mechanics Hilbert space, with $\int dx$ in the $x$-representation.

In this one-dimensional quantum mechanical system, the time evolution of the Krylov complexity for the Gaussian operator
\eqref{eq:GO} is quite nontrivial, as the commutator of
the operator with a given Hamiltonian generically generates infinite kinds of operators consisting of $\hat{x}$ and $\hat{p}$.
As a calculable example we consider the following Hamiltonian
\begin{align}
    H = e (\hat{x}\hat{p} + \hat{p}\hat{x}) 
    \label{eq:xpH}
\end{align}
where $e$ is a real constant.
Taking the commutator of the Gaussian operator ${\cal O}_0$ 
with the Hamiltonian $n$ times generates the following form of the operator
\begin{align}
    {\cal O}_n = 
    \left(\hat{x}^{2n} + \mbox{Lower polynomials in $\hat{x}^2$}\right)\exp[-\hat{x}^2/\alpha],
\end{align}
up to a constant normalization.
Since the Krylov basis is given as an orthonormal set of
operators, we conclude that such a set of the form above is
unique and given by the Hermite polynomials, 
\begin{align}
    {\cal O}_n = 
\left(\frac{2}{\pi\alpha}\right)^{1/4}\frac{1}{\sqrt{(2n)!}\, 2^n} H_{2n}(\sqrt{2/\alpha}\, \hat{x})\exp[- \hat{x}^2/\alpha].
\end{align}
In other words, the Krylov basis in this case is identical to
that of the energy eigenfunctions of a harmonic oscillator.
Using this Krylov basis, we obtain the Lanczos coefficient as
\begin{align}
b_n & =    ({\cal O}_n|{\cal L}|{\cal O}_{n-1}) 
= {\rm Tr}\left[{\cal O}^\dagger_{n} [H, {\cal O}_{n-1}]\right]
\nonumber \\
& =e \sqrt{\frac{2}{\pi\alpha}}\sqrt{n(n-1/2)} .
\end{align}
Therefore at the large $n$ the Lanczos coefficient grows linearly in $n$.

For this form of the Lanczos coefficient, the time evolution of the
Krylov complexity is explicitly obtained. Applying the 
formula given in \cite{Parker:2018yvk}, we find
\begin{align}
    {\cal C}(t) = \frac12 \sinh^2 \left[
     e \sqrt{\frac{2}{\pi\alpha}} t 
    \right]\sim \exp\left[
     2e \sqrt{\frac{2}{\pi\alpha}} t\right]  .
\end{align}
This grows exponentially for $t\gg \sqrt{\alpha}/e$.

In \cite{Parker:2018yvk}, it was argued that the exponential behavior is
related to chaos. However, the Hamiltonian \eqref{eq:xpH}
is too simple that it is integrable and does not give any chaos. Then where can we locate a possible 
relation between the exponential growth of the complexity and the dynamical behavior of the current system?
In fact, 
we can spot a behavior typical to chaotic systems:
it is an exponential 
run-away behavior of the trajectory. 
The classical Hamilton equation is
\begin{align}
    \frac{d}{dt}x(t)= 2 e \, x(t) , \quad
    \frac{d}{dt}p(t)= -2 e \, p(t) , 
\end{align}
whose solution is exponentially growing in time,
\begin{align}
    x(t) = x(0) e^{2e t} , \quad 
    p(t) = p(0) e^{-2e t} .
    \label{eq:exp2et}
\end{align}
Thus the classical motion is exponential in time.
The origin of this exponential behavior can be more easily understood if one makes a canonical transformation\footnote{The generating function of the canonical transformation is
$ W = \frac12 x^2 -\frac12 X^2 - \sqrt{2} x X$.   
}
\begin{align}
    X \equiv \frac{1}{\sqrt{2}}(x-p), \quad
        P \equiv \frac{1}{\sqrt{2}}(x+p).
\end{align}
The transformed Hamiltonian is
\begin{align}
    H = e(P^2-X^2)
\end{align}
which is the inverse harmonic oscillator, giving 
the exponential growth in the motion.
This inverse harmonic oscillator is not chaotic while has been used for the demonstration of scrambling in quantum models \cite{Hashimoto:2020xfr,Bhattacharyya:2020art}.
This suggests the exponential behavior of the 
Krylov complexity of the present integrable system.

\subsubsection{Krylov state complexity for a free particle and a harmonic oscillator
}

As another illustrating analytic examples, we calculate the Lanczos coefficients of the Krylov state complexity of Gaussian quantum mechanical states
for the case of a single free particle and the case of a harmonic oscillator.\footnote{See the similar example of a particle moving in the Heisenberg-Weyl group calculated in Sec.~V.C of \cite{Balasubramanian:2022tpr}. Our derivation differs from that of \cite{Balasubramanian:2022tpr}, so paving another way to look at simple quantum mechanical examples.} 
We will see that they grow linearly in $n$, as $a_n \sim 2 b_n \propto n$ at a large $n$. We also find that the complexity time evolution for the free particle case goes as $t^2$.

Let us start with the free particle in one spatial dimension (the case of the harmonic oscillator is analyzed precisely in the same manner and will be mentioned later). The Hamiltonian is
\begin{align}
    H = \frac{1}{2m}p^2
\end{align}
where $p$ is the momentum operator, and we consider the wave function at $t=0$ as
\begin{align}
    \langle p | \psi \rangle = \left(\frac{2\alpha}{\pi}\right)^{1/4} \exp[-\alpha p^2] 
    \label{eq:gaussian}
\end{align}
where $\alpha$ is a real positive number to parameterize the width of the initial wave function in $p$ space.
The Lanczos algorithm for the Krylov state complexity \cite{Balasubramanian:2022tpr} is applied to this initial state.
In the following we prove that the algorithm generates Hermite polynomials.

The algorithm basically consists of the following procedure: first, multiply the Hamiltonian to the $n$-th state $|K_n\rangle$, and second, subtract a linear combination of $|K_n\rangle$ and $|K_{n-1}\rangle$ to make the generated state be orthogonal to $|K_n\rangle$ and $|K_{n-1}\rangle$, and third, normalize the generated state.
This procedure automatically makes sure that generated states are orthogonal to each other and normalized.
It is easy to notice here that in fact if we start with the Gaussian state \eqref{eq:gaussian} it automatically
generates states represented by Hermite polynomials $H_{2n}$ times the Gaussian. The reason is that the Hamiltonian is simply $p^2$ so the $n$-th state $|K_n\rangle$ need to be of the form
\begin{align}
    \langle p | K_n \rangle = \left(p^{2n} + \mbox{Lower polynomials in $p^2$}\right)\exp[-\alpha p^2], 
\end{align}
up to an overall factor. The only mutually orthogonal set of functions of this form is the Hermite polynomials,
as is well known for energy eigenstates of a harmonic oscillator.
A simple calculation leads to the result
\begin{align}
    \langle p | K_n \rangle = \left(\frac{2\alpha}{\pi}\right)^{1/4}\frac{1}{\sqrt{(2n)!} 2^n} H_{2n}(\sqrt{2\alpha}\, p)\exp[-\alpha p^2].
    \label{eq:Kryfree}
\end{align}
With this explicit Krylov basis, we can easily compute the Lanczos coefficients. 
\begin{align}
    a_n & \equiv \langle K_n | H | K_n\rangle
    = \frac{1}{2m}
    \left(\sqrt{\frac{2\alpha}{\pi}}\frac{1}{(2n)! 2^{2n}}\right) 
    \int dp \, p^2 \left(H_{2n}(\sqrt{2\alpha}\, p)\right)^2\exp[-2\alpha p^2]
\nonumber \\
& = \frac{1}{2m\alpha}\left(n + \frac14\right),
\\
b_n & \equiv \langle K_{n-1} | H | K_n\rangle 
\nonumber \\
&= \frac{1}{2m}
    \left(\sqrt{\frac{2\alpha}{\pi}}\frac{1}{(2n)! 2^{2n}}\right) 2 \sqrt{2n (2n-1)}
    \int dp \, p^2 H_{2n}(\sqrt{2\alpha}\, p)H_{2n-2}(\sqrt{2\alpha}\, p)\exp[-2\alpha p^2]
\nonumber \\
&= 
\frac{1}{4m\alpha}\sqrt{n\left(n + \frac12\right)} .
\end{align}
At a large $n$, we find 
\begin{align*}
    b_n \simeq \frac12 a_n \simeq \frac{1}{4m\alpha}n .
\end{align*}
Therefore the Lanczos coefficients grows linearly in $n$ and satisfies a relation $a_n \simeq 2 b_n$.

As another example, let us consider the harmonic oscillator with the Hamiltonian
\begin{align}
    H = \frac{1}{2m}p^2 + \frac{\omega}{2}x^2 .
    \label{eq:harmonicH}
\end{align}
We find that the logic for the free particle also applies to this harmonic oscillator precisely, and the Krylov basis is just again given exactly by \eqref{eq:Kryfree}. The reason is that the effect of the harmonic oscilator potential $x^2$ on any function of the form \eqref{eq:Kryfree} is exactly the same as the multiplication of $p^2$ 
with some addition of a constant. 
The Lanczos coefficients are obtained in a similar calculation but with the harmonic oscillator Hamiltonian,
\begin{align}
    a_n = \left(\frac{1}{2m\alpha}+2 \omega \alpha\right) \left(n + \frac14\right),
\quad 
b_n = 
\frac12 \left(\frac{1}{2m\alpha}-2 \omega \alpha\right) \sqrt{n\left(n + \frac12\right)} .
\end{align}
At a large $n$, we again find the linear growth of the Lanczos coefficients as
\begin{align}
    b_n \simeq \left(\frac12\frac{1-4m\omega\alpha^2}{1+4m\omega\alpha^2} \right)
     a_n \simeq \frac12 \left(\frac{1}{2m\alpha}-2 \omega \alpha\right) n .
\end{align}
Note that for $\alpha= 1/\sqrt{4m\omega}$ all the Lanczos coefficients disappear. This is because the
initial wave function becomes the ground state wave function for that value of $\alpha$ and thus the Krylov
basis consists just only of the ground state wave function.

Finally let us calculate the time evolution of the Krylov state complexity for the case of the free particle. The time evolution of the wave function is
\begin{align}
    \langle p | \psi(t)\rangle = 
    \left(\frac{2\alpha}{\pi}\right)^{1/4}
    \exp\left[\frac{-i p^2}{2m}
    {t}
    \right] \exp[-\alpha p^2] .
\end{align}
Then we find the Krylov components of this wave function as 
\begin{align}
    \langle K_n | \psi(t)\rangle
    =\left(\frac{2\alpha}{\pi}\frac{1}{(2n)! \, 2^{2n}}\right)^{1/2}\int \! dp \, 
    H_{2n}(\sqrt{2\alpha} p)
    \exp\left[\frac{-i p^2}{2m}
    {t}
    -2\alpha p^2
    \right] .
\end{align}
Putting $\tilde{t}\equiv t/(4m\alpha)$, the explicit integration gives
\begin{align}
    \bigm| \langle K_n | \psi(t)\rangle \bigm|^2
    =\frac{(2n-1)!!}{n!\, 2^n} \frac{\tilde{t}^{2n}}{\left(1+\tilde{t}^2\right)^{n+(1/2)}}.
\end{align}
Using this, the Krylov state complexity is calculated 
as\footnote{The nontrivial summation in \eqref{compsum} can be performed in the following manner. First one should note that the consistency of the probability conservation $1 = \sum_n \bigm| \langle K_n | \psi(t)\rangle \bigm|^2$ is proven by the expansion
\begin{align}
    \frac{1}{\sqrt{1-y}} = \sum_n \frac{(2n-1)!!}{n!\, 2^n} y^n 
\end{align}
and the identification $y \equiv (\tilde{t})^2/(1 + (\tilde{t})^2)$. Then the desired summation \eqref{compsum}
is rephrased  in terms of $y$ as
\begin{align}
    \sum_n n \frac{(2n-1)!!}{n!\, 2^n} y^n 
    =   y \frac{\partial}{\partial y} \sum_n  \frac{(2n-1)!!}{n!\, 2^n} y^n  = y \frac{\partial}{\partial y}\frac{1}{\sqrt{1-y}} = \frac{y}{2(1-y)^{3/2}} .
\end{align}
} 
\begin{align}
    {\cal C}(t) \equiv \sum_{n=0}^\infty n
    \bigm| \langle K_n | \psi(t)\rangle \bigm|^2 = \frac12 \left(\tilde{t}\right)^2 = \frac{1}{32 (m \alpha)^2} t^2.
    \label{compsum}
\end{align}
Thus the Krylov state complexity of the Gaussian state for a free particle evolves as $t^2$, with the typical time scale $m\alpha$
where $m$ is the mass of the particle and $\sqrt{\alpha}$ is the spatial Gaussian width.
In summary, the Lanczos coefficients are linearly growing, while the system is not chaotic, and the Krylov state complexity is not exponentially growing in time. 

In this subsection we have performed analytic evaluation of the Krylov complexities. Unfortunately, systems in which the Krylov complexity can be analytically calculated are quite limited, and in particular, chaotic systems allow only numerical evaluations. In the next section, we explain our numerical method.


\section{Preliminary for numerical calculations}
\label{sec:3}
We consider numerical calculations of Krylov complexity.
Since the dimension of the Hilbert space is often infinite, some regularization is necessary to perform numerical calculations. In our later calculations, we consider only a finite number of levels and ignore the others. For a Hamiltonian of the form
\begin{equation}
    H = p_1^2+p_2^2+V(x,y),
\label{eq:Hamiltonian}
\end{equation}
we perform our numerical calculation by the following procedure.\footnote{In \cite{Guo:2022hui}, finite-temperature Krylov operator complexity is numerically calculated.} We assume that the system is bounded from below. Generalization to systems with more degrees of freedom is straightforward.

\subsection*{Krylov operator complexity}
Suppose that we want to calculate the Krylov complexity for the momentum operator $p_1$.\footnote{
See App.~\ref{app:3} for other choice of the initial operator.
} For that purpose, we first numerically solve the Schr\"odinger equation
\begin{equation}
    H|n\rangle = E_n|n\rangle\,.
\end{equation}
We define energy eigenfunctions as $\phi_n(x,y)\equiv \langle x,y|n\rangle ~(n=1,2,\cdots)$. We compute them by numerical calculations.\footnote{To find the solutions numerically, we used NDEigensystem of Mathematica. The numerical solution depends on the fineness of discretization of the domain of the potential. See App.~\ref{app:1}.}
The matrix representation of the position operator $x$ is given by
\begin{equation}
    x_{mn}= \langle m | \,x\, |n\rangle = \int dx\,dy~ \phi_m^*(x,y)x\phi_n(x,y)\ .
\end{equation}
We can evaluate above by the numerical integration.
Although we can also obtain the matrix representation of the momentum operator using derivatives of eigenfunctions, the derivative of the numerical solution tends to lose  numerical accuracy. 
Thus, we employ the technique used in \cite{Hashimoto:2017oit}, as described below. With \eqref{eq:Hamiltonian}, the momentum operator $p_1$ can be expressed as
\begin{equation}
    p_1 = \frac{i}{2}[H,x]\,.
\label{eq:momentum}
\end{equation}
Then using \eqref{eq:momentum}, we can find the matrix element of $p_1$ as
\begin{equation}
    P_{mn} \equiv \langle m | \,p_1\, |n\rangle = \frac{i}{2}L_{mn}x_{mn}\ ,
\end{equation}
where 
\begin{equation}
    L_{mn} \equiv E_m-E_n \ .
\end{equation}
We now define the truncated momentum operator as
\begin{equation}
 P=\sum_{m,n=1}^{N_\textrm{max}} |m\rangle 
 P_{mn}
 \langle n|
=
\begin{pmatrix}
P_{11} & P_{12} & \cdots & P_{1 N_\textrm{max}} \\
P_{21} & P_{22} &\cdots & P_{2 N_\textrm{max}} \\
 \vdots & \vdots & & \vdots \\
P_{N_\textrm{max}1} & P_{N_\textrm{max}2} &\cdots & P_{N_\textrm{max} N_\textrm{max}} 
\end{pmatrix}
\ ,
\label{truncP}
\end{equation}
where $N_{\rm max}$ is an truncation number.
Then, we perform the Lanczos algorithm for $P$.
Technically, in the numerical calculation, the naive Lanczos algorithm does not work well near the end of the algorithm due to numerical errors.
In our numerical calculation, we use the full orthogonalization method \cite{Rabinovici:2020ryf,parlett1998symmetric} instead of the naive Lanczos algorithm to ensure orthogonality.
In the current case, the algorithm is as follows:
\begin{itemize}
\item[1.~] $\mathcal{O}_0\equiv P/\|P\|$, where $\|P\|\equiv\sqrt{(P|P)}$ with the inner product defined by \eqref{eq:inner product}.
\item[2.~] For $n\geq1$: $\mathcal{A}_n=L \odot \mathcal{O}_{n-1}$, where $\odot$ represents the Hadamard product.\footnote{
The Hadamard product is defined by $(A\odot B)_{ij} = A_{ij}B_{ij}$ (no sum over) for matrices $A$ and $B$. 
}
\item[3.~] Replace as $\mathcal{A}_n\to\mathcal{A}_n-\sum_{m=0}^{n-1}\mathcal{O}_m(\mathcal{O}_m|\mathcal{A}_n)$.
\item[4.~] Repeat step 3.
\item[5.~] Set $b_n=\|\mathcal{A}_n\|$.
\item[6.~] If $b_n=0$ stop; otherwise set $\mathcal{O}_n=\mathcal{A}_n/b_n$ and go to step 2.
\end{itemize}
Let $K_P$ be the dimension of the Krylov space, which is determined by $b_{K_P}=0$.\footnote{Since the algorithm is run numerically, $b_n$ does not become exactly zero. In each calculation, one needs to look for an appropriate $K_P$.} Then, we numerically solve \eqref{eq:Krylov chain for operator} and obtain the Krylov operator complexity \eqref{eq:def Krylov operator complexity}. Also, the variance \eqref{eq:variance} can be obtained from the Lanczos coefficients. In our calculation, the variance \eqref{eq:variance} is evaluated for the region $N_{\rm max}^2/200\leq i\leq N_{\rm max}^2/40$.

\subsection*{Krylov state complexity}
Let us describe how to calculate the Krylov complexity for a state $|\psi\rangle$. Similarly to the calculation of Krylov operator complexity, Krylov state complexity can be calculated by expanding $|\psi\rangle$ in terms of the energy eigenstates $|n\rangle$ as a basis. Solving the Schr\"odinger equation numerically, we can prepare the state vector as
\begin{equation}
    \Psi \equiv (\langle1|\psi\rangle,\cdots,\langle N_{\rm max}|\psi\rangle)^{\rm T}\,,
\end{equation}
where $N_{\rm max}$ is an arbitrarily chosen truncation number. Each component is obtained by the integral
\begin{equation}
    \langle n|\psi\rangle = \int dx\,dy\,\phi_n^*(x,y)\psi(x,y)\,,
\end{equation}
where $\psi(x,y)\equiv \langle x,y|\psi\rangle$. In our numerical calculation, however, we simply use the following vector as the initial state:
\begin{equation}
    \Psi \equiv \left(\frac{1}{\sqrt{N_{\rm max}}},\cdots,\frac{1}{\sqrt{N_{\rm max}}}\right)^{\rm T}\,,
    \label{flatinitialstate}
\end{equation}
while the other initial states give qualitatively similar numerical results.
Then, we perform the Lanczos algorithm for the state $\Psi$.
As we did for the operator case, we need to modify the naive algorithm to avoid numerical errors during the orthogonalization.
Defining
\begin{equation}
    D\equiv {\rm diag}(E_1,\cdots,E_{N_{\rm max}})\,,
\end{equation}
the algorithm goes as follows \cite{parlett1998symmetric}:
\begin{itemize}
\item[1.~] $b_0\equiv0\,, \quad K_{-1}\equiv0$
\item[2.~] $K_0 \equiv \Psi\,,\quad a_0=K_0^\dagger D K_0$
\item[3.~] For $n\geq1$: $\mathcal{A}_n=(D-a_{n-1})K_{n-1}-b_{n-1}K_{n-2}$
\item[4.~] Replace as $\mathcal{A}_n\to\mathcal{A}_n-\sum_{m=0}^{n-1}(\mathcal{A}_m^\dagger\Psi)\mathcal{A}_m$.
\item[5.~] Set $b_n=\sqrt{\mathcal{A}_n^\dagger \mathcal{A}_n}$.
\item[6.~] If $b_n=0$ stop; otherwise set $K_n=\frac{1}{b_n}\mathcal{A}_n\,,~a_n=K_n^\dagger D K_n$, and go to step 3.
\end{itemize}
Let $K_\Psi$ be the dimension of the Krylov space, which is determined by $b_{K_\Psi}=0$. Then, we numerically solve \eqref{eq:Krylov chain for state} and obtain the Krylov state complexity \eqref{eq:Krylov state complexity}.

We set $N_\text{max}=500$ in numerical calculations of Krylov state complexity shown in the following sections.
The variance in \eqref{eq:varianceab} is calculated for $a_n, b_n$ with $50\leq n\leq 250$, where this range is chosen to avoid transient feature at small $n$ and possible numerical error at large $n \sim N_\text{max}$.


\section{Krylov operator complexity in the stadium billiard}
\label{sec:4}

\begin{figure}[t]
\centering
    \includegraphics[width=7cm]{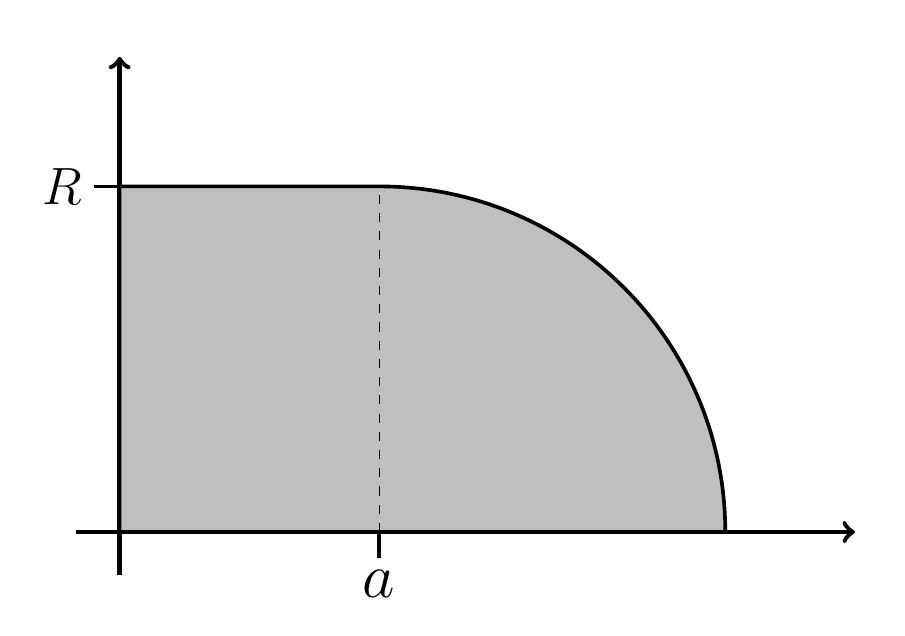}
    \caption{Geometry of the stadium billiard. Dirichlet boundary conditions are imposed on the boundaries.}
    \label{fig:stadium shape}
\end{figure}

\begin{figure}[t]
\centering
    \subfigure[The Lyapunov exponent as a function of $a/R$. The area of the billiard and the velocity of the particle are normalized to the unity.]
     {\includegraphics[width=7cm]{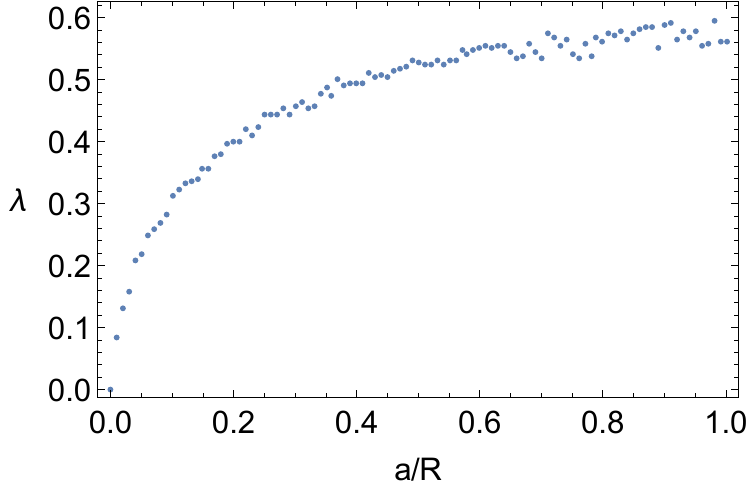} \label{fig:stadium Lyapunov}}
    \hspace{3mm}
    \subfigure[The ratio $\langle\tilde{r}\rangle$ as a function of $a/R$. The number of levels used in this calculation is 100. The green and orange lines correspond to the values for the Poisson and 
 the Wigner-Dyson statistics respectively.]
     {\includegraphics[width=7cm]{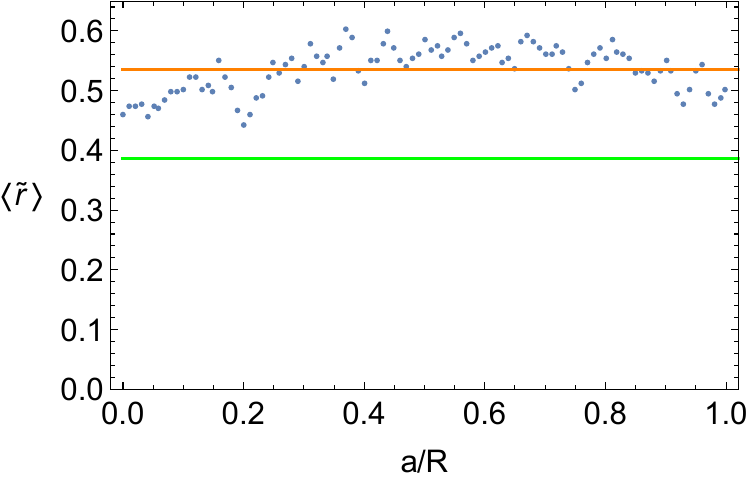} \label{fig:stadium ratio}}
    \caption{The $a/R$ dependence of the Lyapunov exponent and the ratio.}
\label{fig:stadium Lyapunov and ratio}
\end{figure}

It is well known that the stadium (circular) billiard is chaotic (integrable) both classically and quantum mechanically \cite{Benettin:1978,McDonald:1979zz,Casati:1980}.
In Fig.~\ref{fig:stadium shape}, we show the geometry of the stadium billiard. The shape of the billiard is determined by $a/R$. Considering the one-parameter deformation of the billiard, we study the integrable/chaos transition and correlation between the variance of Lanczos coefficients and classical/quantum chaoticity.

Classical chaoticity can be measured by the Lyapunov exponent. In Fig.~\ref{fig:stadium Lyapunov}, we show the $a/R$ dependence of the Lyapunov exponent, which is consistent with \cite{Benettin:1978}. In the calculation, we set the area of the billiard and the velocity of the particle to the unity. When $a/R=0$, the system becomes an integrable circular billiard. As $a/R$ increases, so does the Lyapunov exponent and the system becomes chaotic. 

Quantum chaoticity is traditionally distinguished by the level statistics, or more conveniently by the ratio \eqref{eq:ratio}. In Fig.~\ref{fig:stadium ratio}, we show the $a/R$ dependence of the ratio.\footnote{In this calculation, we used 100 levels. The result becomes clearer if one uses more levels since the statistics become better.} The orange line corresponds to the value for the Poisson statistics and the green line corresponds to that for the Wigner-Dyson (GOE) statistics \eqref{eq:canonical values of ratio}. We can see that the transition similar to the classical case takes place as the $a/R$ increases, which is consistent with \cite{McDonald:1979zz}.

\subsection{The early time dependence of Krylov operator complexity}
\label{sec:4-1}
\begin{figure}[t]
\centering
    \includegraphics[width=15cm]{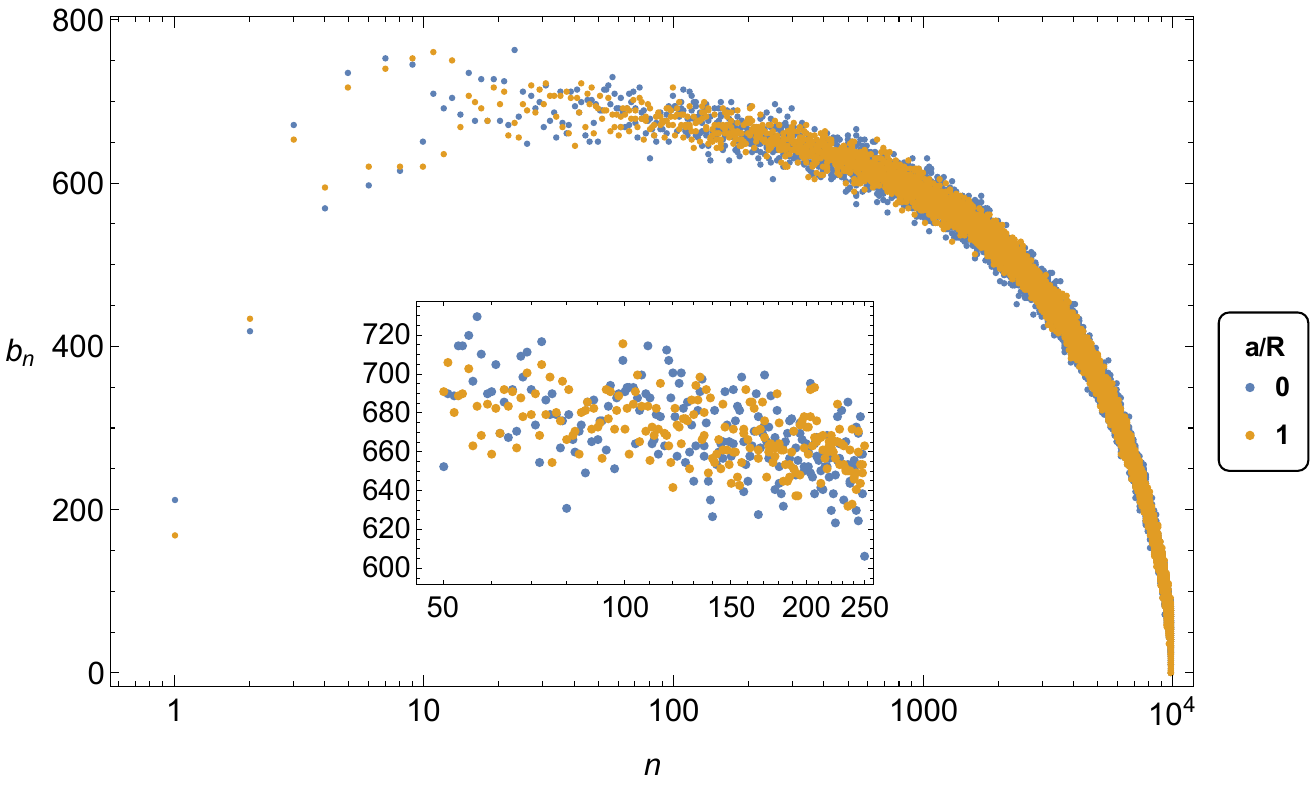}
    \caption{The Lanczos coefficients for the truncated momentum operator $P$ in stadium billiards with $a/R=0$ (blue dots) and $a/R=1$ (orange dots). Note that the horizontal axis is in log scale. The inset is the enlarged version, where the data are used to calculate the variance.}
    \label{fig:stadium op Lanczos}
\end{figure}

Using the method in Sec.~\ref{sec:3}, we numerically compute Krylov operator complexity for the truncated momentum operator $P$ defined in Eq.~(\ref{truncP}), with truncation $N_{\rm max}=100$.\footnote{This number of levels for our calculation was decided by the numerical cost.
The numerical results (e.g.~variance of Lanczos coefficients) depend on $N_{\rm max}$, while their physical properties are expected to be universal for any large $N_{\rm max}$. 
See App.~\ref{app:2} for details.
The area of the billiard is normalized to the unity.} In Fig.~\ref{fig:stadium op Lanczos}, we show the Lanczos coefficients for $a/R=0$ and $a/R=1$. Although the former case is integrable and the latter is chaotic, the initial behaviors of the corresponding Lanczos coefficients are almost identical. We identify the dimension of the Krylov space $K_P$ as $K_P=9900$. Note that the horizontal axis in Fig.~\ref{fig:stadium op Lanczos} is in log scale.

\begin{figure}[t]
\centering
    \includegraphics[width=8cm]{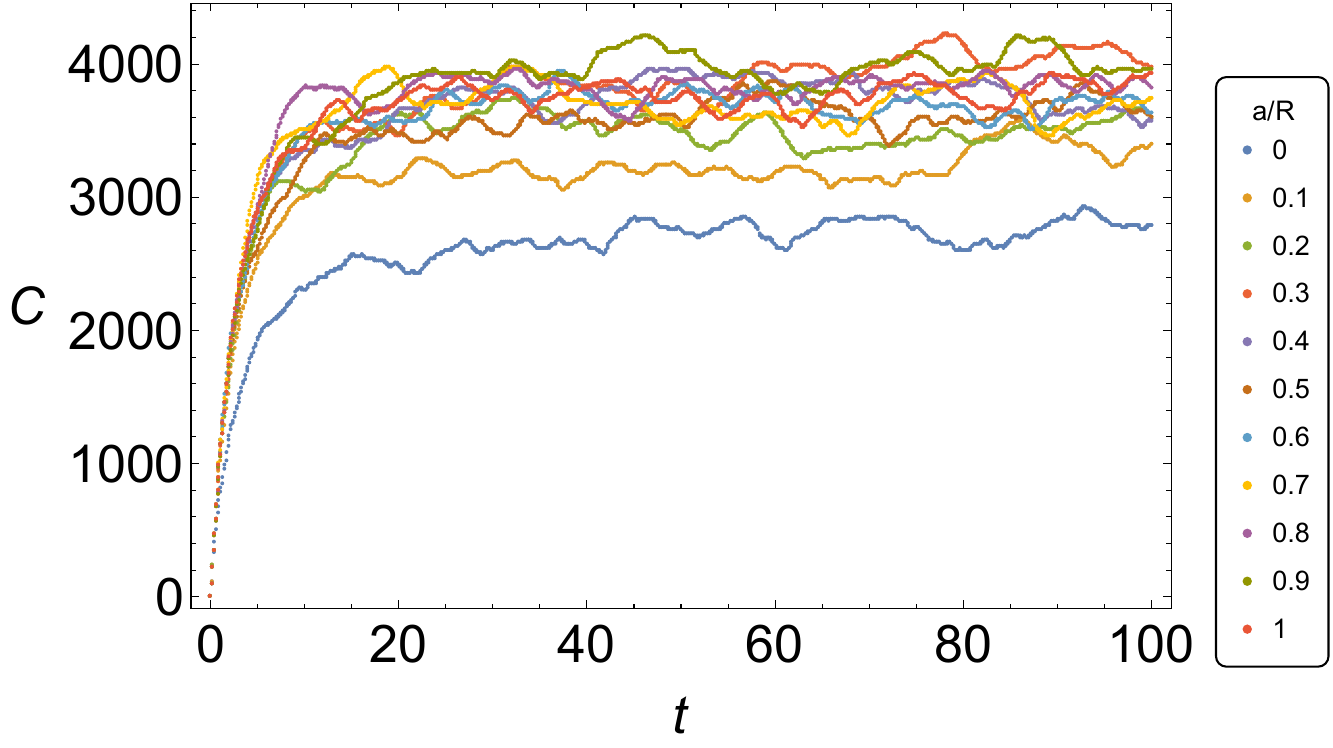}
    \caption{The time dependence of Krylov operator complexity for various values of $a/R$.}
    \label{fig:stadium KOC}
\end{figure}

In Fig.~\ref{fig:stadium KOC}, we show the Krylov operator complexities as functions of $t$ for the stadium billiards with $a/R=0,0.1,0.2,\cdots,1$.
While the stadium billiard with $a/R>0$ is chaotic, the early time growth of the Krylov operator complexity is not apparently exponential.\footnote{
At least the data obtained did not confirm a clear exponential growth,
which may be due to the lack of the typical time scale since we used an infinite-temperature inner product in our analysis. 
}

\begin{figure}[t]
\centering
    \includegraphics[width=7cm]{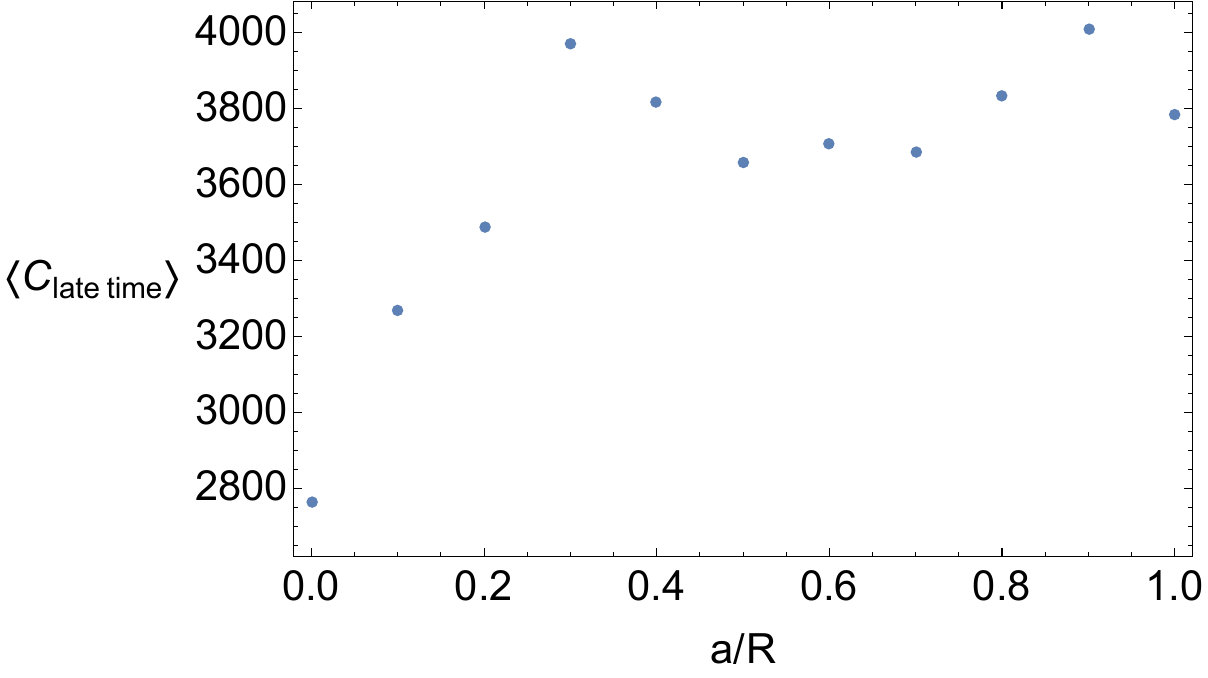}
    \caption{The $a/R$ dependence of the late-time value of Krylov operator complexity.}
    \label{fig:stadium asym KOC}
\end{figure}

Figure~\ref{fig:stadium KOC} shows that $C(t)$ saturates by $t\lesssim 30$ and reaches asymptotic values that depend on $a/R$.
In Fig.~\ref{fig:stadium asym KOC}, we show the dependence of saturation value of $C(t)$ (the average of $C(t)$ taken over $40\leq t \leq 100$). We discuss its implications in Sec.~\ref{sec:7}.

\subsection{Correlation between Lanczos coefficients and chaos}
\label{sec:4-2}

\begin{figure}[t]
\centering
    \includegraphics[width=7cm]{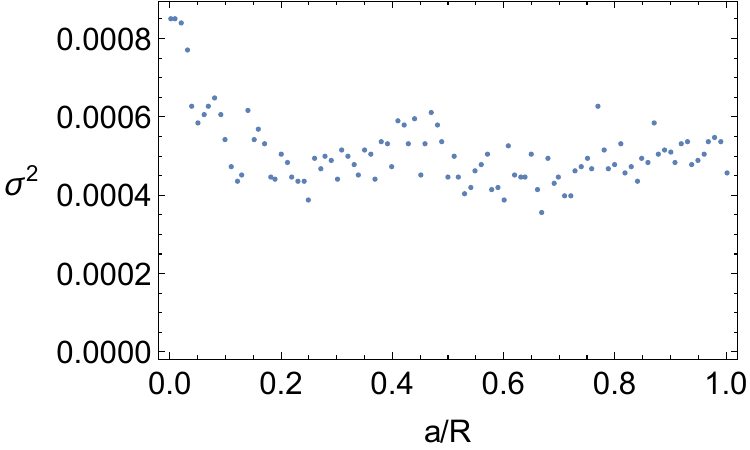}
    \caption{The variance $\sigma^2$ as a function of $a/R$.}
    \label{fig:stadium variance}
\end{figure}

\begin{figure}[t]
    \centering
    \subfigure[$\lambda$ vs $\sigma^2$]
    {\includegraphics[width=7cm]{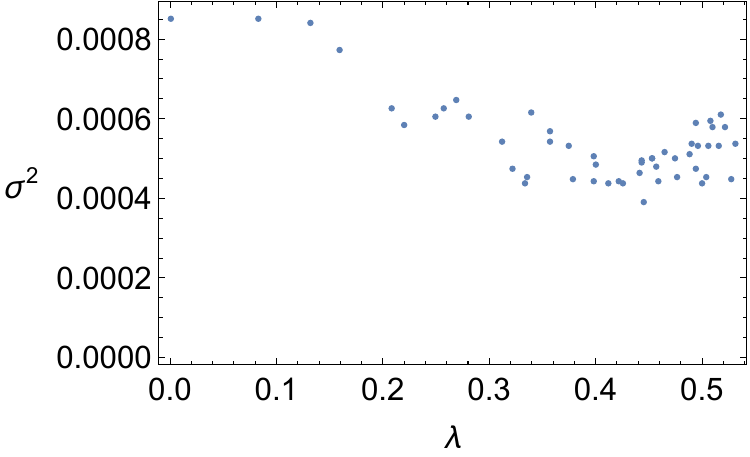}\label{fig:stadium_op_Lyapunov_vs_variance}}
    \hspace{5mm}
    \subfigure[$\langle\tilde{r}\rangle$ vs $\sigma^2$]{\includegraphics[width=7cm]{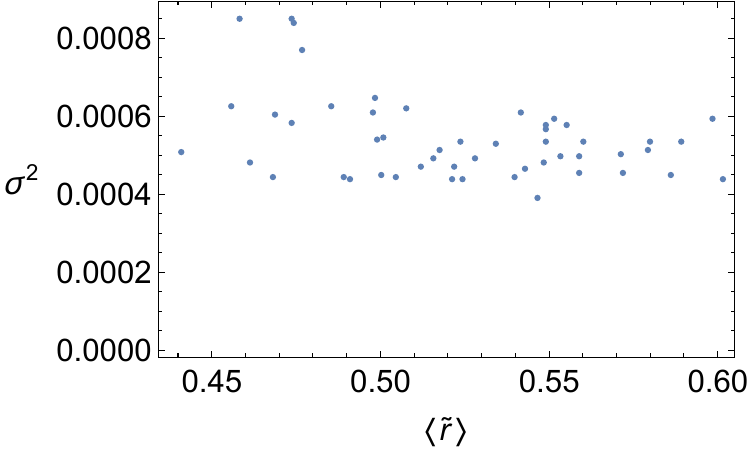}\label{fig:stadium_op_ratio_vs_variance}}
    \subfigure[$\lambda$ vs $\langle\tilde{r}\rangle$]{\includegraphics[width=7cm]{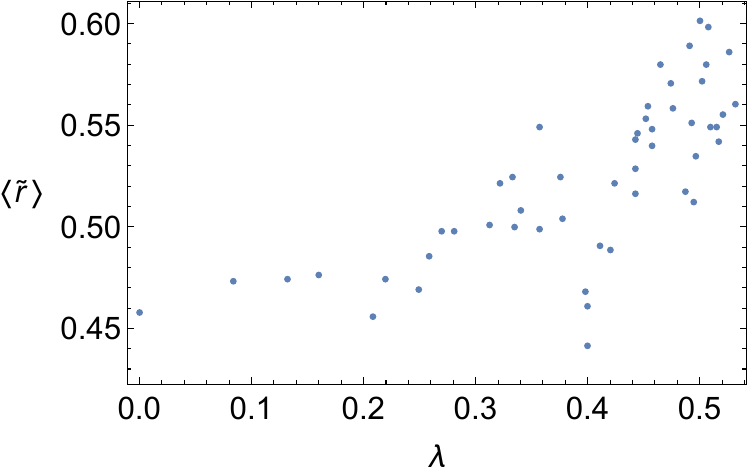}\label{fig:stadium_op_Lyapunov_vs_ratio}}
    \caption{The scatter plots of (a) the Lyapunov exponents and the variances, (b) the ratios and the variances, (c) the Lyapunov exponents and the ratios. The points are sampled from $0\leq a/R \leq 0.5$.}
    \label{fig:Stadium_scatter_plot}
\end{figure}

The Lanczos coefficients for $a/R=0$ distribute broader compared to $a/R=1$ in Fig.~\ref{fig:stadium op Lanczos}. In Fig.~\ref{fig:stadium variance} we show the variance of Lanczos coefficients \eqref{eq:variance} as a function of $a/R$. The variance becomes larger in the integrable regime compared to the chaotic regime. To compare with other chaos indicators, we show scatter plots in Fig.~\ref{fig:Stadium_scatter_plot}, where the points are sampled from $0\leq a/R \leq 0.5$.\footnote{Changing the sampling region, for example, to $0\leq a/R \leq 1$ does not affect the correlation much.} We can see that there are correlations between $\lambda,\sigma^2,$ and $\langle\tilde{r}\rangle$. Quantitatively, a correlation between two sets of data $A$ and $B$ can be evaluated by the correlation coefficient defined by
\begin{equation}
    \frac{{\rm E}[(A-{\rm E}[A])(B-{\rm E}[B])]}{\sqrt{{\rm E}[(A-{\rm E}[A])^2]\,{\rm E}[(B-{\rm E}[B])^2]}}\,,
\label{eq:correlation coefficient}
\end{equation}
where $E[\,\cdot\,]$ means the average value. If there is no correlation between two sets of data, the correlation coefficient will be close to zero.\footnote{The correlation coefficient takes values between $-1$ and $1$. If the correlation coefficient is close to $1$ or $-1$, a linear relationship is likely to exist between the two sets of data.}
\noindent In Table~\ref{table:stadium_correlation_coefficients}, we show the calculated correlation coefficients. Since the coefficients are far from zero, there should be some correlations between $\lambda,\sigma^2,$ and $\langle\tilde{r}\rangle$.
We can see that the quantity $\sigma^2$ is as good as a possible indicator of quantum chaos as the ratio $\langle\tilde{r}\rangle$.

\begin{table}[t]
\begin{center}
    \begin{tabular}{|c|l|}
    \hline
    $\lambda$ vs $\sigma^2$ & -0.720372\\
    $\langle \tilde{r}\rangle$ vs $\sigma^2$ & -0.391709\\
    $\lambda$ vs $\langle \tilde{r}\rangle$ & \hphantom{-}0.741396 \\ \hline
    \end{tabular}
    \caption{The correlation coefficients between $\lambda$, $\langle \tilde{r}\rangle$, $\sigma^2$ for the stadium billiards.}
    \label{table:stadium_correlation_coefficients}
\end{center}
\end{table}


\section{Krylov state complexity in the stadium billiard}
\label{sec:5}

In this section, we summarize the numerical results on the Krylov state complexity for the stadium billiard.
In Sec.~\ref{sec:4} we found that the Lyapunov exponent, variation of the Lanczos coefficients, and the ratio $\langle \tilde r \rangle$ are correlated with each other for the Krylov operator complexity. We will observe similar tendency for the state complexity.
The time dependence of the state complexity, however, shows behavior qualitatively different from that of the operator complexity.

\begin{figure}[t]
\centering
    \subfigure[$a_n$]{\includegraphics[height=4.45cm]{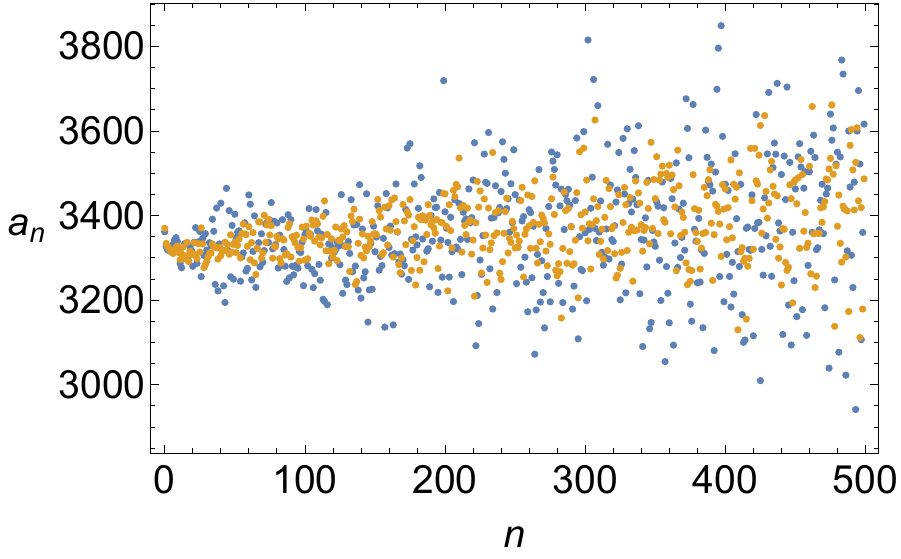}}
    \subfigure[$b_n$]{\includegraphics[height=4.45cm]{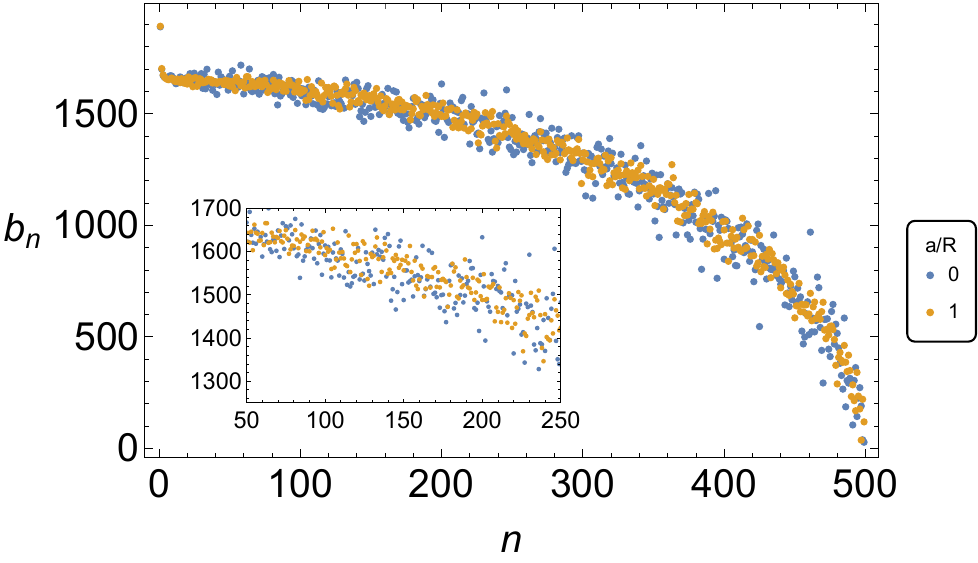}}
    \caption{The Lanczos coefficients of the state complexity for equally-distributed initial state in stadium billiards with $a/R=0$ (blue dots) and $a/R=1$ (orange dots). The inset is the enlarged version, where the data are used to calculate the variance.}
    \label{fig:stadium state Lanczos}
\end{figure}

Figure~\ref{fig:stadium state Lanczos} shows the Lanczos coefficient $a_n, b_n$ for the state complexity on the stadium billiard
in the integrable case ($a/R=0$) and chaotic case ($a/R=1$).
We find tendency similar to that of the operator complexity, that is, the variation of the Lanczos coefficients becomes small in a chaotic system.
We also find that not only $b_n$ but also $a_n$, which appears only for the state complexity, has a smaller variation when the system is chaotic.
Later in Sec.~\ref{sec:5-2}, we will quantitatively show the correlation of the variation of the Lanczos coefficients with various indicators of the chaos.

\begin{figure}[t]
\centering
    \subfigure[The time dependence of Krylov state complexity for various values of $a/R$.]
     {\includegraphics[width=7cm]{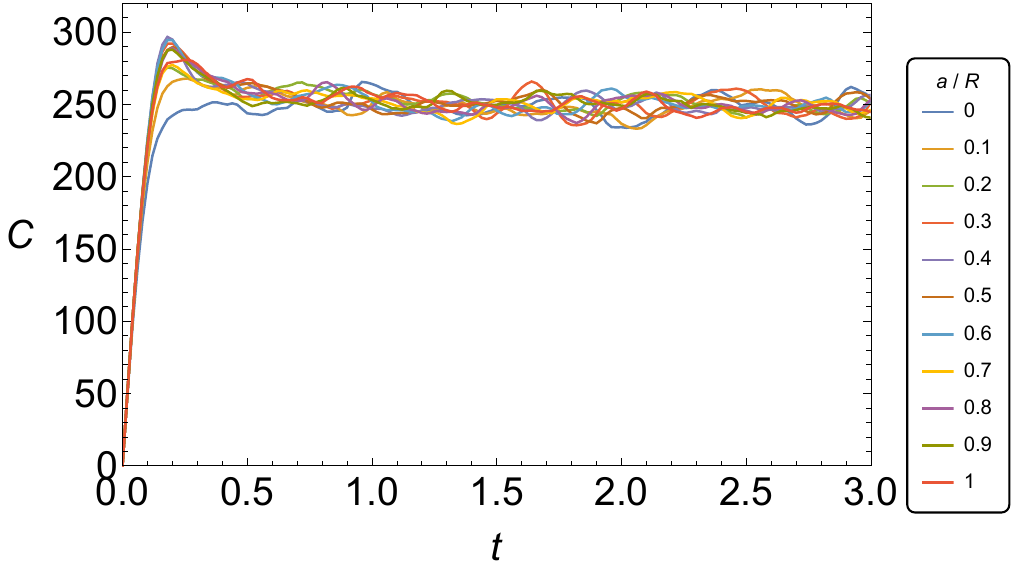}\label{fig:stadium KSC}}
    \hspace{3mm}
    \subfigure[The $a/R$ dependence of the late-time value of Krylov state complexity.]
     {\includegraphics[width=7cm]{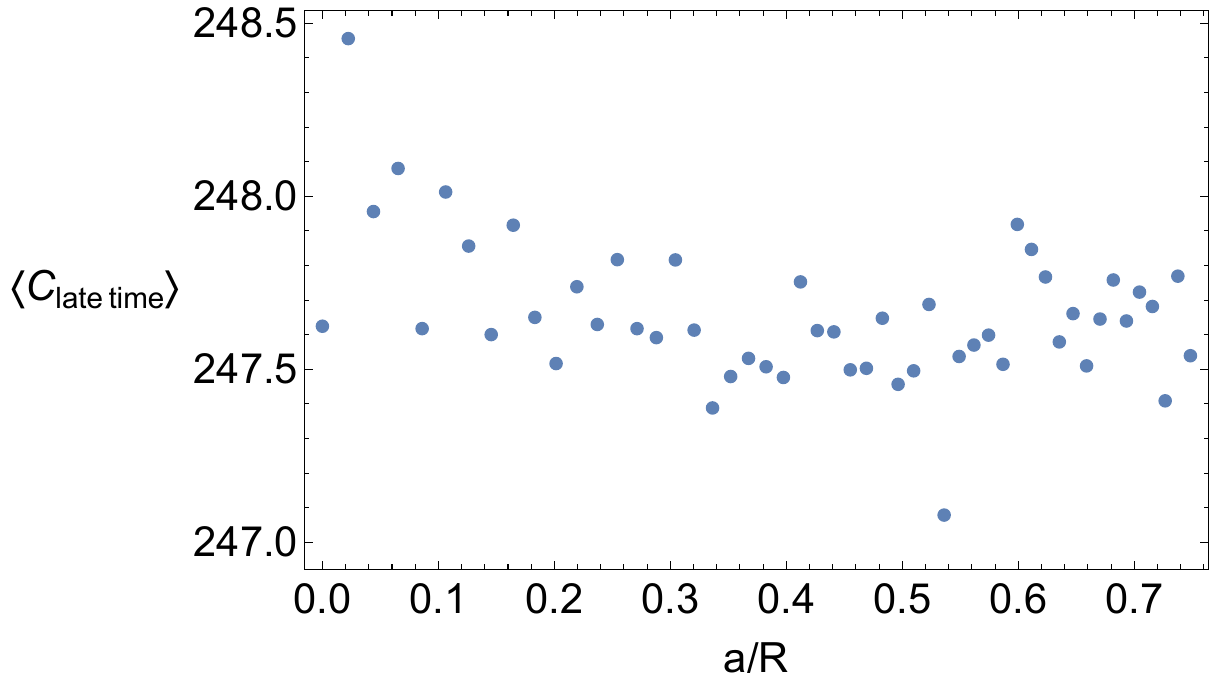} \label{fig:stadium asym KSC}}
    \subfigure[The $a/R$ dependence of the peak value of Krylov state complexity.]
     {\includegraphics[width=7cm]{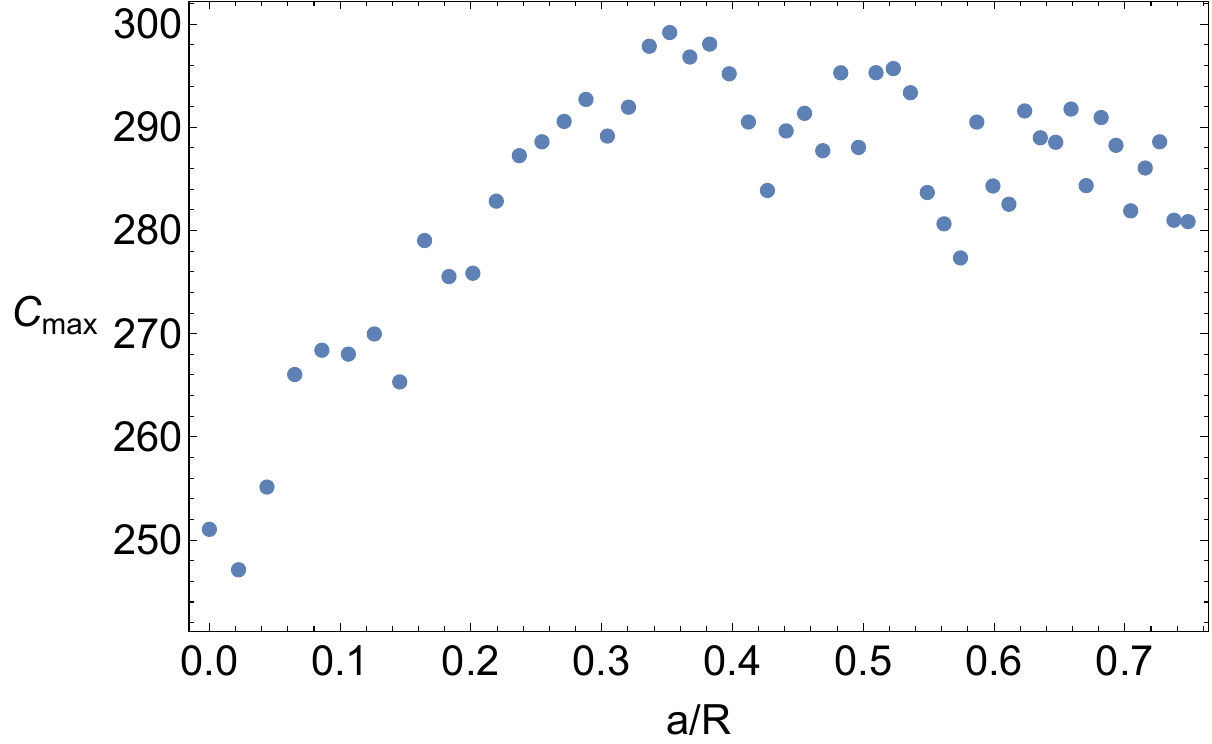} \label{fig:stadium peak KSC}}
    \caption{The $a/R$ dependence of Krylov state complexity. Panel (a): time dependence of the complexity. Panels (b), (c): the late-time average and the peak value of the complexity. The late-time average of the complexity is taken over the time range $1<t<20$.}
\label{fig:stadium KSC and asym KSC}
\end{figure}

\subsection{The time dependence of Krylov state complexity}
\label{sec:5-1}

Figure~\ref{fig:stadium KSC} shows the time dependence of the state complexity for the stadium billiard with $0\leq a/R \leq 1$.
Despite the system is chaotic for $a/R\neq 0$,
the state complexity grows almost linearly in time at early time, neither exponentially nor polynomially.

The growth pattern of the complexity at early time, rather than the growth rate however, shows a clear correlation with the chaos. The complexity reaches the maximum value at $t\sim 0.2$ before it settles down to the asymptotic value.
Although the asymptotic value is insensitive to the shape ($a/R$) of the stadium billiard (Fig.~\ref{fig:stadium asym KSC}), the peak value $C_\text{max}$ of the complexity depends on $a/R$ rather smoothly (Fig.~\ref{fig:stadium peak KSC}).
Hence, for the state complexity, the variation of the Lanczos coefficients is reflected not in the asymptotic value but in the peak value at early time. 

\begin{figure}[t]
\centering
    \includegraphics[width=7cm]{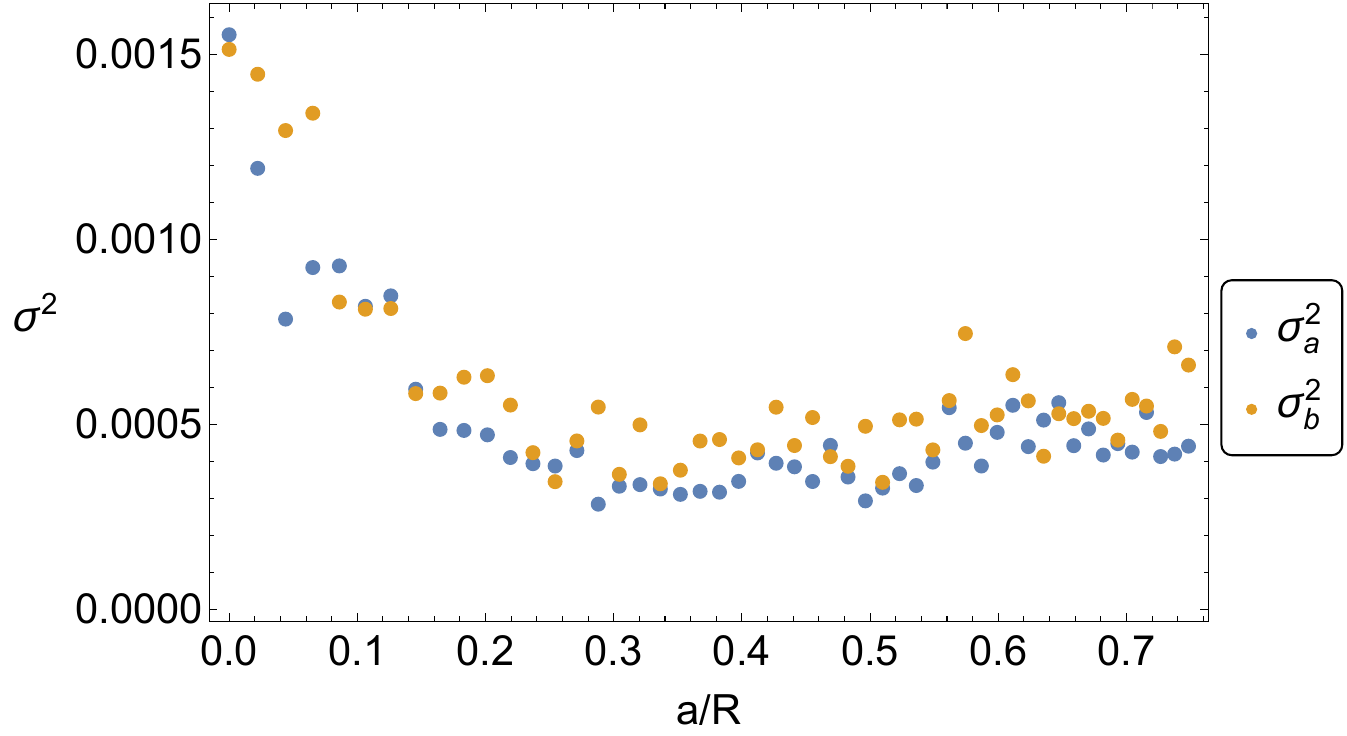} 
    \caption{The variance $\sigma_{a,b}^2$ as functions of $a/R$.}
    \label{fig:stadium state variance}
\end{figure}

\begin{figure}[t]
    \centering
    \subfigure[$\lambda$ vs $\sigma_{a,b}^2$]
    {\includegraphics[width=7cm]{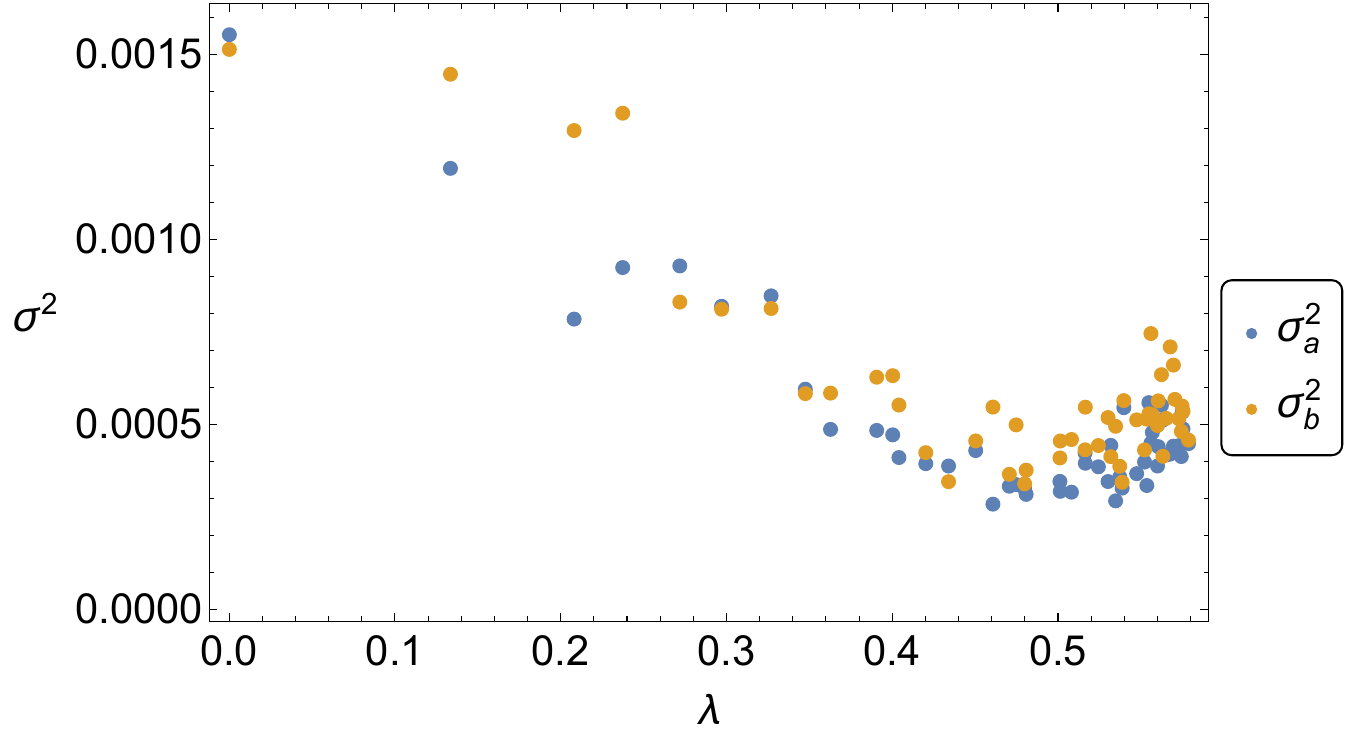}\label{fig:Stadium-state_lambda-varxab}}
    \hspace{5mm}
    \subfigure[$\langle\tilde{r}\rangle$ vs $\sigma_{a,b}^2$]{\includegraphics[width=7cm]{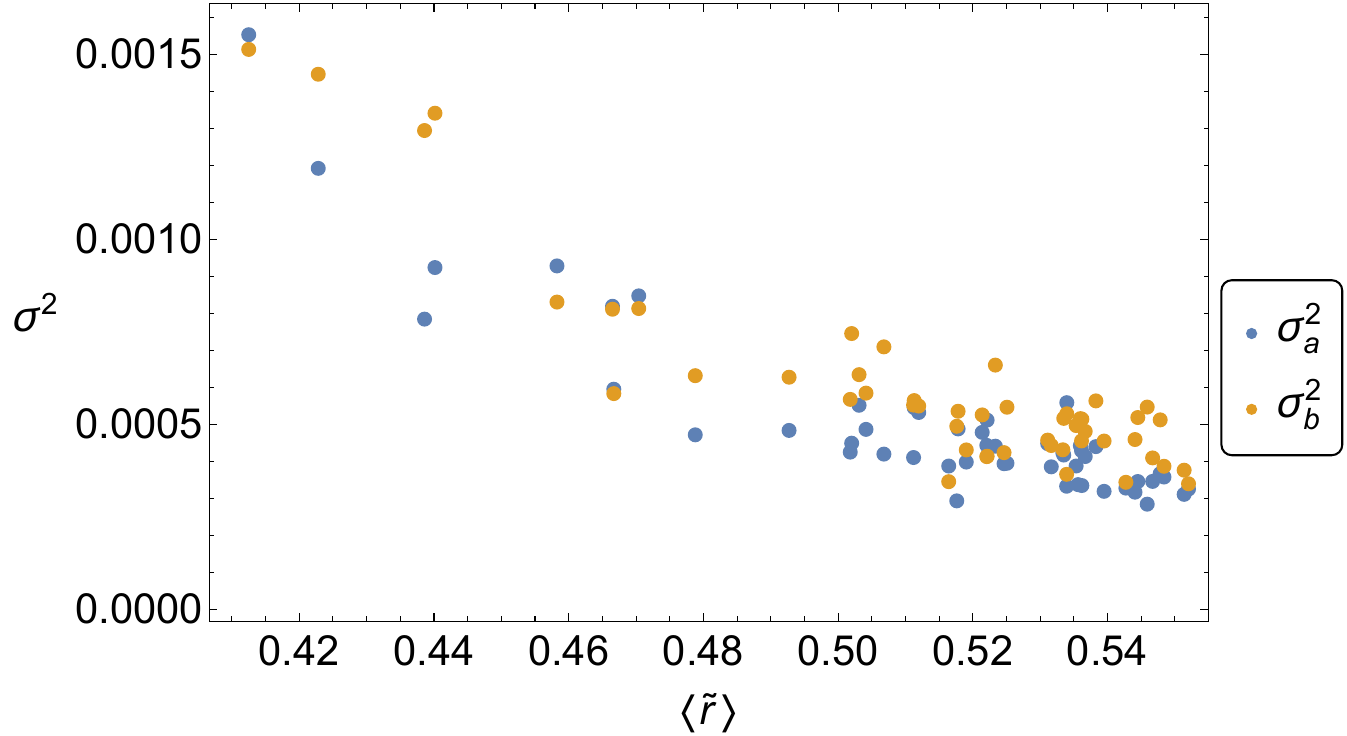}\label{fig:Stadium-state_r-varxab}}
    \caption{The scatter plots of (a) the Lyapunov exponents and the variances, and (b) the ratios and the variances.
    The points are sampled from $0\leq a/R \leq 0.75$.
    }
    \label{fig:Stadium_scatter_plot_state}
\end{figure}

\begin{table}[t]
 \begin{center}
  \begin{tabular}{|c|l|}
  \hline
    $\lambda$ vs $\sigma_a^2$ & -0.832395\\
    $\lambda$ vs  $\sigma_b^2$ & -0.806238\\
    $\langle \tilde{r}\rangle$ vs $\sigma_a^2$ & -0.891642\\
    $\langle \tilde{r}\rangle$ vs $\sigma_b^2$ & -0.893569\\
    \hline
  \end{tabular}
     \caption{Correlations between $\lambda$, $\langle \tilde{r}\rangle$, $\sigma_{a,b}^2$ for the state complexity of stadium billiard.}
   \label{table:Stadium_correlations}
 \end{center}
\end{table}

\subsection{Correlation between Lanczos coefficients and chaos}
\label{sec:5-2}

Now we turn to the correlation between the variation of the Lanczos coefficients and other indicators of the chaos, namely, the Lyapunov exponent $\lambda$ and the ratio $\langle \tilde r \rangle$.
Prior to such an analysis, let us show the dependence of the variance $\sigma^2_{a,b}$ of the Lanczos coefficients on the parameter $a/R$ in Fig.~\ref{fig:stadium state variance}.
This figure shows that the variances (both $\sigma_a^2$ and $\sigma_b^2$) decrease as $a/R$ increases in the range $0<a/R\lesssim 0.3$, then it stays at the asymptotic value for $a/R\gtrsim 0.3$.
Since the Lyapunov exponent is monotonically increasing with respect to $a/R$, it is expected that $\sigma_{a,b}^2$ and $\lambda$ are negatively correlated.

Such an expectation can be confirmed by explicitly plotting the relationship between 
$\sigma_{a,b}^2$ and $\lambda$.
Figure~\ref{fig:Stadium-state_lambda-varxab} shows the relationship between $\lambda$ and $\sigma_{a,b}^2$, in which the negative correlation between these quantities can be observed. Since $\lambda$ and $\langle \tilde r \rangle$ are positively correlated as explained in the previous section, $\langle \tilde r \rangle$ and $\sigma_{a,b}^2$ are negatively correlated as shown in Fig.~\ref{fig:Stadium-state_r-varxab}.
Quantitative values of the correlations are summarized in Table~\ref{table:Stadium_correlations}, which clearly shows the negative correlations of $\sigma_{a,b}^2$ with $\lambda$ and $\langle \tilde r \rangle$.


\section{Universality: the case of the Sinai billiard}
\label{sec:6}
In this section, we analyze Krylov complexity of the Sinai billiard, which is another typical chaotic system, to show that our results in the previous section is universal. 

\begin{figure}[t]
    \centering
    \includegraphics[width=5cm]{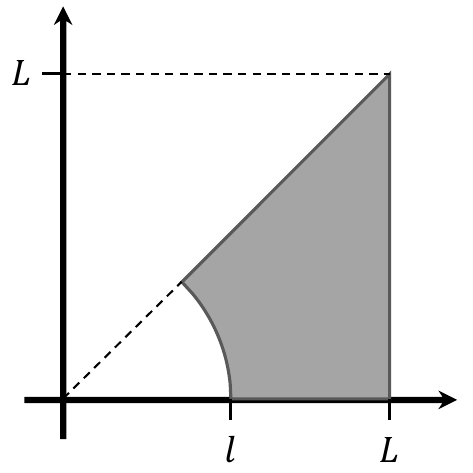}
    \caption{Geometry of the Sinai billiard. Dirichlet boundary conditions are imposed on the boundaries.}
    \label{fig:Sinai-billiard}
\end{figure}

\begin{figure}[t]
\centering
    \subfigure[The Lyapunov exponent as a function of $l/L$. The area of the billiard and the velocity of the particle are normalized to the unity.]
     {\includegraphics[width=7cm]{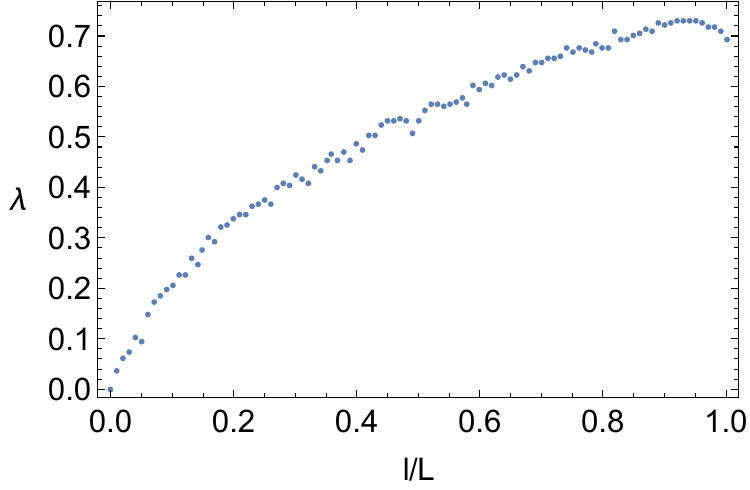} \label{fig:sinai Lyapunov}}
    \hspace{3mm}
    \subfigure[The ratio $\langle\tilde{r}\rangle$ as a function of $l/L$. The number of levels used in this calculation is 100. The green and orange lines correspond to the values for the Poisson and the Wigner-Dyson statistics respectively.]
     {\includegraphics[width=7cm]{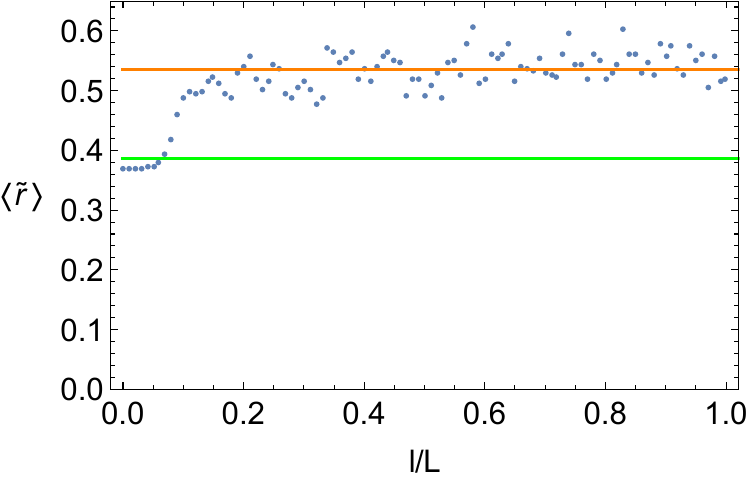} \label{fig:sinai ratio}}
    \caption{The $l/L$ dependence of the Lyapunov exponent and the ratio.}
\label{fig:sinai Lyapunov and ratio}
\end{figure}

In Fig.~\ref{fig:Sinai-billiard}, we show the shape of the Sinai billiard which we consider. The system is chaotic for $l>0$ while the system becomes integrable for $l=0$ \cite{Berry:1981b}. In Fig.~\ref{fig:sinai Lyapunov and ratio}, we show the $l/L$ dependence of the Lyapunov exponent $\lambda$ and the ratio $\langle\tilde{r}\rangle$. Although the Lyapunov exponent increases gradually with $l/L$, the ratio increases rapidly around $l/L=0.1$.

\begin{figure}[t]
\centering
    \includegraphics[width=15cm]{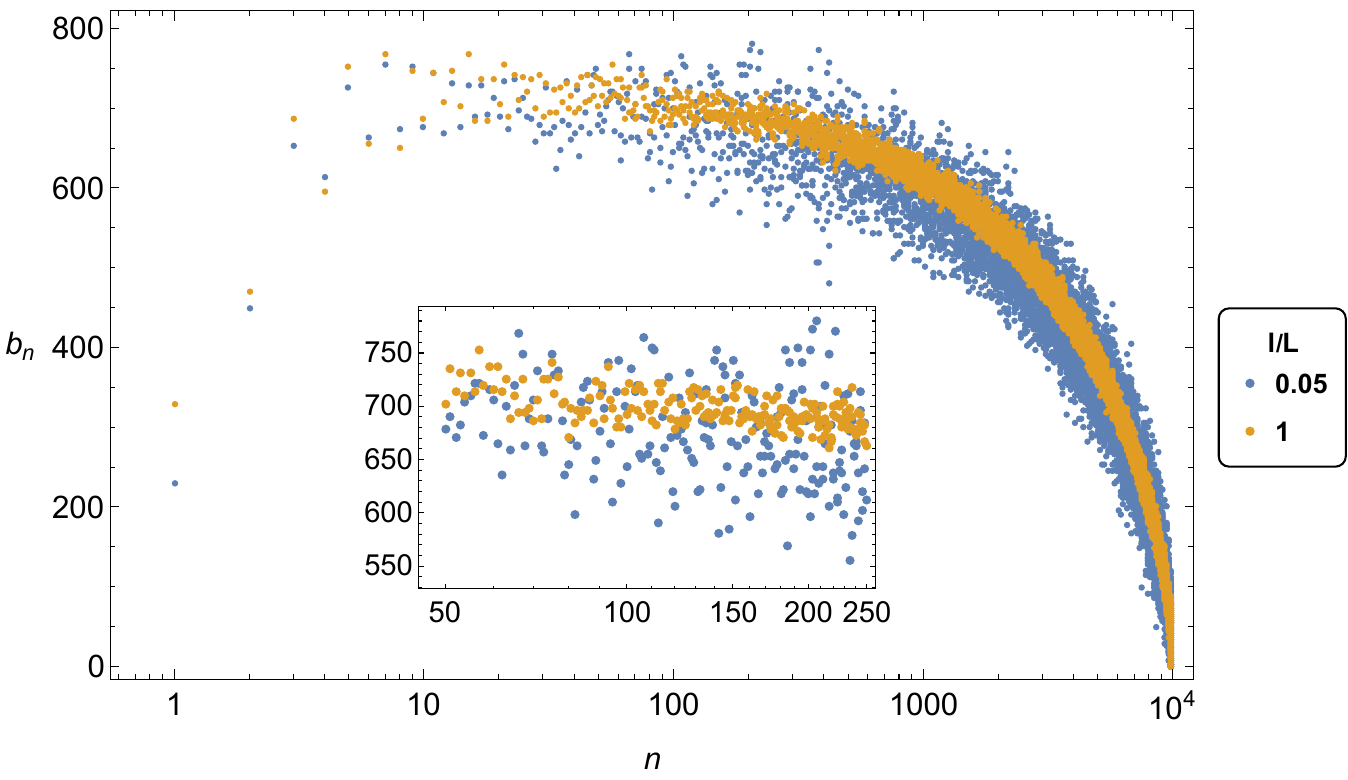}
    \caption{The Lanczos coefficients for the truncated momentum operator $P$ in Sinai billiards with $l/L=0.05$ (blue dots) and $l/L=1$ (orange dots). Note that the horizontal axis is in log scale. The inset is the enlarged version, where the data are used to calculate the variance.}
    \label{fig:sinai op Lanczos}
\end{figure}

\subsection{Krylov operator complexity}

\begin{figure}[t]
\centering
    \includegraphics[width=8cm]{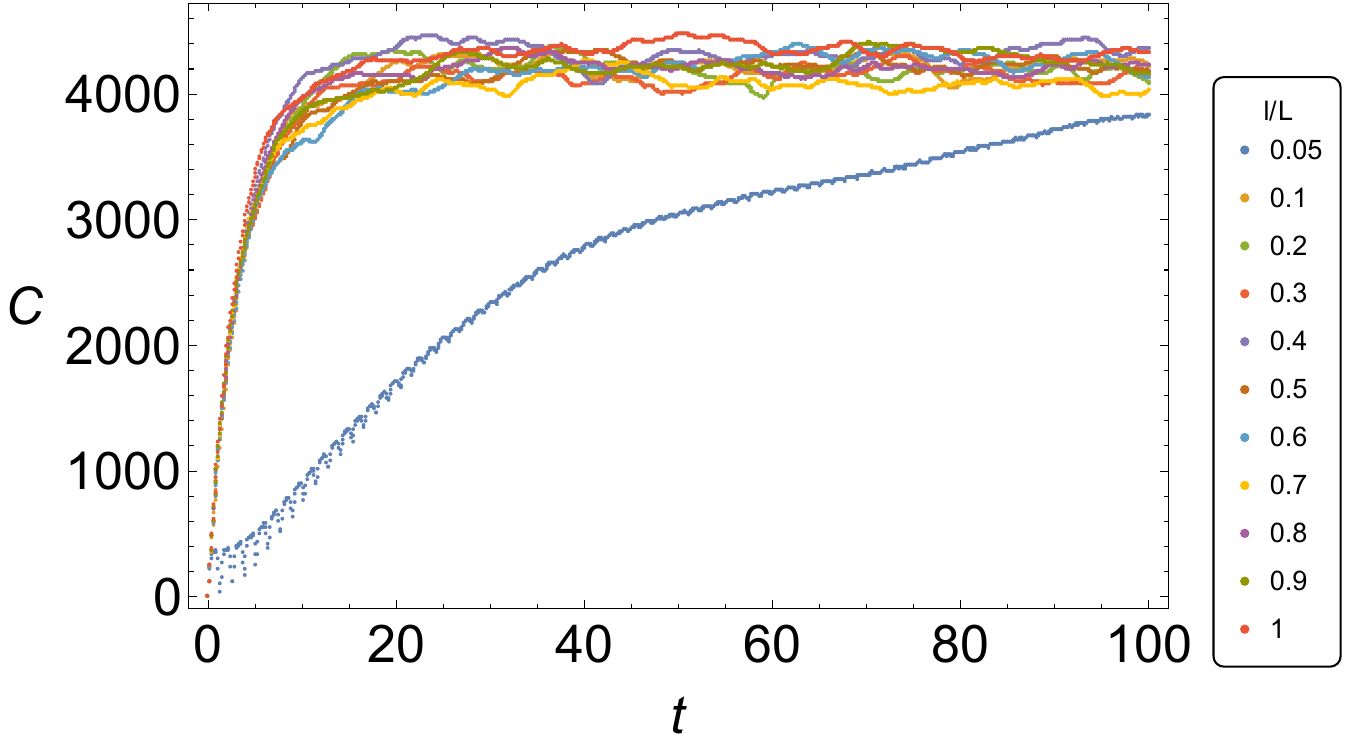}
    \caption{The time dependence of Krylov operator complexity for various values of $l/L$.}
    \label{fig:sinai KOC}
\end{figure}

\begin{figure}[t]
\centering
    \includegraphics[width=7cm]{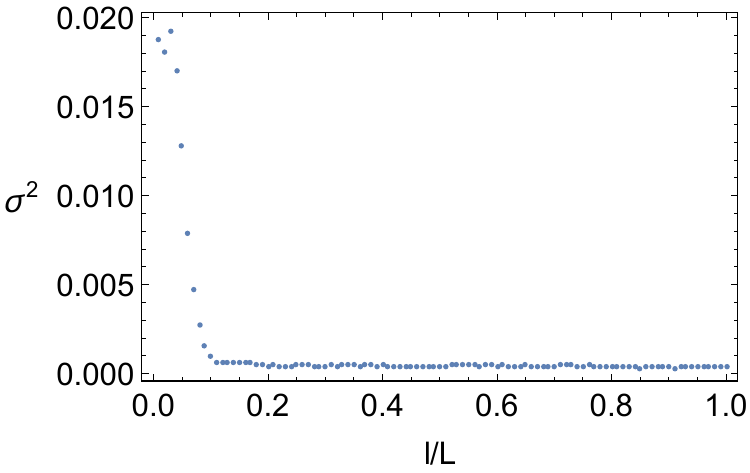}
    \caption{The variance $\sigma^2$ as a function of $l/L$ of the Krylov operator complexity for the Sinai billiard.}
    \label{fig:sinai variance}
\end{figure}

Using the method described in Sec.~\ref{sec:3}, we numerically compute Krylov operator complexity for the truncated momentum operator $P$, with truncation $N_{\rm max}=100$.\footnote{Again, the area of the billiard is normalized to the unity.} In Fig.~\ref{fig:sinai op Lanczos}, we show the Lanczos coefficients for $l/L=0.05$ and $l/L=1$.\footnote{In the rest of this section, we concentrate on $l>0$ regime since the integrable case ($l/L=0$) is numerically unstable. We will discuss this issue in App.~\ref{app:1}.} We identify the dimension of the Krylov space $K_P$ as $K_P=9900$. Obviously, the Lanczos coefficients for $l/L=0.05$ distribute much broader compared to $l/L=1$. On the other hand, their initial behaviors are almost identical.

In Fig.~\ref{fig:sinai KOC}, we show the Krylov operator complexities as functions of $t$ for the stadium billiards with $l/L=0.05,0.1,0.2,\cdots,1$. While the Sinai billiard is chaotic for $l/L>0$, the early time growth of the Krylov operator complexity is not exponential. The result for $l/L=0.05$ is different from the others. This is because the behavior of the Lanczos coefficients changes abruptly 
below
$l/L=0.1$. 
In Fig.~\ref{fig:sinai variance}, we show the variance \eqref{eq:variance} as a function of $l/L$. The variance becomes larger in the integrable regime compared to the chaotic regime. Similarly to the ratio given in Fig.~\ref{fig:sinai ratio}, the variance changes rapidly with $l/L$.

\begin{figure}[t]
    \centering
    \subfigure[$\lambda$ vs $\sigma^2$]
    {\includegraphics[width=7cm]{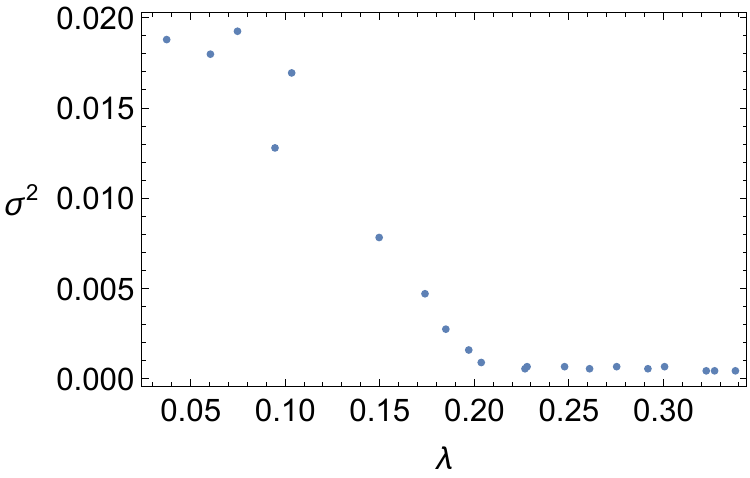}\label{fig:sinai_op_Lyapunov_vs_variance}}
    \hspace{5mm}
    \subfigure[$\langle\tilde{r}\rangle$ vs $\sigma^2$]{\includegraphics[width=7cm]{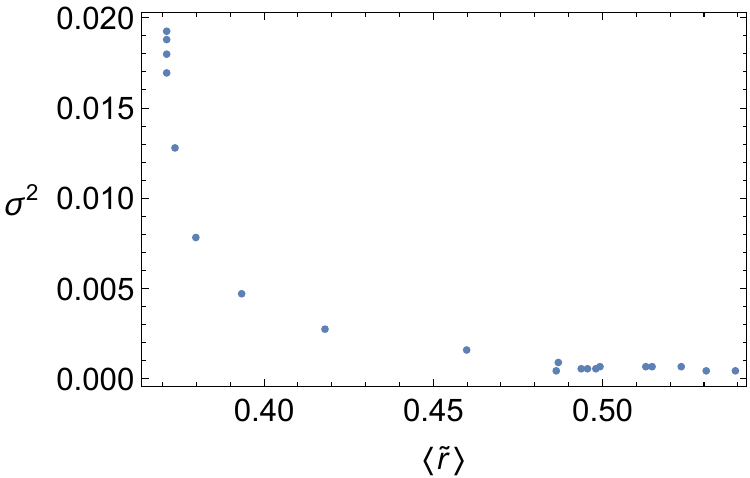}\label{fig:sinai_op_ratio_vs_variance}}
    \subfigure[$\lambda$ vs $\langle\tilde{r}\rangle$]{\includegraphics[width=7cm]{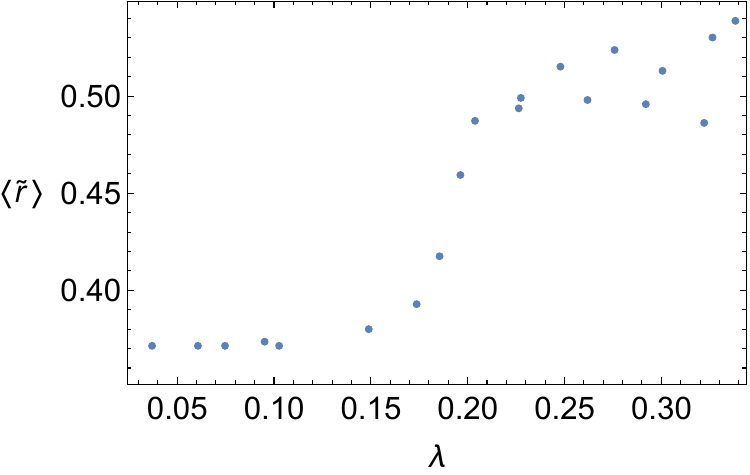}\label{fig:sinai_op_Lyapunov_vs_ratio}}
    \caption{The scatter plots of (a) the Lyapunov exponents and the variances, (b) the ratios and the variances, (c) the Lyapunov exponents and the ratios of the Krylov operator complexity for the Sinai billiard. The points are sampled from $0.01\leq l/L \leq 0.2$.}
    \label{fig:sinai_scatter_plot}
\end{figure}

In Fig.~\ref{fig:sinai_scatter_plot}, we show scatter plots, where the points are sampled from $0.01\leq l/L \leq 0.2$.\footnote{Changing the sampling region, for example, to $0\leq l/L \leq 1$ does not affect the correlation much.} 
In Table~\ref{table:sinai_correlation_coefficients}, we show the correlation coefficients \eqref{eq:correlation coefficient} calculated from these plots. Since the correlation coefficients are far from zero as easily expected directly from the plots, there should be some correlations between $\lambda,\sigma^2,$ and $\langle\tilde{r}\rangle$. Again, we see that the quantity $\sigma^2$ works as an indicator of quantum chaos as good as the ratio $\langle\tilde{r}\rangle$ does.

\begin{table}[t]
\begin{center}
    \begin{tabular}{|c|l|}
    \hline
    $\lambda$ vs $\sigma^2$ & -0.899970\\
    $\langle \tilde{r}\rangle$ vs $\sigma^2$ & -0.872723 \\
    $\lambda$ vs $\langle \tilde{r}\rangle$ & \hphantom{-}0.924828 \\ \hline
    \end{tabular}
    \caption{The correlation coefficients between $\lambda$, $\langle \tilde{r}\rangle$, $\sigma^2$
   of the Krylov operator complexity for the Sinai billiard. }
    \label{table:sinai_correlation_coefficients}
\end{center}
\end{table}

\subsection{Krylov state complexity}

We briefly summarize the results for the Krylov state complexity of the Sinai billiard in this section. The results are in parallel with those for the stadium billiard, which show the universality of the properties of the chaos indicators studied in this work. The numerical setup is the same as that for the stadium billiard analyzed in Sec.~\ref{sec:5} except that the billiard table is switched to that given in Fig.~\ref{fig:Sinai-billiard}.

\begin{figure}[t]
\centering
    \subfigure[$a_n$]{\includegraphics[height=4.45cm]{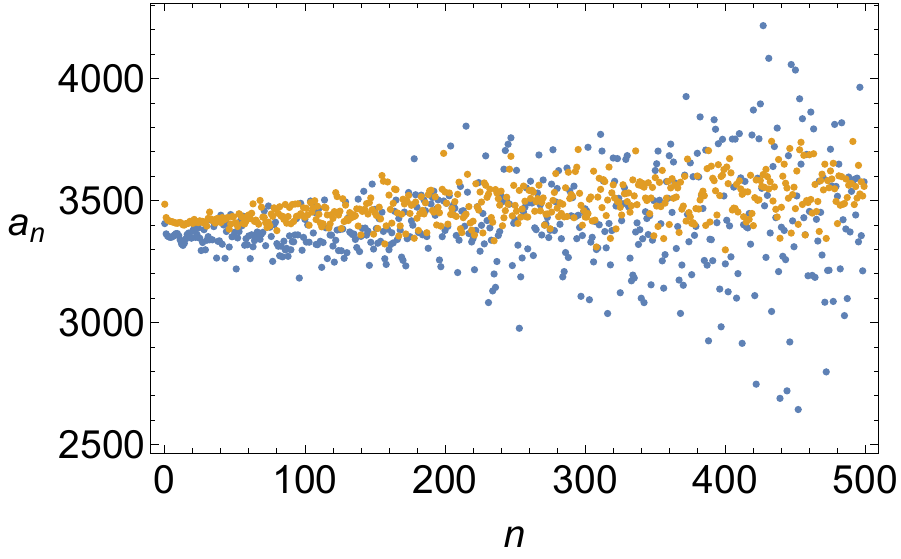}}
    \subfigure[$b_n$]{\includegraphics[height=4.45cm]{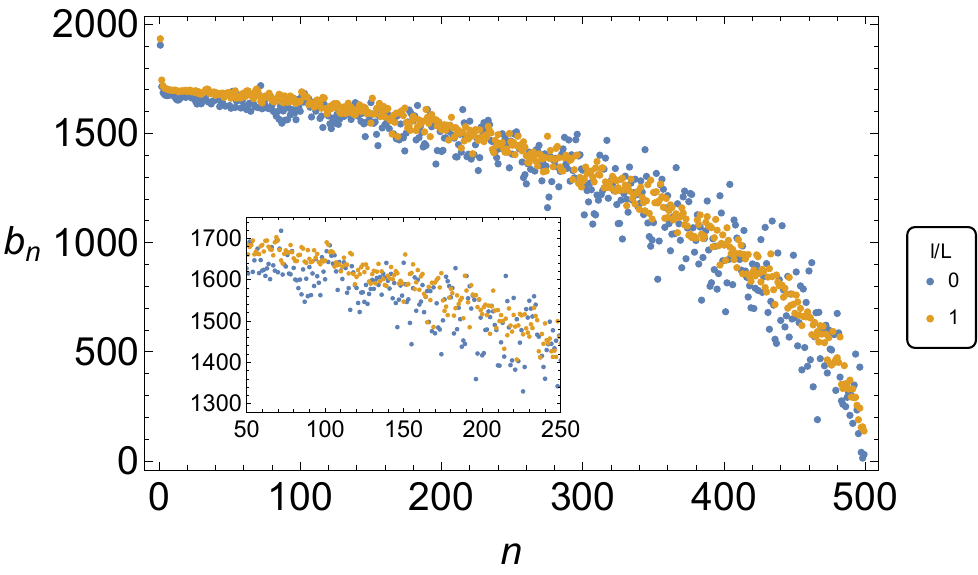}}
    \caption{The Lanczos coefficients of the Krylov state complexity of the Sinai billiard with $l/L=0$ (blue dots) and $l/L=1$ (orange dots). The inset is the enlarged version, where the data are used to calculate the variance.}
    \label{fig:Sinai state Lanczos}
\end{figure}

\begin{figure}[t]
\centering
    \subfigure[The time dependence of Krylov operator complexity for various values of $l/L$.]
     {\includegraphics[width=7cm]{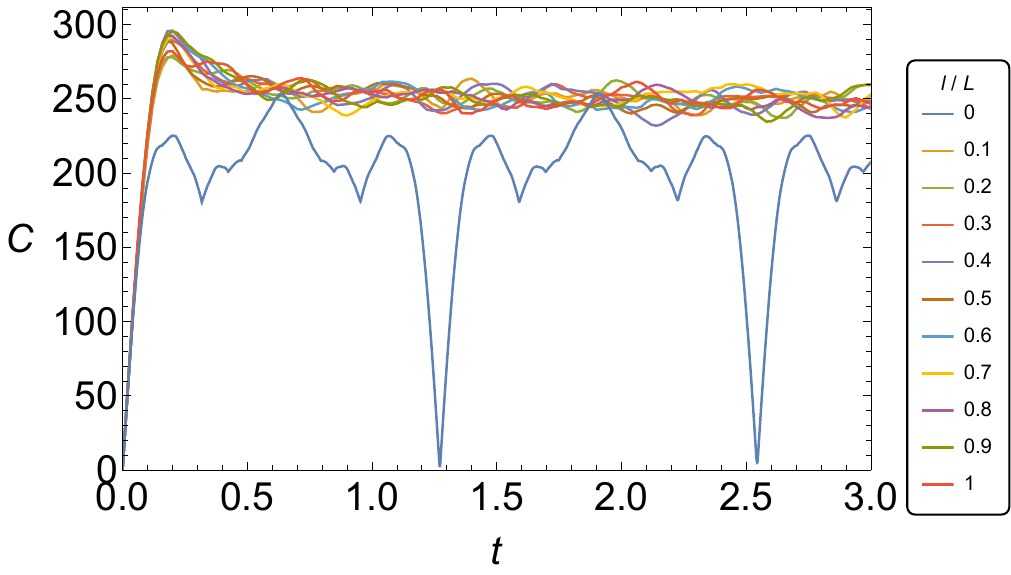}\label{fig:sinai KSC}
     }
    \hspace{3mm}
    \subfigure[The $l/L$ dependence of the late-time value of Krylov state complexity.]
     {\includegraphics[width=7cm]{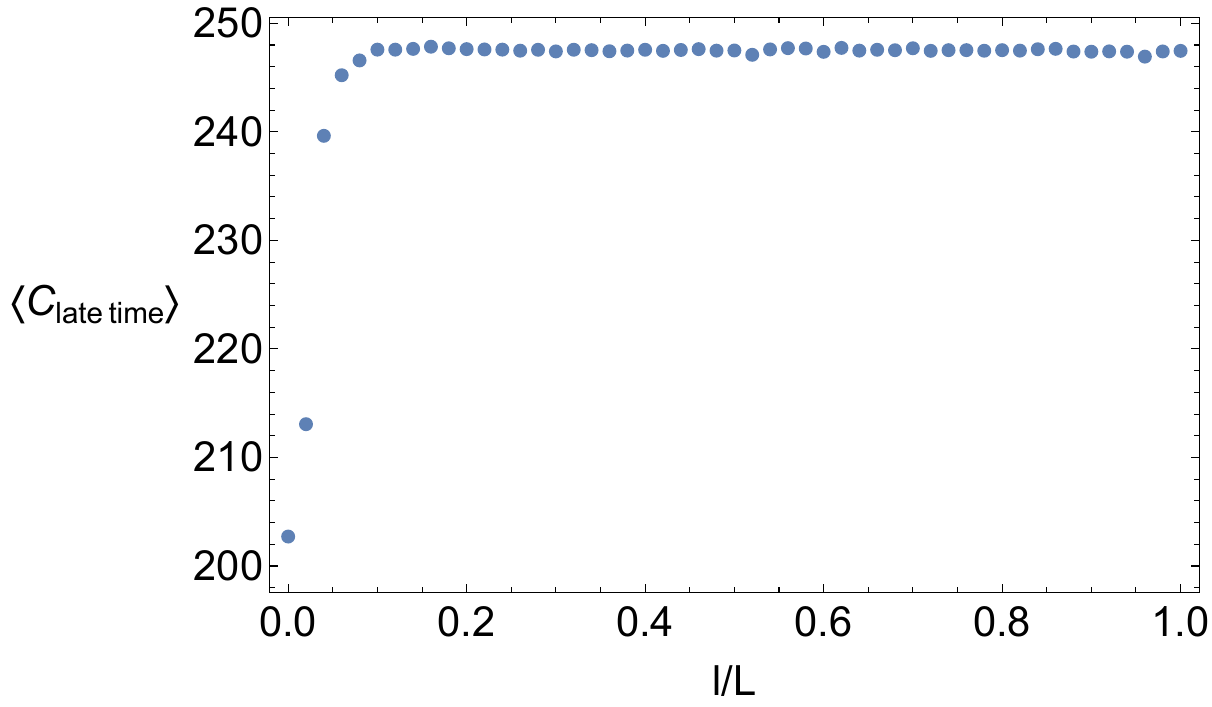} \label{fig:sinai asym KSC}}
    \subfigure[The $l/L$ dependence of the peak value of Krylov state complexity.]
     {\includegraphics[width=7cm]{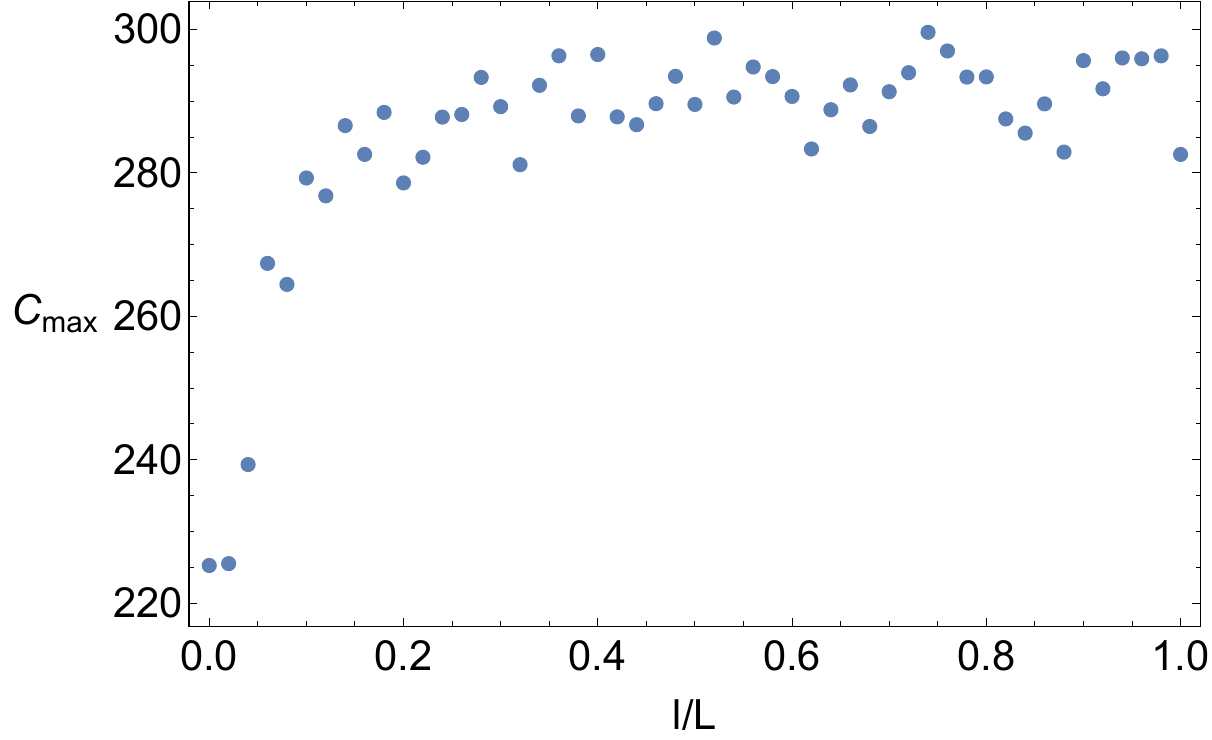} \label{fig:sinai peak KSC}}
    \caption{The $l/L$ dependence of Krylov state complexity for the Sinai billiard. Panel (a): time dependence of the Krylov state complexity. Panels (b), (c): the late-time average and the peak value of the  Krylov state complexity. The late-time average of the complexity is taken over $1<t<20$.}
\label{fig:sinai KSC and asym KSC}
\end{figure}

\begin{figure}[t]
\centering
    \includegraphics[width=7cm]{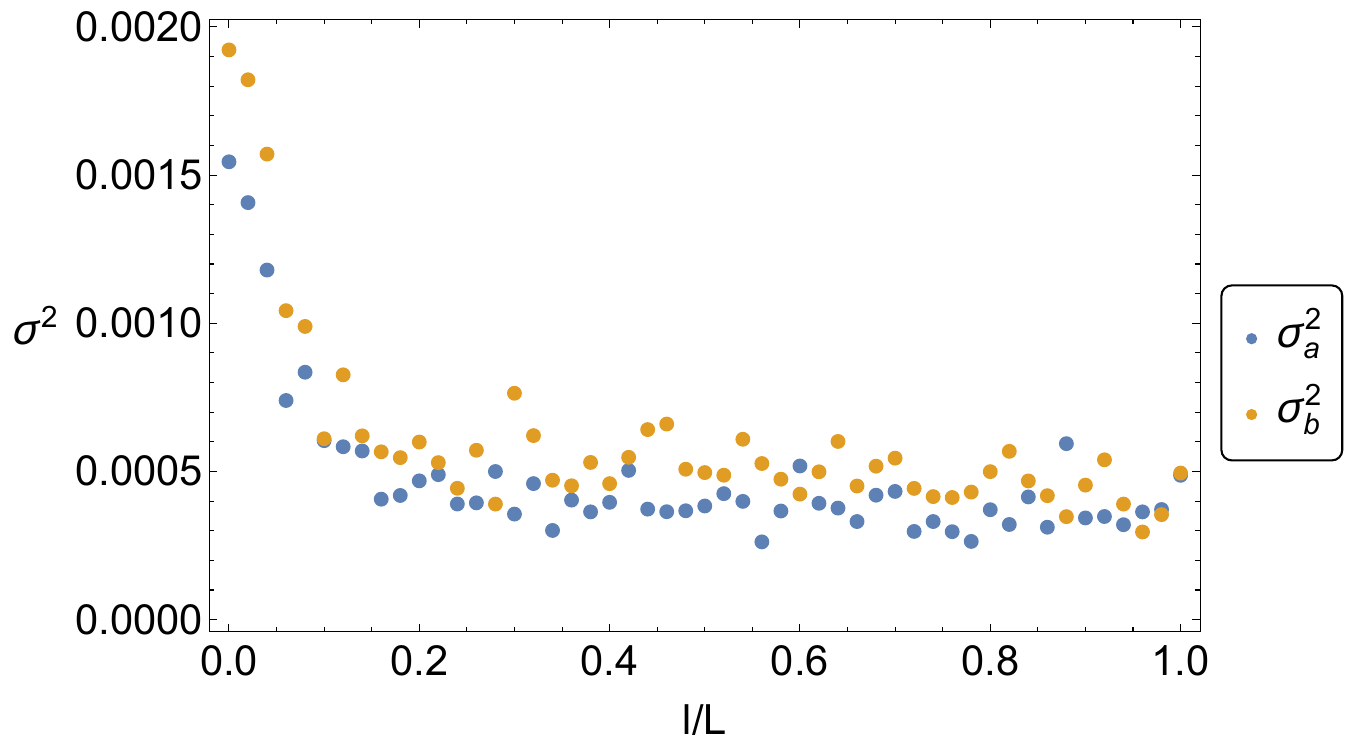} 
    \caption{The variance $\sigma_a^2, \sigma_b^2$ as functions of $l/L$.}
    \label{fig:sinai state variance}
\end{figure}

The Lanczos coefficients $a_n, b_n$ for the Krylov state complexity (Fig.~\ref{fig:Sinai state Lanczos}), the time dependence of the Krylov state complexity (Fig.~\ref{fig:sinai KSC and asym KSC}), and the variance $\sigma^2_{a,b}$ of the Lanczos coefficients (\ref{fig:sinai state variance}) for the Sinai billiard are qualitatively similar to those for the stadium billiard given in Sec.~\ref{sec:5}.
One of the differences is that, in Fig.~\ref{fig:sinai KSC}, the complexity $C(t)$ becomes oscillatory when the billiard is integrable ($l/L=0$), while such a periodicity was not observed in the stadium billiard (Fig.~\ref{fig:stadium KSC}). 
This feature is attributed to the fact that the energy spectrum of the Sinai billiard with $l/L=0$ is commensurable, which is not the case for the stadium billiard with $a/R=0$.

\begin{figure}[t]
    \centering
    \subfigure[$\lambda$ vs $\sigma_{a,b}^2$]
    {\includegraphics[width=7cm]{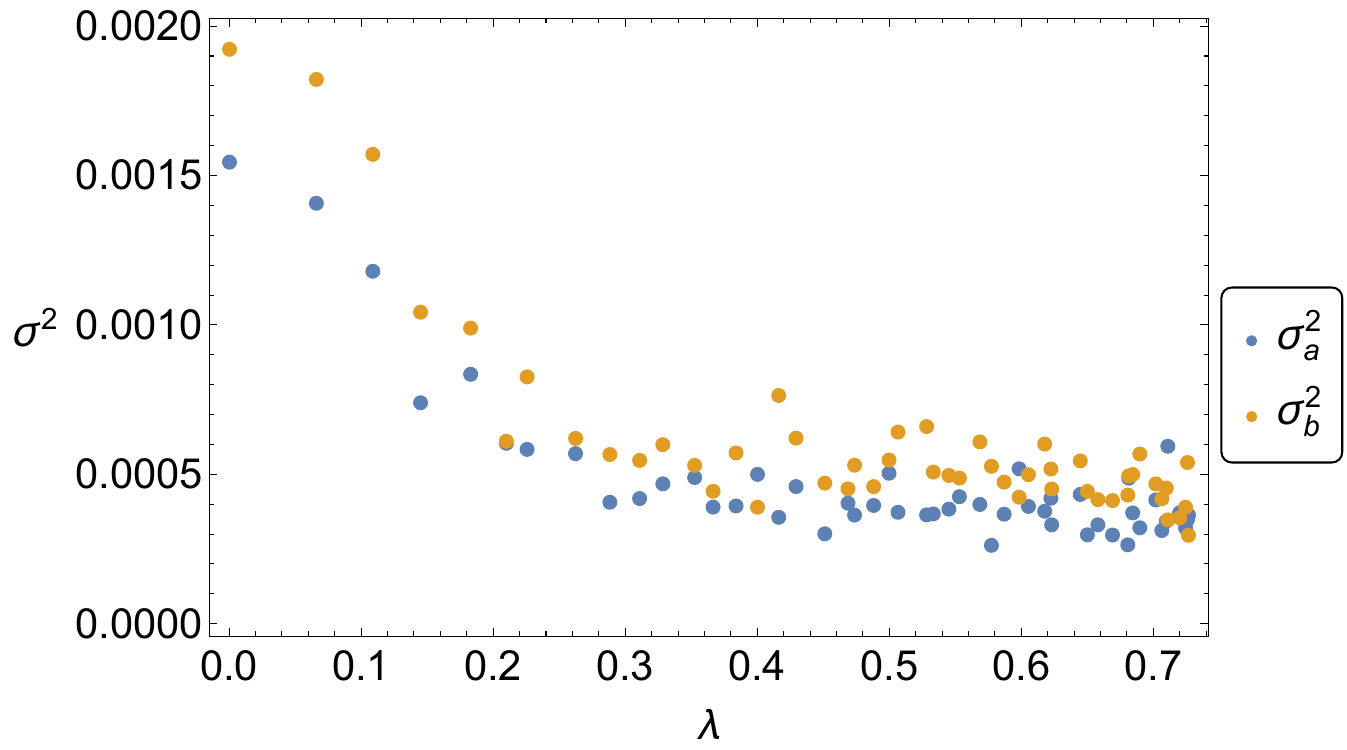}\label{fig:Sinai-state_lambda-varxab}}
    \hspace{5mm}
    \subfigure[$\langle\tilde{r}\rangle$ vs $\sigma_{a,b}^2$]{\includegraphics[width=7cm]{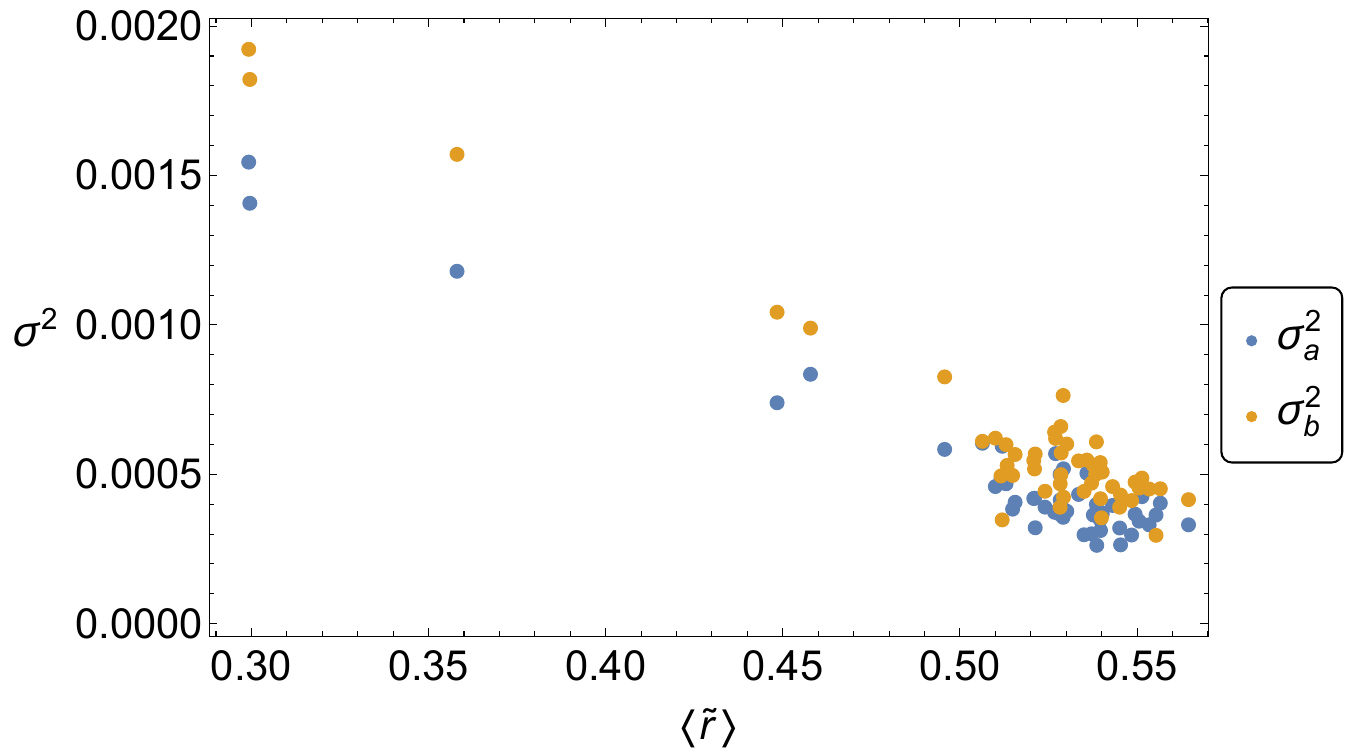}\label{figSinai-state_r-varxab}}
    \caption{The scatter plots of (a) the Lyapunov exponents and the variances and (b) the ratios and the variances.
    }
    \label{fig:sinai_scatter_plot_state}
\end{figure}

As shown in Fig.~\ref{fig:sinai_scatter_plot_state} and Table \ref{table:Sinai_correlations_state}, the variances $\sigma^2_{a,b}$ are correlated with the indicators of classical and quantum chaos, namely $\lambda$ and $\langle\tilde r \rangle$. Hence the variance of Lanczos coefficients is a faithful indicator of chaos also for the Sinai billiard.

On top of it, the peak value $C_\text{max}$ of the complexity at early time in the time evolution (Fig.~\ref{fig:sinai peak KSC}) depends rather smoothly on the parameter $l/L$ compared to the average of the complexity at late time $\langle C_\text{late time} \rangle$ (Fig.~\ref{fig:sinai asym KSC}). In this sense, $C_\text{max}$ is more faithful compared to $\langle C_\text{late time} \rangle$ as an indicator of quantum chaos.
This tendency is in common with the stadium billiard, hence it is suggested that this feature is universal.


\section{Summary and discussions}
\label{sec:7}

\begin{table}[t]
 \begin{center}
  \begin{tabular}{|c|l|}
  \hline
    $\lambda$ vs $\sigma_a^2$ & -0.741803\\
    $\lambda$ vs $\sigma_b^2$ & -0.757869\\
    $\langle \tilde{r}\rangle$ vs $\sigma_a^2$ & -0.965785\\
    $\langle \tilde{r}\rangle$ vs $\sigma_b^2$ & -0.962833\\
 \hline
  \end{tabular}
  \caption{Correlations between $\lambda$, $\langle \tilde{r}\rangle$, and $\sigma_{a,b}^2$
   for the Krylov state complexity of Sinai billiard.
   }
   \label{table:Sinai_correlations_state}
 \end{center}
\end{table}

In this paper, we studied the Krylov complexity in quantum mechanics.
To investigate the relationship between complexity and chaos, we numerically studied Krylov complexity in billiard systems. 
The stadium billiard, which is chaotic, allows a one-parameter deformation of its shape, and in a limit it reduces to the circular billiard which is integrable.
We observed that under this deformation there exists a significant correlation between the classical Lyapunov exponents and
the variances of the Lanczos coefficients for both the Krylov operator complexity and the Krylov state complexity.
We also found a significant correlation between the variances of the Lanczos coefficients and the statistical distribution of the adjacent spacings of the quantum energy levels.
This suggests that the variances of the Lanczos coefficients are a good indicator of quantum chaos as well as the energy level statistics.
Similar results were confirmed for the one-parameter family of the Sinai billiard, which suggests that in more general quantum mechanical systems the variances of the Lanczos coefficients can be a good indicator of quantum chaos.
Note that in our numerical analysis a finite cutoff $N_{\rm max}$ in the quantum energy spectrum was necessary.
The descent part of the sequence of the Lanczos coefficients, which we used in the evaluation of the variance, would be lost in the limit where $N_{\rm max}$ goes to infinity since the Lanczos algorithm would not always terminate in that limit.
In this point of view, the variance of the Lanczos coefficients measures the response of the system to a regularization $N_{\rm max}$, and this response distinguishes the chaoticity.

In viewing the structure of the resultant correlations, we observe one issue:
when the billiard is deformed, there is an abrupt change in variances of the Lanczos coefficients and the energy level statistics, while the change in classical Lyapunov exponents is mild. This suggests that there may be some discrepancy between the notion of classical chaos and the currently proposed quantum chaos. 
We leave to a future work the quest for  
the cause of this discrepancy and 
a search for a better indicator of quantum chaos that more accurately reflects classical chaoticity, with a consistent understanding of how classical chaoticity manifests itself through the classical limit from quantum theory.

Here let us discuss the late-time behavior of the complexity in
more detail.
As briefly mentioned in Sec.~\ref{sec:2}, there was an expectation that the late-time value of the Krylov operator complexity would be larger in chaotic systems compared to integrable systems \cite{Rabinovici:2021qqt}.
This behavior may be understood \cite{Rabinovici:2021qqt} in terms of the Anderson localization on the one-dimensional chain \eqref{eq:Krylov chain for operator}.\footnote{The Krylov complexity in a many-body localization system was studied in \cite{Trigueros:2021rwj}.}
Intuitively, when the Lanczos coefficients behave erratically, it becomes difficult for the amplitude to spread smoothly along the chain and the amplitude will be localized around the initial position.
Since the Krylov complexity represents the expectation value of the position along the chain, this means that the complexity stays at a small value.
By the same argument, when variances of the Lanczos coefficients are small, the Krylov complexity is expected to take a larger value.

In the stadium billiard, as shown in Fig.~\ref{fig:stadium asym KOC}, the saturation value of the complexity becomes smaller as the system approaches the integrable regime ($a/R\to0$).
However, this late-time behavior is not so obvious in the Sinai billiard. 
In Fig.~\ref{fig:sinai KOC}, the complexity for $l/L=0.05$ appears to be small, but it does not show any saturation in the displayed range.
It also seems that the complexity for $l/L\geq 0.05$ approaches the same value.
In App.~\ref{app:1}, we also consider the Krylov operator complexity for $l/L=0$, whose time dependence is shown in Fig.~\ref{fig:sinai resolution KOC integrable}.
It is possible that the complexity remains small and oscillates for $l/L=0$, but this result should not be seriously trusted because of the numerical instabilities.
Therefore, our tentative conclusion is that it may be difficult to distinguish the chaoticity from the late-time behavior of the Krylov operator complexity.
This also implies that the effect of the erratic behavior of the Lanczos coefficients on the one-dimensional chain \eqref{eq:Krylov chain for operator} is more subtle.
The authors of \cite{Espanol:2022cqr} have reported similar results in Ising spin chains that the late-time behavior of the complexity strongly depends on the choice of the operator and is not correlated with chaoticity although there was a correlation between the variance of the Lanczos coefficients and chaoticity.

Even for the Krylov state complexity, its asymptotic value is 
insensitive to the billiard table shape, hence its relationship with the variance of the Lanczos coefficients is subtle as it was for the Krylov operator complexity. One difference is that the peak value of the Krylov state complexity at early time shows some correlation with the classical Lyapunov exponent, which was discussed in \cite{Balasubramanian:2022tpr,Erdmenger:2023shk}.
It would be interesting to study this property in more detail and to examine how universal this behavior is for various systems showing quantum chaos.

As concluding words of this paper, let us make a remark from a broad perspective.
The numerical method in this paper is valid for general quantum mechanical systems.
As in the case of OTOCs, it is important to investigate the universal nature of chaos by performing similar analyses specifically on various quantum mechanical systems such as, for example, polygonal billiards \cite{Rozenbaum:2019nwn}, coupled harmonic oscillators \cite{Matinyan:1981dj,Matinyan:1981ys,Savvidy:1982jk,Biro:1994bi,Akutagawa:2020qbj}\footnote{The finite-temperature Krylov operator complexity was numerically studied in \cite{Guo:2022hui}.}, inverted harmonic oscillators \cite{Hashimoto:2020xfr,Bhattacharyya:2020art}, and anharmonic oscillators \cite{Romatschke:2020qfr}. Evaluating the variances of the Lanczos coefficients in those popular examples, one can compare them with classical/quantum Lyapunov exponents, leading to a broad and universal view of quantum chaos and complexity.

Readers are reminded of the history that the chaos and the complexity have been intensively studied in the context of the AdS/CFT correspondence and black holes.
In that viewpoint, clarifying the relationship between conventional computational complexity and Krylov complexity is another important issue.
There were proposals for the gravity dual interpretation of the conventional complexity, and recently, a dual interpretation of Krylov complexity has been studied \cite{Rabinovici:2023yex}.
A further exploration of the Krylov complexity in various quantum mechanics may reveal the mystery of the interior of black holes.


\section*{Acknowledgments}

The work of K.~H.~was supported in part by JSPS KAKENHI Grant No.~JP22H01217, JP22H05111 and JP22H05115.
The work of K.~M.~was supported in part by JSPS KAKENHI Grant No.~JP20K03976, JP21H05186 and JP22H01217.
The work of N.~T.~was supported in part by JSPS KAKENHI Grant No.~JP18K03623 and JP21H05189.
The work of R.~W.~was supported by Grant-in-Aid for JSPS Fellows No.~JP22KJ1940.


\appendix
\section{Resolution dependence of numerical results}
\label{app:1}

To solve the Schr\"odinger equations numerically, we used the Mathematica command, NDEigensystem. This is based on the finite element method, which divides the region into smaller cells. To improve the accuracy of the numerical calculation, it is desirable to 
make the area of a cell as small as possible.
In this appendix, we discuss the dependence of the numerical results on the area of the cell.
We focus on the Krylov operator complexity in the following, while qualitatively similar results follow also for the Krylov state complexity.
In the following discussion, the area of the billiard is normalized to the unity. We will call the area of the largest cell ``resolution".

\subsection{Stadium billiard}

\begin{figure}[t]
\centering
    \subfigure[The resolution dependence of the Frobenius norm of $P$.]
    {\includegraphics[width=4cm]{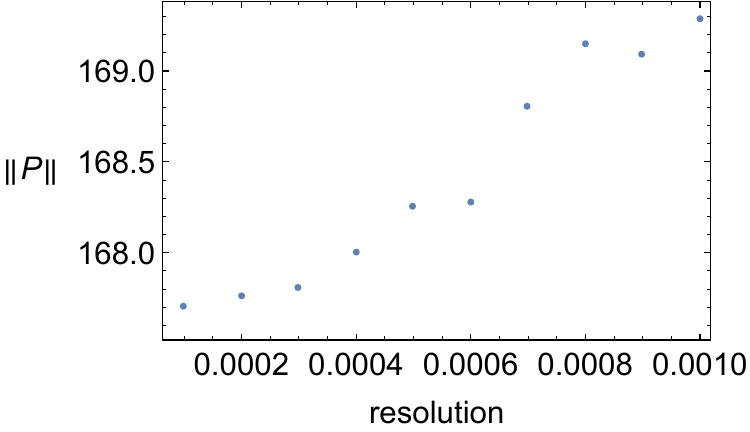}
    \label{fig:stadium resolution norm}}
    \hspace{3mm}
    \subfigure[The resolution dependence of the variance $\sigma^2$.]
    {\includegraphics[width=4cm]{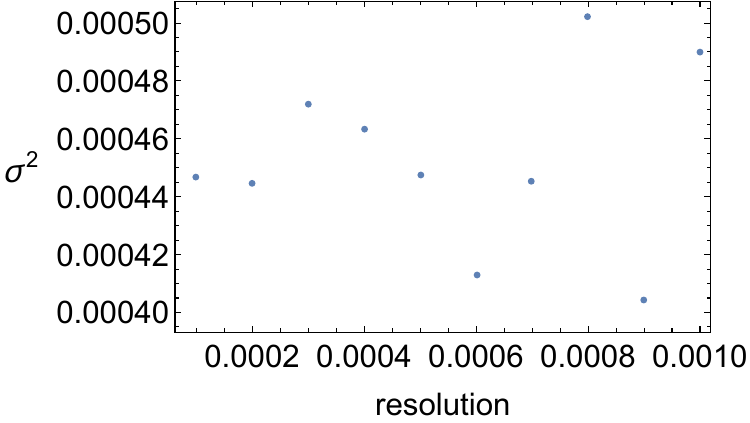}
    \label{fig:stadium resolution variance chaos}}
    \hspace{3mm}
    \subfigure[The Krylov operator complexities as functions of time for $\text{resolution} = 0.0001$ and $0.0002$.]{\includegraphics[width=4.5cm]{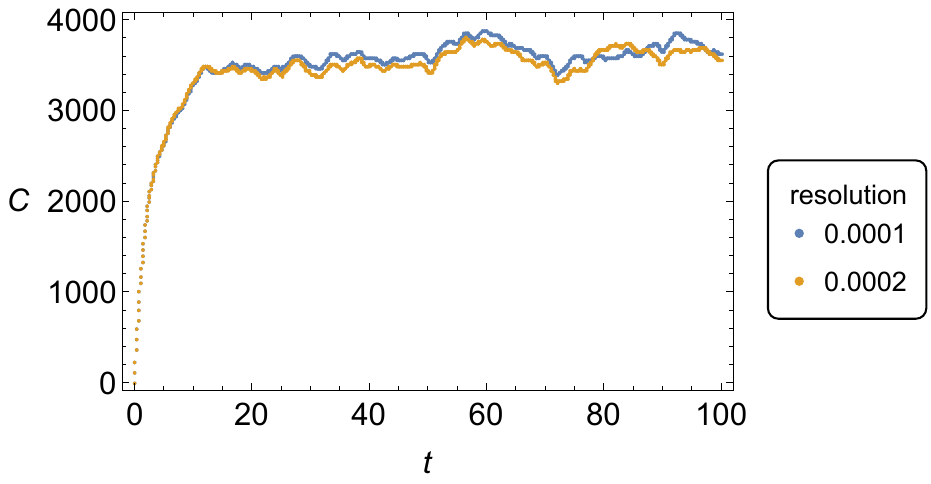}
    \label{fig:stadium resolution KOC}}
\caption{The resolution dependence of the Frobenius norm of $P$, the variance $\sigma^2$ and the Krylov operator complexity for $a/R=0.5$.}
\label{fig:stadium resolution norm variance KOC}
\end{figure}

In Sec.~\ref{sec:3}, we summarized the numerical method for the calculation of Krylov complexity. In order to obtain reliable numerical results, the truncated matrix representation $P$ of the momentum operator should not depend on the resolution sensitively. Define the norm of $P$ by
\begin{equation}
    \|P\| \equiv \sqrt{P^\dagger P} = \sqrt{\sum_{i,j=1}^{N_{\rm max}}|P_{ij}|^2},
\end{equation}
which is called the Frobenius norm of the matrix $P$. The shape of the stadium billiard is determined by $a/R$ (see Fig.~\ref{fig:stadium shape}).
In Fig.~\ref{fig:stadium resolution norm variance KOC}, we show the resolution dependence of the numerical results for $a/R=0.5$.
We can see that $\text{resolution} = 0.0001$ is sufficient for convergence. Therefore, we used $\text{resolution} = 0.0001$ 
in the numerical calculations presented in the main text.

\begin{figure}[t]
\centering
    \subfigure[The resolution dependence of the Frobenius norm of $P$.]
    {\includegraphics[width=4cm]{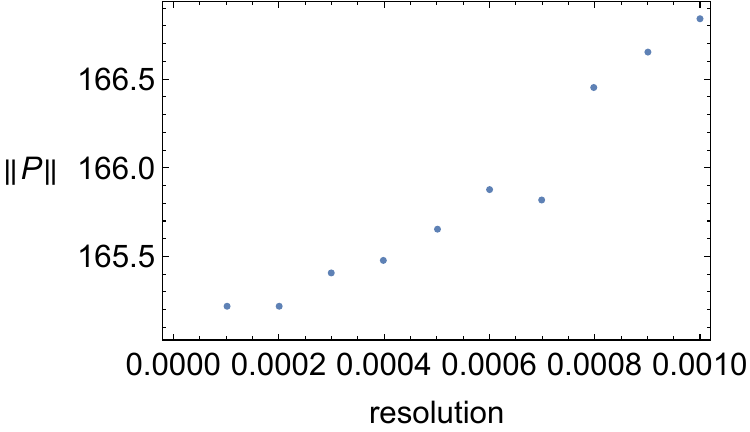}
    \label{fig:stadium resolution norm integrable}}
    \hspace{3mm}
    \subfigure[The resolution dependence of the variance $\sigma^2$.]
    {\includegraphics[width=4cm]{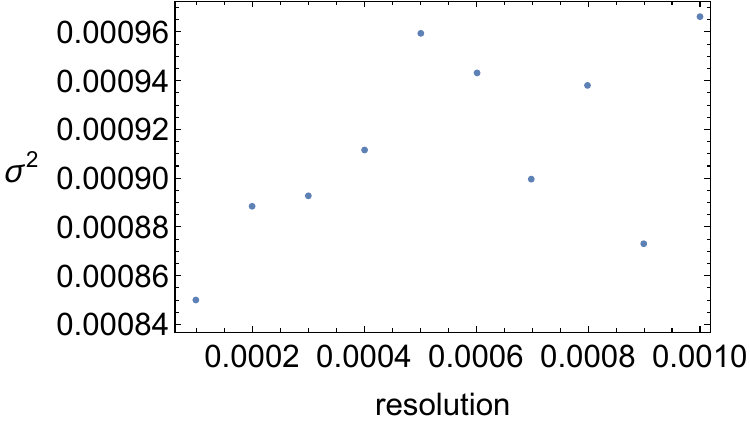}
    \label{fig:stadium resolution variance integrable}}
    \hspace{3mm}
    \subfigure[The Krylov operator complexities as functions of time for $\text{resolution} = 0.0001$ and $0.0002$.]{\includegraphics[width=4.5cm]{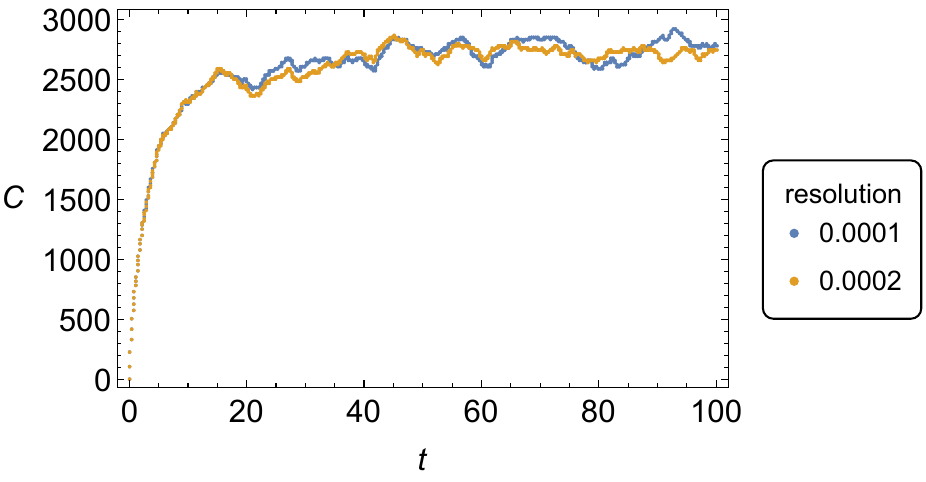}
    \label{fig:stadium resolution KOC integrable}} 
\caption{The resolution dependence of the Frobenius norm of $P$, the variance $\sigma^2$ and the Krylov operator complexity for $a/R=0$.}
\label{fig:stadium resolution norm variance KOC integrable}
\end{figure}

In Fig.~\ref{fig:stadium resolution norm variance KOC integrable}, we show the resolution dependence of the numerical results for $a/R=0$.
We see again that the convergence is sufficient around $\text{resolution} = 0.0001$. The stadium billiard system is integrable for $a/R=0$. Indeed, the energy eigenvalues and the energy eigenfunctions can be obtained analytically:
\begin{align}
    E &= \frac{1}{R^2}\rho_{kl}^2 \quad (k\in 2\mathbb{Z}_{>0},\,l\in\mathbb{Z}_{>0})\,, \\
    \phi &= \mathcal{N}J_k\left(\rho_{kl}\frac{r}{R}\right)\sin(k\theta)\,,\quad \mathcal{N}^{-1} \equiv \sqrt{\frac{\pi}{8}}RJ_{k+1}(\rho_{kl})\,,
\end{align}
where $J_k$ is the Bessel function of the first kind and $\rho_{kl}$ is the $l$-th zero of $J_k$. The matrix element of the momentum operator can be obtained by numerical integration.

\subsection{Sinai billiard}

\begin{figure}[t]
\centering
    \subfigure[The resolution dependence of the Frobenius norm of $P$.]
    {\includegraphics[width=4cm]{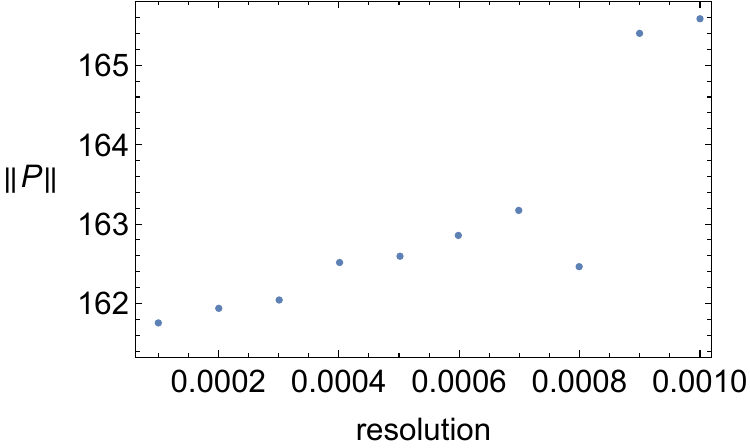}
    \label{fig:sinai resolution norm}}
    \hspace{3mm}
    \subfigure[The resolution dependence of the variance $\sigma^2$.]
    {\includegraphics[width=4cm]{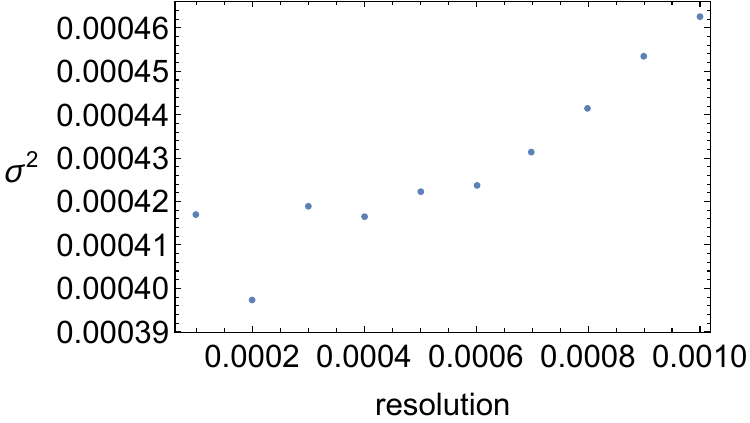}
    \label{fig:sinai resolution variance chaos}}
    \hspace{3mm}
    \subfigure[The Krylov operator complexities as functions of time for $\text{resolution} = 0.0001$ and $0.0002$.]{\includegraphics[width=4.5cm]{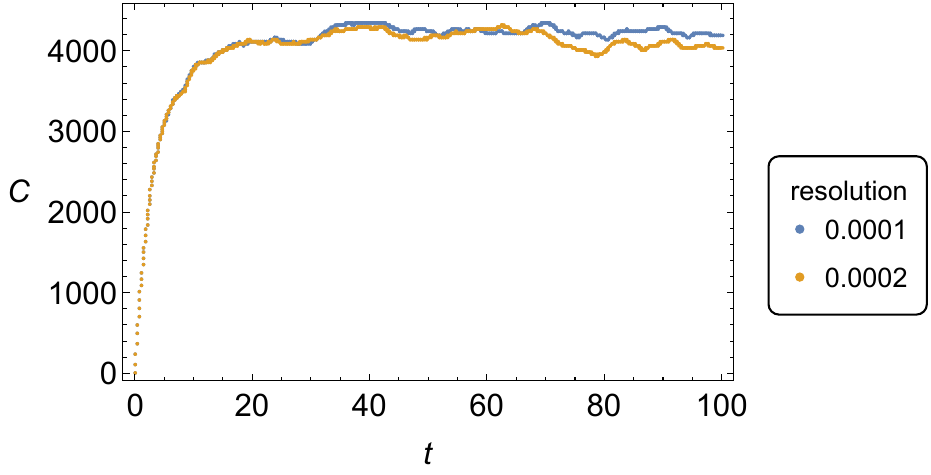}
    \label{fig:sinai resolution KOC}}   
\caption{The resolution dependence of the Frobenius norm of $P$, the variance $\sigma^2$ and the Krylov operator complexity for $l/L=0.5$.}
\label{fig:sinai resolution norm variance KOC}
\end{figure}

The shape of the Sinai billiard is determined by $l/L$ (see Fig.~\ref{fig:Sinai-billiard}).
In Fig.~\ref{fig:sinai resolution norm variance KOC}, we show the resolution dependence of the numerical results for $l/L=0.5$.
Although $\sigma^2$ slightly drops at $\text{resolution}=0.0002$,  $\text{resolution} = 0.0001$ is sufficient for convergence as a whole. 

\begin{figure}[t]
\centering
    \subfigure[The resolution dependence of the Frobenius norm of $P$.]
    {\includegraphics[width=4cm]{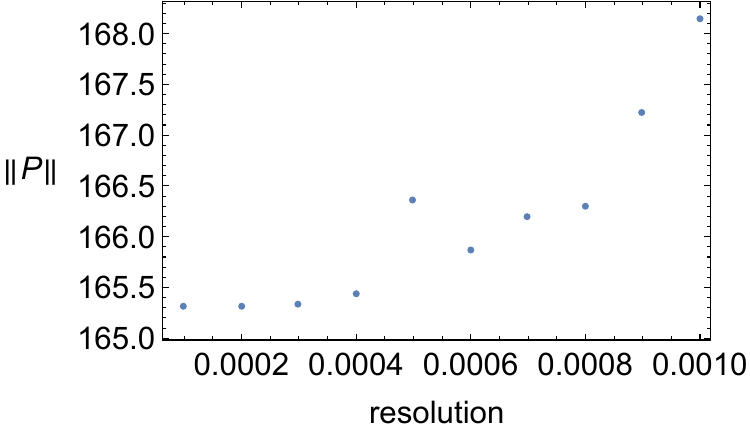}
    \label{fig:sinai resolution norm integrable}}
    \hspace{3mm}
    \subfigure[The resolution dependence of the variance $\sigma^2$.]
    {\includegraphics[width=4cm]{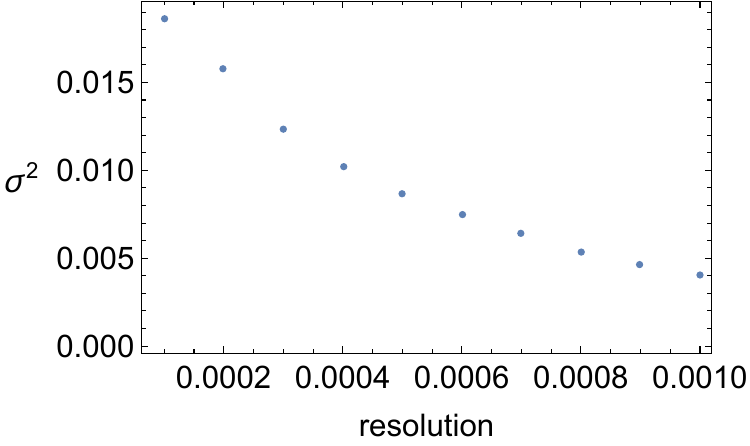}
    \label{fig:sinai resolution variance integrable}}
    \hspace{3mm}
    \subfigure[The Krylov operator complexities as functions of time for $\text{resolution} = 0.0001$ and $0.0002$.]{\includegraphics[width=4.5cm]{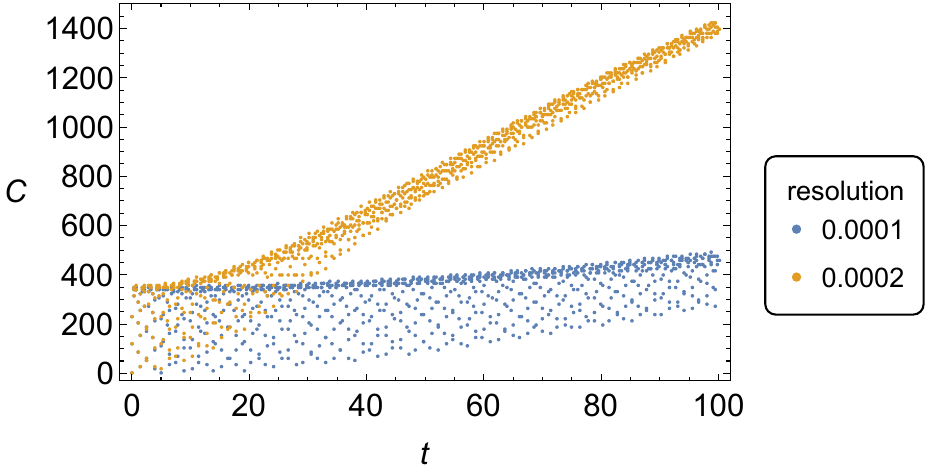}
    \label{fig:sinai resolution KOC integrable}}   
\caption{The resolution dependence of the Frobenius norm of $P$, the variance $\sigma^2$ and the Krylov operator complexity for $l/L=0$.}
\label{fig:sinai resolution norm variance KOC integrable}
\end{figure}

In Fig.~\ref{fig:sinai resolution norm variance KOC integrable}, we show the resolution dependence of the numerical results for $l/L=0$, at which the system becomes integrable.
Although the Frobenius norm of $P$ is well convergent, the variance $\sigma^2$ does not converge and increases as the resolution decreases. Correspondingly, the Krylov operator complexity does not also converge. This behavior is different from that of the stadium billiard. Notice that although the numerical calculation is unstable for $l/L=0$, the variance increases and the Krylov operator complexity decreases, respectively, when the resolution becomes smaller. This is consistent with the general expectation that for integrable systems, the variance becomes large and the late-time asymptotic value of the Krylov operator complexity becomes small.

To explore the possible causes of the irregular dependence on the resolution, we consider the analytic solution of the Schr\"odinger equation. We can regard the analytic solution as $\text{resolution} = 0$. The energy eigenvalues and the energy eigenfunctions for $l/L=0$ are
\begin{align}
    E &= \frac{\pi^2}{L^2}(m^2+n^2) \quad (m,n\in\mathbb{Z}_{>0},\,m>n)\,, \\
    \phi &= \frac{2}{L}\left\{\sin\left(\frac{m\pi x}{L}\right)\sin\left(\frac{n\pi y}{L}\right)-\sin\left(\frac{n\pi x}{L}\right)\sin\left(\frac{m\pi y}{L}\right)\right\}\,.
\end{align}
Then, the matrix element of the momentum operator can be obtained by analytic integration. Due to the orthogonality of the trigonometric functions, there may be many zero elements in the analytically obtained $P$. In numerical calculations, these elements can have small non-zero values due to numerical errors. These errors do not significantly affect the Frobenius norm of $P$. However, when the numerically obtained $P$ is input to the Lanczos algorithm, small numerical errors on elements that should be zero can have a significant impact on the final result due to the nonlinear nature of the algorithm.


\section{Truncation dependence}
\label{app:2}
In our numerical calculation of the Krylov complexity, we considered only a finite number of energy levels by introducing a truncation $N_{\rm max}$ to the spectrum. In this section, we will investigate the dependence of the numerical results on the different choice of $N_{\rm max}$. For simplicity, we will concentrate on the stadium billiards. 

\subsection{Krylov operator complexity}
\begin{figure}[t]
\centering
    \subfigure[The time dependence of Krylov operator complexity for various values of $a/R$.]{\includegraphics[height=4.2cm]{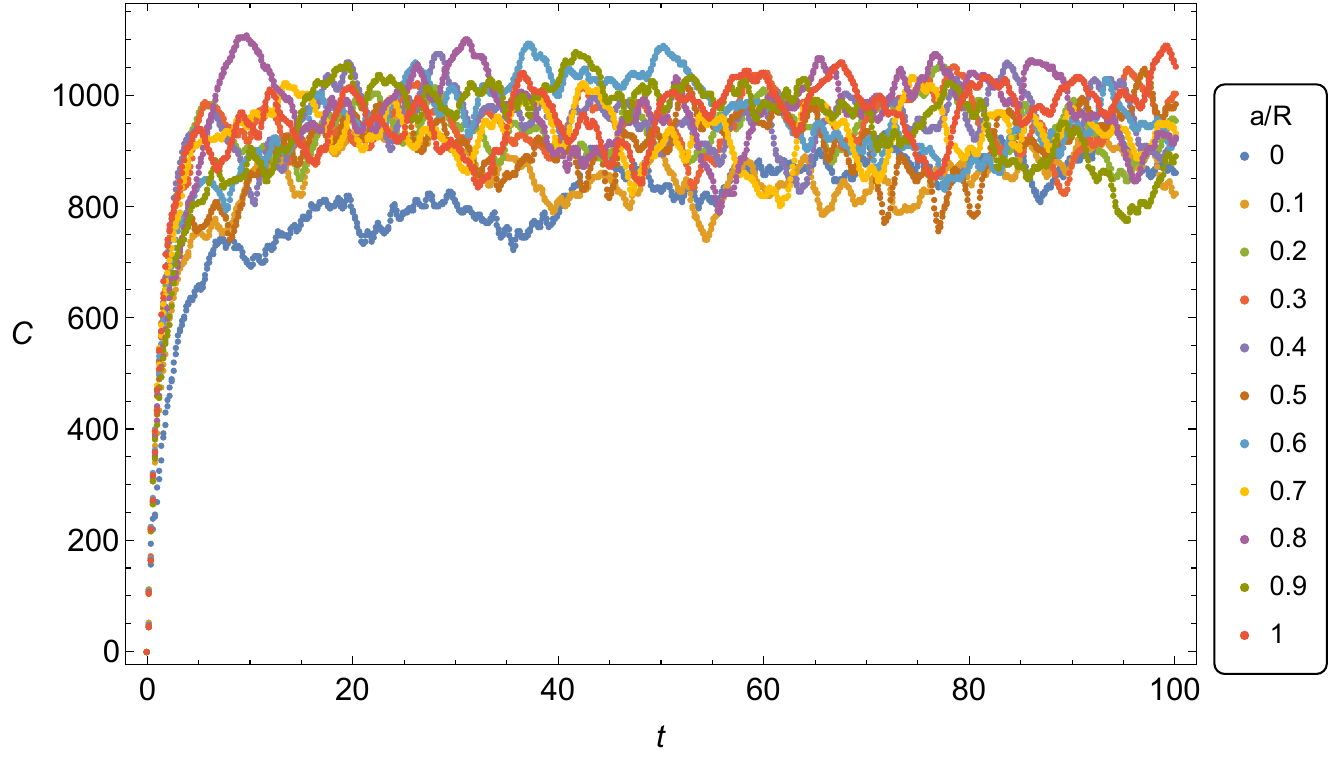}}
    \subfigure[The variance $\sigma^2$ as a function of $a/R$.]{\includegraphics[height=4.2cm]{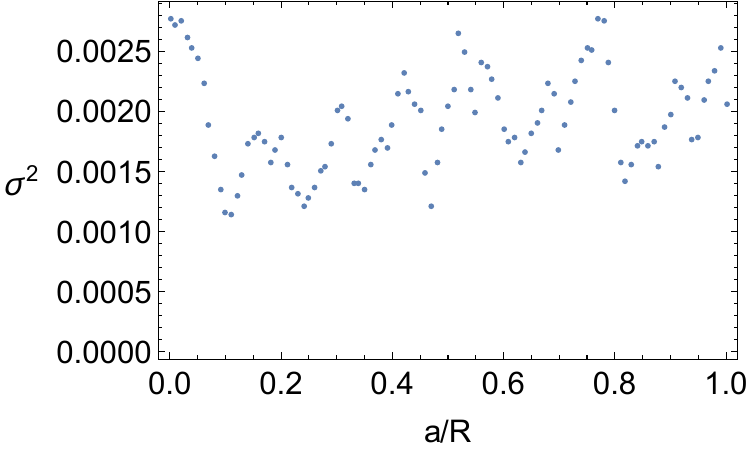}}
    \caption{The $a/R$ dependence of Krylov operator complexity and the variance of the Lanczos coefficients.}
    \label{fig:KOC and variance for 50 levels}
\end{figure}

\begin{figure}[t]
\centering
    \subfigure[$\lambda$ vs $\sigma^2$]{\includegraphics[height=4.2cm]{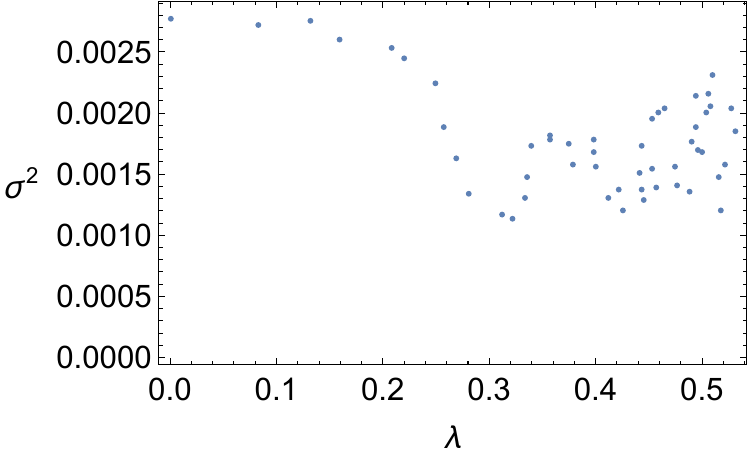}}
    \hspace{3mm}
    \subfigure[$\langle\tilde{r}\rangle$ vs $\sigma^2$]{\includegraphics[height=4.2cm]{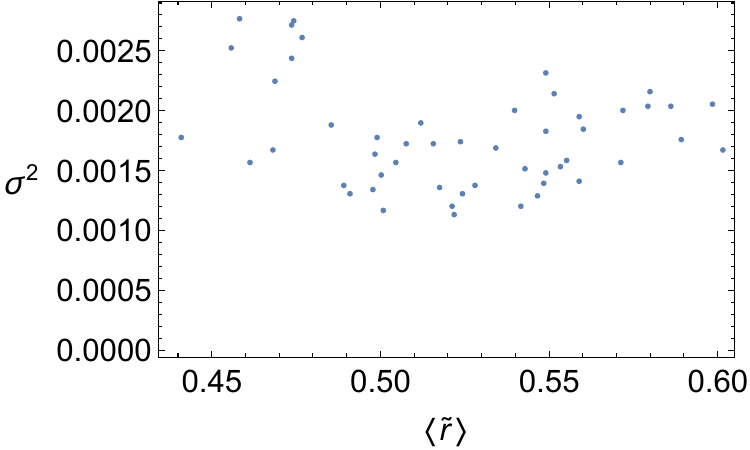}}
    \caption{The scatter plots of (a) the Lyapunov exponents and the variances and (b) the ratios
and the variances.}
    \label{fig:KOC correlation for 50 levels}
\end{figure}

The numerical setting is the same as Sec.~\ref{sec:3}. Performing the numerical calculation for $N_{\rm max}=50$, we obtained the Krylov operator complexity and the variance of the Lanczos coefficients as shown in Fig.~\ref{fig:KOC and variance for 50 levels}. Compared to the results for $N_{\rm max}=100$, the time dependence of Krylov operator complexity is less smooth. As for the $a/R$ dependence of the variance, while there is a dip around $a/R=0.1$, the difference between chaoticity and integrability is less clear. Nevertheless, there are still weak correlations as shown in Fig.~\ref{fig:KOC correlation for 50 levels} and by the correlation coefficients in Table~\ref{table:correlation coefficients for 50 levels}. Although the behavior of the variance is dependent on $N_{\rm max}$, it will be useful to distinguish chaoticity and integrability once $N_{\rm max}$ is chosen sufficiently large.

\begin{table}[t]
\begin{center}
    \begin{tabular}{|c|l|}
    \hline
    $\lambda$ vs $\sigma^2$ & -0.497215\\
    $\langle \tilde{r}\rangle$ vs $\sigma^2$ & -0.214267\\
    \hline
    \end{tabular}
    \caption{The correlation coefficients between $\lambda$, $\langle \tilde{r}\rangle$, $\sigma^2$ for the Krylov operator complexity.}
    \label{table:correlation coefficients for 50 levels}
\end{center}
\end{table}

\subsection{Krylov state complexity}

We summarize the numerical results for the Krylov state complexity with reduced truncation order $N_\text{max}=250$ and compare them with those shown in Sec.~\ref{sec:5} for $N_\text{max}=500$.

\begin{figure}[t]
\centering
    \subfigure[Krylov state complexity]{\includegraphics[height=3.7cm]{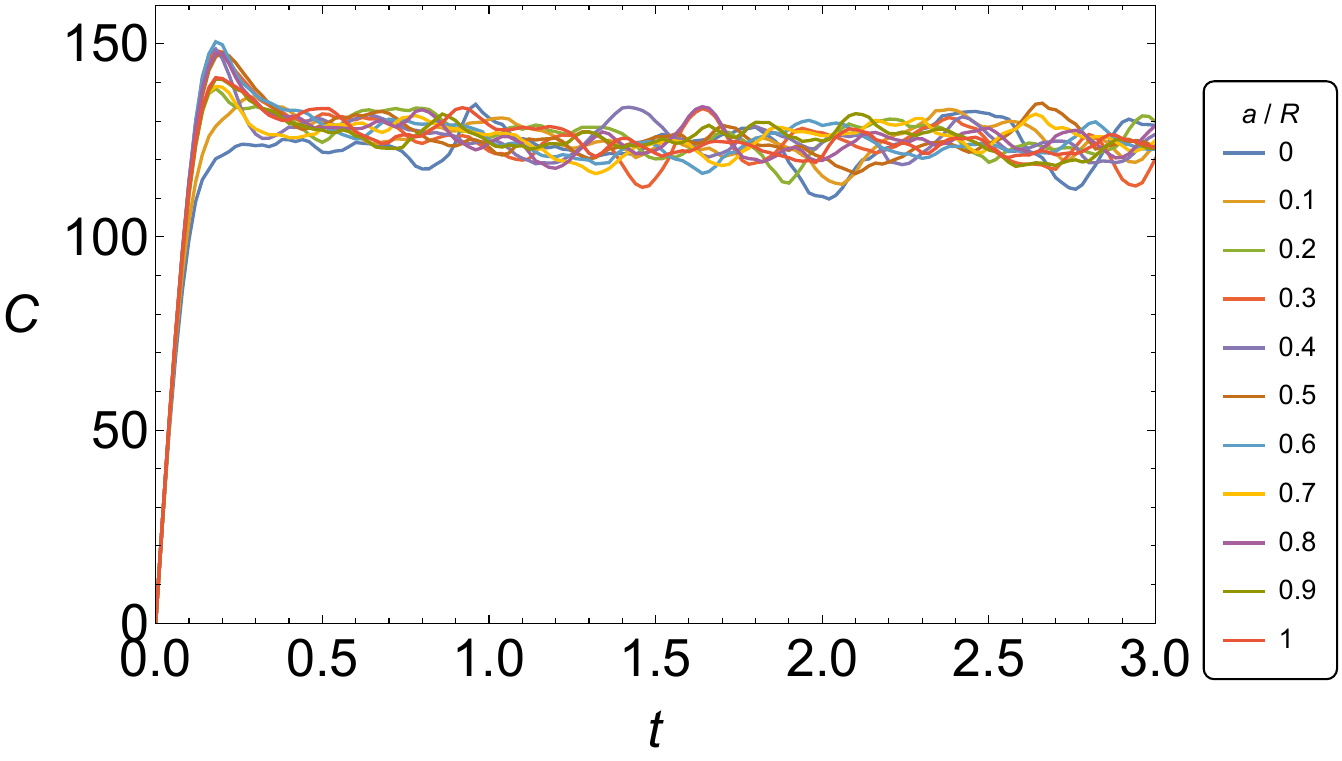}}
    \subfigure[Variances $\sigma_{a,b}^2$ as functions of $a/R$.]{\includegraphics[height=3.7cm]{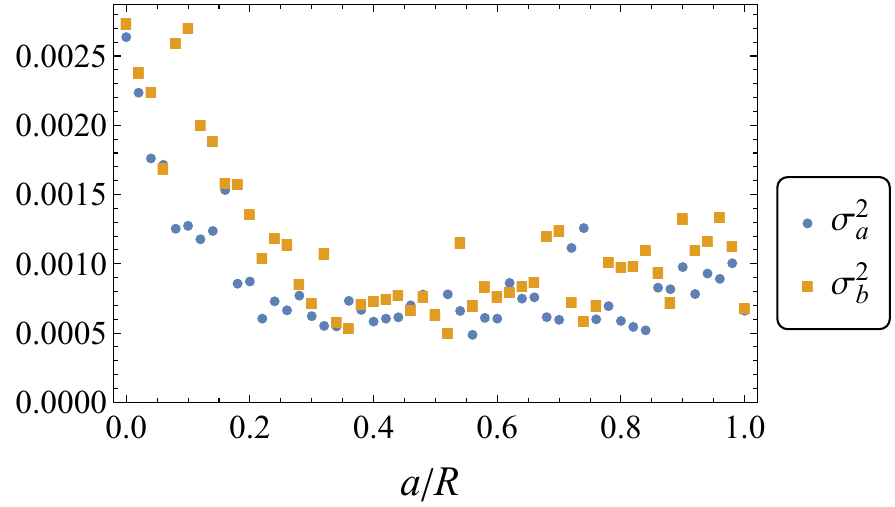}}
    \subfigure[The peak value of Krylov state complexity.]{\includegraphics[height=3.7cm]{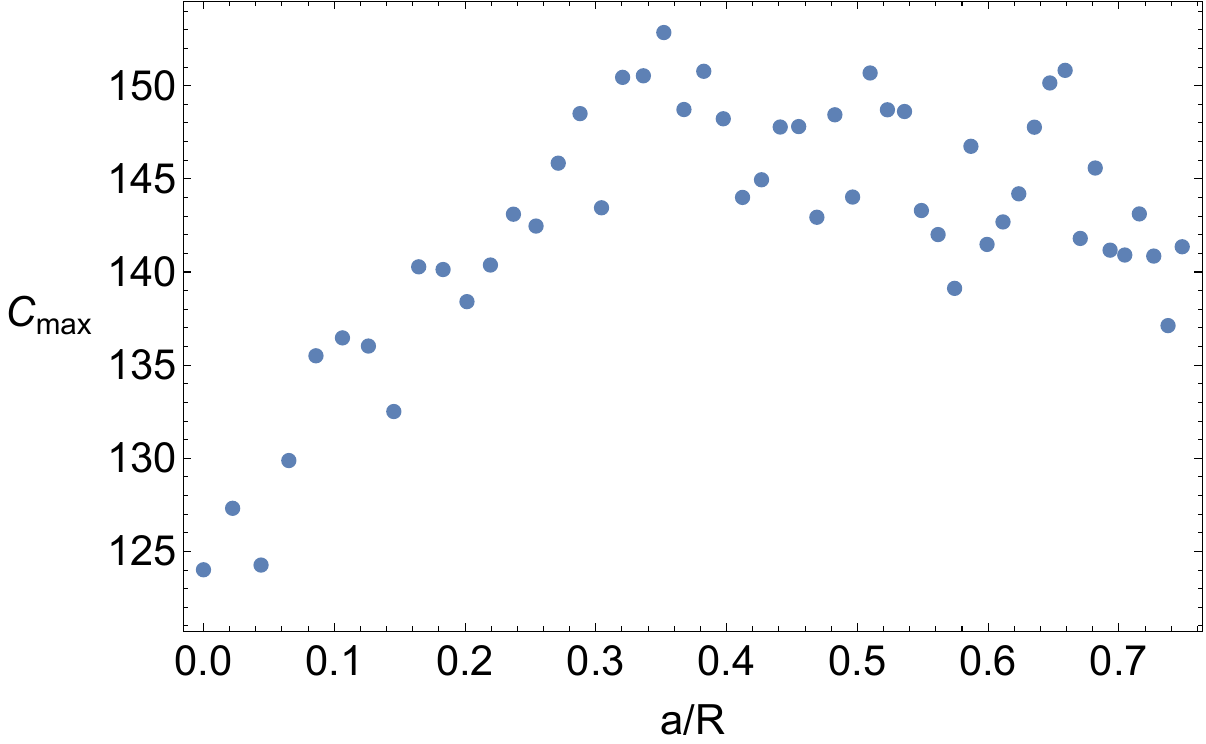}}
    \caption{
    The $a/R$ dependence of Krylov state complexity. 
    Panel (a): time dependence of the complexity. Panel (b): variance $\sigma_a^2, \sigma_b^2$ of the Lanczos coefficients. (c): the peak value of the complexity.}
    \label{fig:KSC and variance for 250 levels}
\end{figure}

Figure~\ref{fig:KSC and variance for 250 levels} shows the time dependence and peak value of the Krylov state complexity and also the variance of the Lanczos coefficients for $N_\text{max} = 250$. Qualitative features are the same as those in Figs.~\ref{fig:stadium KSC and asym KSC} and \ref{fig:stadium state variance}, while the asymptotic value of the complexity is reduced to $\sim 125 = N_\text{max}/2$.
Other difference is that the variance $\sigma_a^2,\sigma_b^2$ of the Lanczos coefficients 
and also the fluctuation of the complexity
slightly increase compared to the case with $N_\text{max}=500$.

\begin{figure}[t]
    \centering
    \subfigure[$\lambda$ vs $\sigma_{a,b}^2$]
    {\includegraphics[width=7cm]{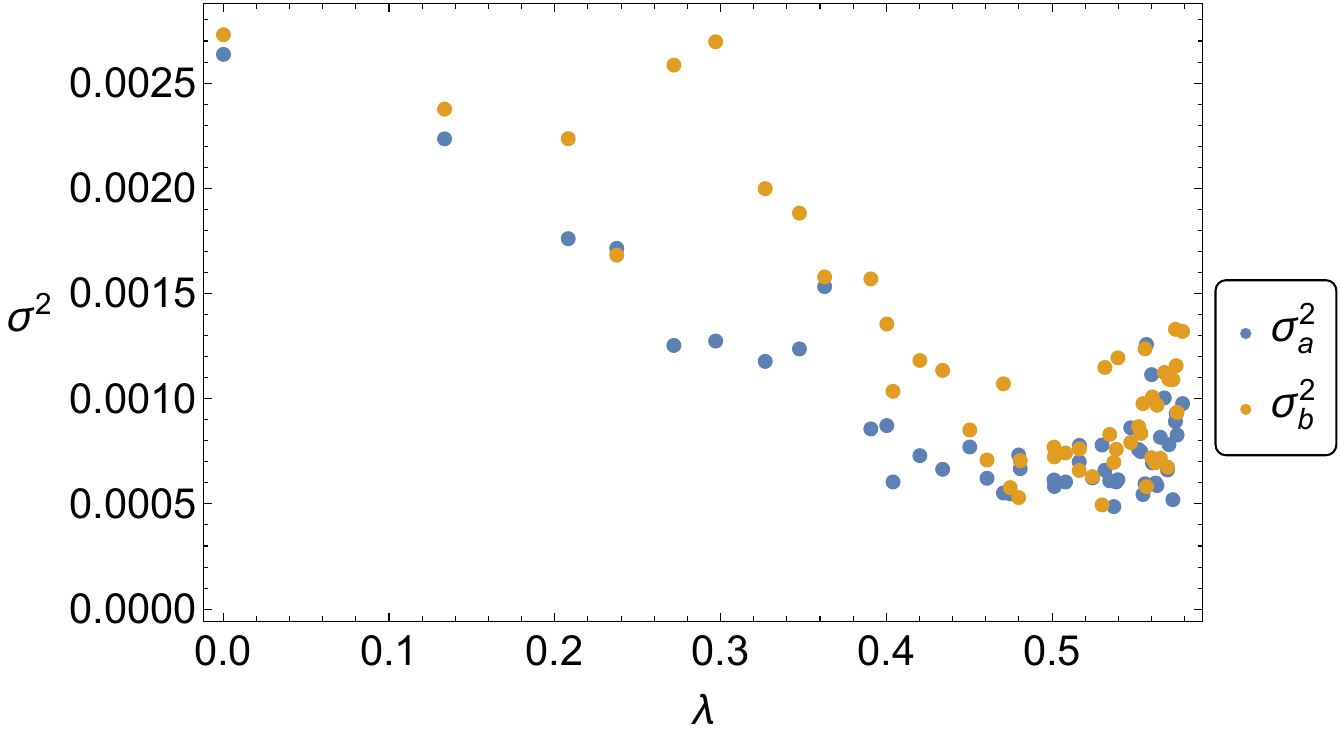}\label{fig:Stadium-state_lambda-varxab Nmax=250}}
    \hspace{5mm}
    \subfigure[$\langle\tilde{r}\rangle$ vs $\sigma_{a,b}^2$]{\includegraphics[width=7cm]{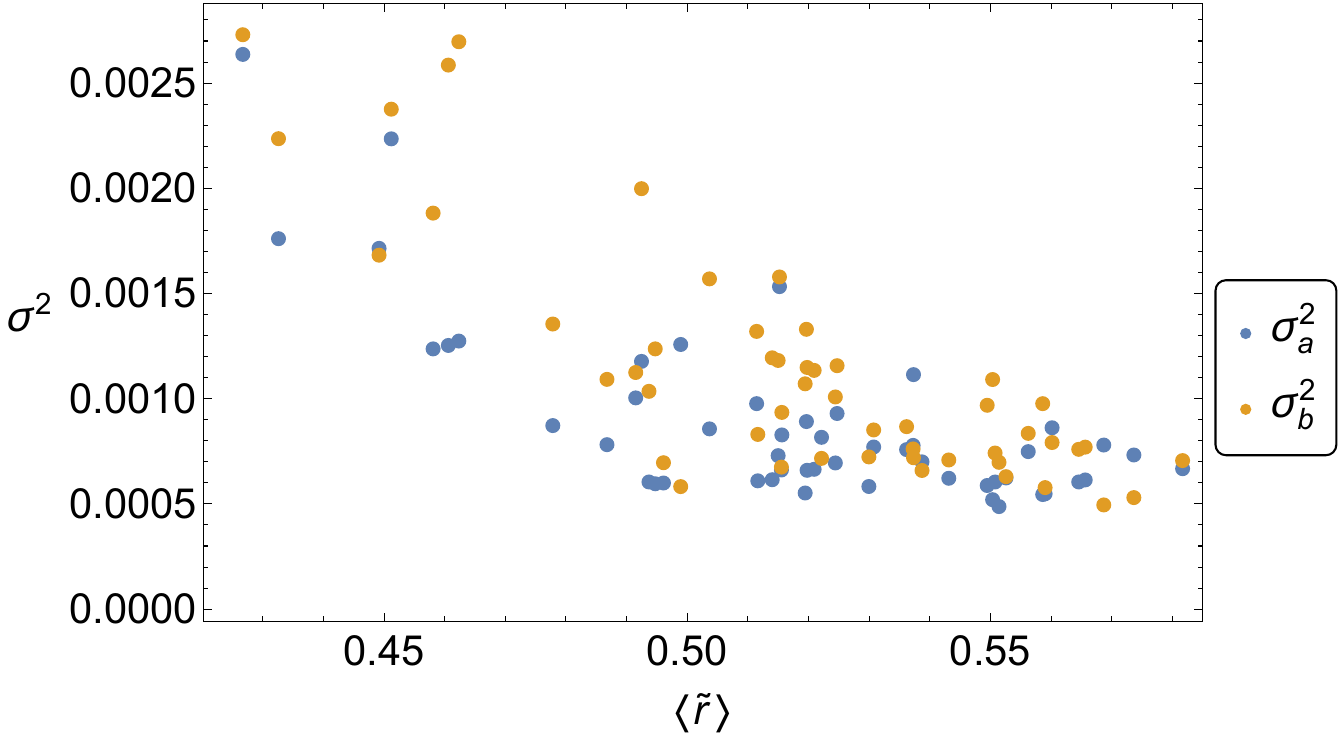}\label{fig:Stadium-state_r-varxab Nmax=250}}
    \caption{The scatter plots of (a) the Lyapunov exponents and the variances of the Krylov state complexity, and (b) the ratios and the variances.
    The points are sampled from $0\leq a/R \leq 0.75$.
    }
    \label{fig:Stadium_scatter_plot_state Nmax=250}
\end{figure}

Correlation between the indicators of classical and quantum chaos are summarized in Fig.~\ref{fig:Stadium_scatter_plot_state Nmax=250} and Table~\ref{table:Stadium_correlations Nmax=250}. The correlation coefficients in Table~\ref{table:Stadium_correlations Nmax=250} are as large as those for $N_\text{max} = 500$ in Table~\ref{table:Stadium_correlations}, which the variance of the Lanczos coefficients $\sigma_a^2, \sigma_b^2$ can be used as an indicator of chaos as long as $N_\text{max}$ is moderately large.

\begin{table}[t]
 \begin{center}
  \begin{tabular}{|c|l|}
  \hline
    $\lambda$ vs $\sigma_a^2$ & -0.825666\\
    $\lambda$ vs  $\sigma_b^2$ & -0.819637\\
    $\langle \tilde{r}\rangle$ vs $\sigma_a^2$ & -0.734249\\
    $\langle \tilde{r}\rangle$ vs $\sigma_b^2$ & -0.817428\\
    \hline
  \end{tabular}
     \caption{Correlations between $\lambda$, $\langle \tilde{r}\rangle$, $\sigma_{a,b}^2$ for the state complexity of stadium billiard.}
   \label{table:Stadium_correlations Nmax=250}
 \end{center}
\end{table}


\section{Initial operator/state dependence of the Krylov complexity}
\label{app:3}
In this section, we consider the dependence of the Krylov complexity on the choice of the initial operator/state.

\subsection{Krylov operator complexity}
\begin{figure}[t]
\centering
    \subfigure[The time dependence of Krylov operator complexity for various values of $a/R$.]{\includegraphics[height=4.2cm]{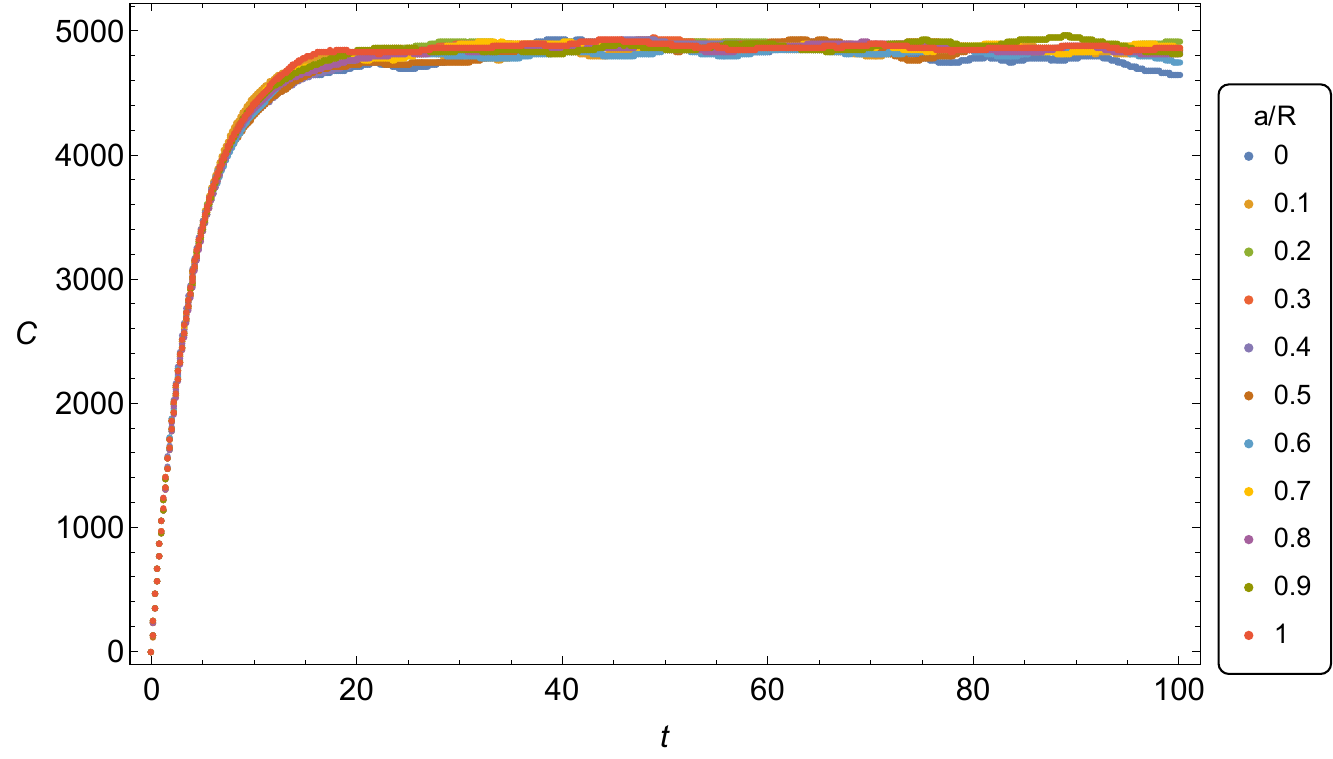}}
    \subfigure[The variance $\sigma^2$ as a function of $a/R$.]{\includegraphics[height=4.2cm]{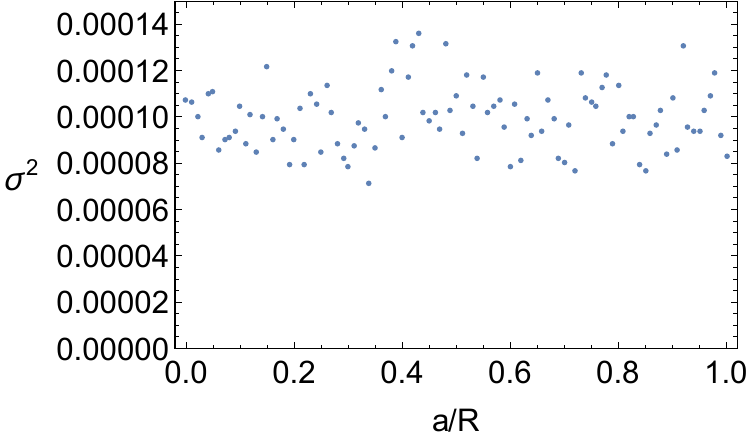}}
    \caption{The $a/R$ dependence of Krylov operator complexity and the variance of the Lanczos coefficients.}
    \label{fig:KOC and variance for flat operator}
\end{figure}

\begin{figure}[t]
\centering
    \subfigure[$\lambda$ vs $\sigma^2$]{\includegraphics[height=4.2cm]{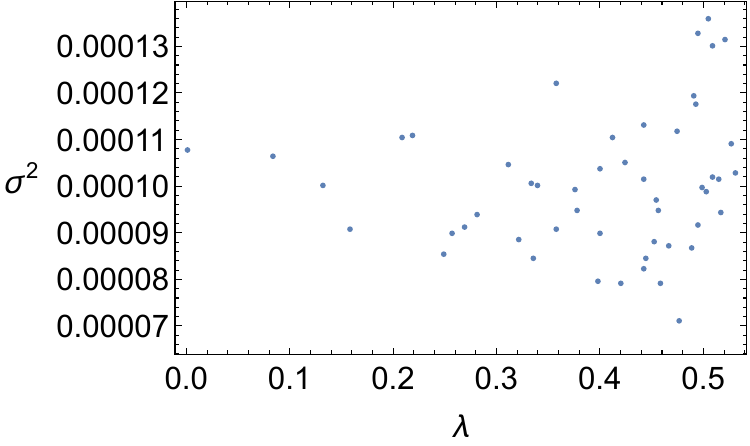}}
    \hspace{3mm}
    \subfigure[$\langle\tilde{r}\rangle$ vs $\sigma^2$]{\includegraphics[height=4.2cm]{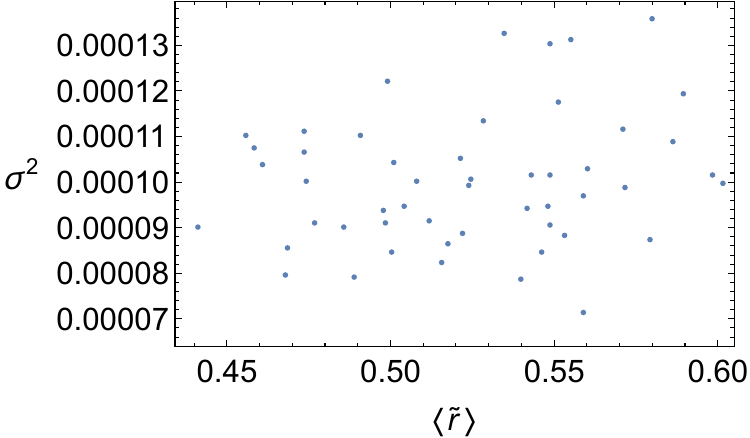}}
    \caption{The scatter plots of (a) the Lyapunov exponents and the variances and (b) the ratios
and the variances.}
    \label{fig:KOC correlation for flat operator}
\end{figure}

Since we studied Krylov state complexity for the flat initial state,
we consider the flat operator, which is considered in appendix B of \cite{Rabinovici:2022beu},
\begin{equation}
    \mathcal{O}_{mn}=1 \quad \text{for all} \quad n,m=1,\cdots,N_{\rm max}\,,
\end{equation}
with $N_{\rm max}=100$ in this section. In Fig.~\ref{fig:KOC and variance for flat operator}, we show the Krylov operator complexity and the variance of the Lanczos coefficient. For the flat operator, the Krylov operator complexity, especially its saturation value, hardly depends on $a/R$ \cite{Rabinovici:2022beu}. Also, the variance of the Lanczos coefficient
cannot distinguish chaoticity and integrability in this case. In Fig.~\ref{fig:KOC correlation for flat operator}, we cannot find any correlation. The correlation coefficients take positive values as shown in Table~\ref{table:correlation coefficients for flat operator} even though they should be negative. Thus, the Krylov operator complexity and the behavior of the Lanczos coefficients depend on the choice of the initial operator. In order to distinguish chaoticity, we need to choose the initial operator properly. Although we do not know the criterion for this choice, we expect that the fundamental operators, which appear in the definition of the Hamiltonian, should be proper.

\begin{table}[t]
\begin{center}
    \begin{tabular}{|c|l|}
    \hline
    $\lambda$ vs $\sigma^2$ & 0.118923\\
    $\langle \tilde{r}\rangle$ vs $\sigma^2$ & 0.204392\\
    \hline
    \end{tabular}
    \caption{The correlation coefficients between $\lambda$, $\langle \tilde{r}\rangle$, $\sigma^2$ for the Krylov operator complexity.}
    \label{table:correlation coefficients for flat operator}
\end{center}
\end{table}

\subsection{Krylov state complexity}

\begin{figure}[t]
\centering
    \subfigure[$\Psi(x,y)$]{\includegraphics[height=4.0cm]{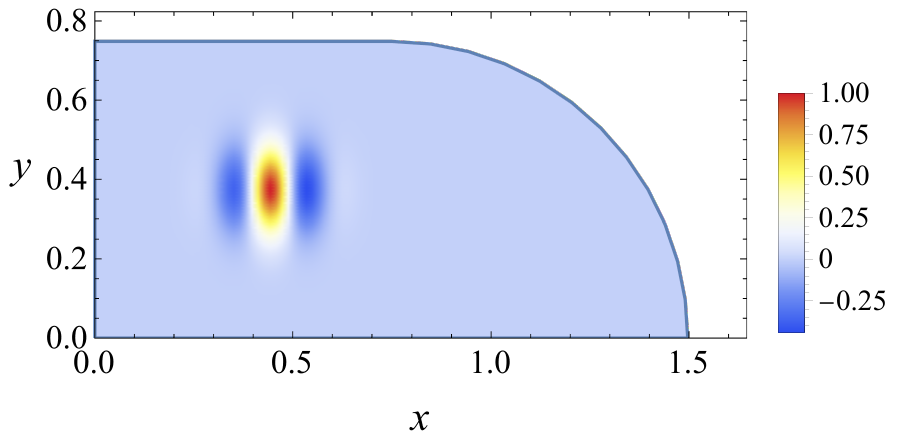}}
    \hspace{3mm}
    \subfigure[$\Psi_n$]{\includegraphics[height=4.0cm]{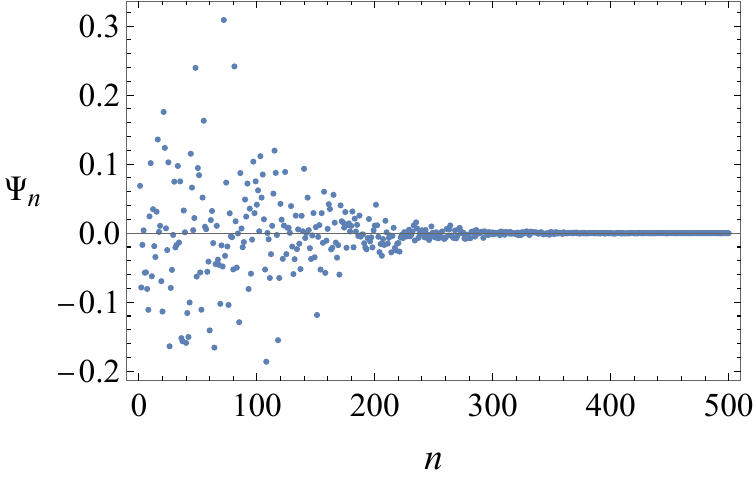}}
    \caption{Panel (a): the profile of the wave packet $\Psi(x,y)$. Peak value is normalized to the unity. Panel (b) the components of the initial state $\Psi=\left(\Psi_1,\ldots,\Psi_{N_\text{max}}\right)^\text{T}$ for $a/R=1$. The components are normalized so that $|\Psi|^2=1$.}
    \label{fig:wavepacket}
\end{figure}

The results on the Krylov state complexity are based on the flat initial state (\ref{flatinitialstate}).
To examine
the dependence of the results on the choice of the initial state, in this appendix we show the results for the initial state corresponding to a wave packet on the Stadium billiard table.
We use a wave packet defined by
\begin{equation}
\Psi(x,y) = A \exp\left(
-\frac{(x-x_0)^2+(y-y_0)^2}{2\sigma^2}
\right)\cos\left(
p_0 x
\right)~.
\label{wavepacket}
\end{equation}
For numerical calculation below, we chose $(x_0,y_0)=(0.45,0.37)$, $\sigma = 0.075$ and $p_0 = 40$ so that the distribution of the wave packet is contained well within the billiard table for any $a/R$.
We show the profile of the wave packet and the corresponding components of the initial state 
$\Psi=\left(\Psi_1,\ldots,\Psi_{N_\text{max}}\right)^\text{T}$ 
for $a/R=1$
in Fig.~\ref{fig:wavepacket}.
We use Eq.~(\ref{wavepacket}) for any $a/R$, and the energy eigenstates and the components $\Psi_{n=1,\ldots,N_\text{max}}$ with respect to them are calculated for each $a/R$. 
For any $a/R$, the components $\Psi_n$ take nonzero value fluctuating around zero for $n\lesssim 300$, while they tend to zero for $n\gtrsim 300$.

\begin{figure}[t]
\centering
    \subfigure[$a_n$]{\includegraphics[height=4.2cm]{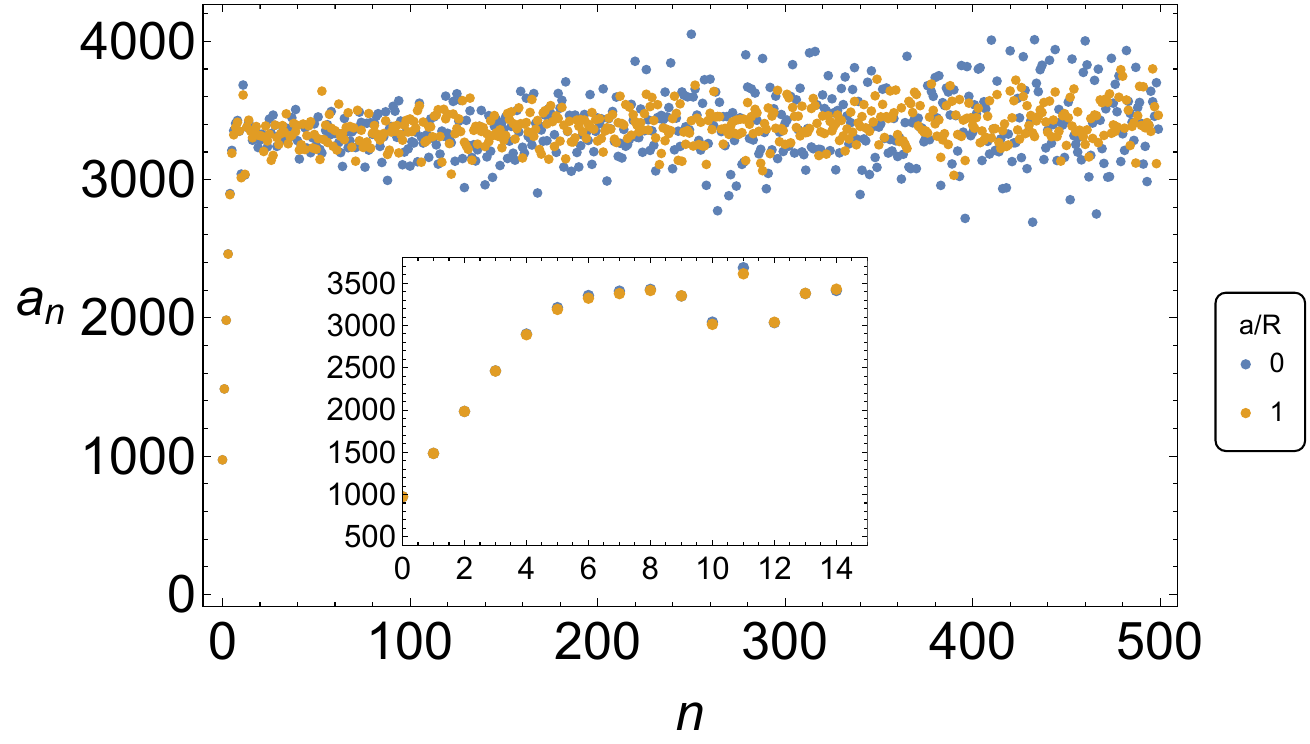}}
    \subfigure[$b_n$]{\includegraphics[height=4.2cm]{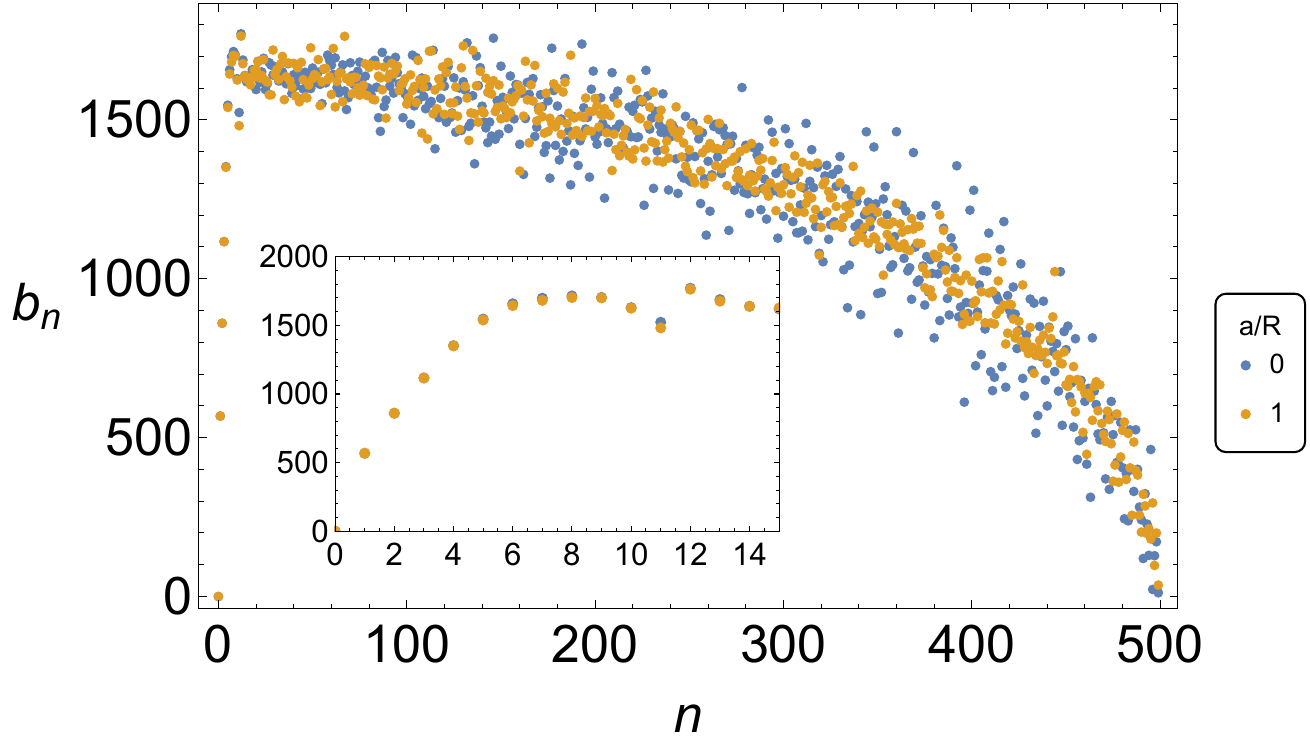}}
    \caption{The Lanczos coefficients of the Krylov state complexity of the wave packet initial state for $a/R=0$ (blue dots) and $a/R=1$ (orange dots). The insets show $a_n, b_n$ at low $n$.}
    \label{fig:wavepacket_ab}
\end{figure}

\begin{figure}[t]
\centering
    \subfigure[Krylov state complexity]{\includegraphics[height=3.7cm]{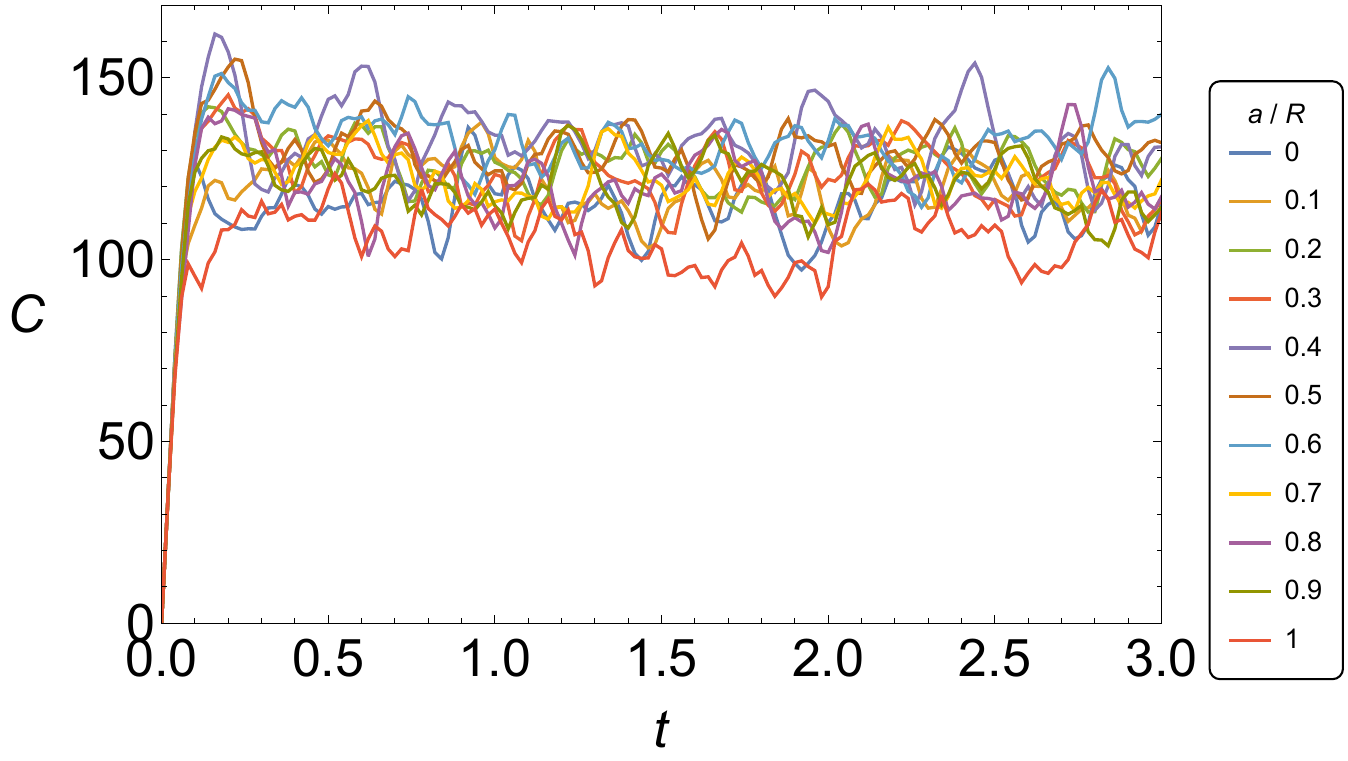}\label{fig:KSC wavepacket}}
    \subfigure[The variances $\sigma_{a,b}^2$ as functions of $a/R$]{\includegraphics[height=3.7cm]{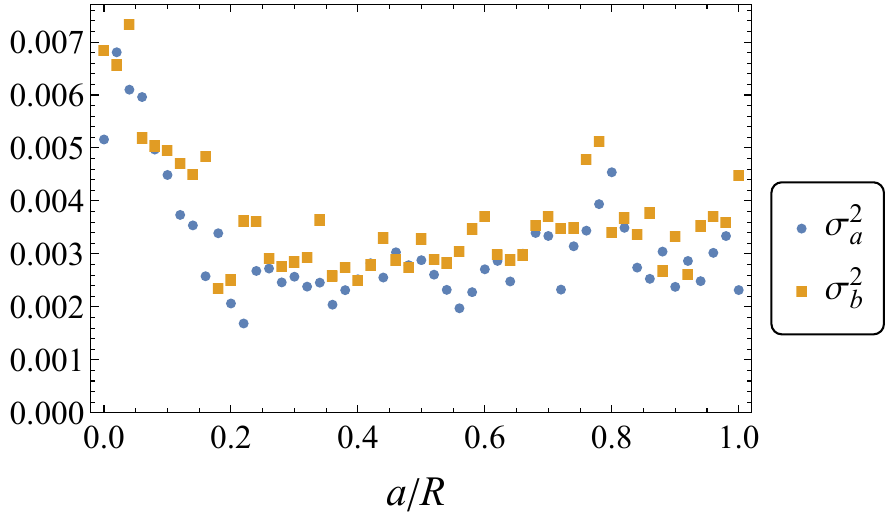}}
    \caption{
    The $a/R$ dependence of Krylov state complexity for the wave packet initial state (\ref{wavepacket}). 
    Panel (a): time dependence of the complexity. Panel (b): variance $\sigma_a^2, \sigma_b^2$ of the Lanczos coefficients.}
    \label{fig:wavepacket_C-varab}
\end{figure}

\begin{figure}[t]
    \centering
    \subfigure[$\lambda$ vs $\sigma_{a,b}^2$]
    {\includegraphics[width=7cm]{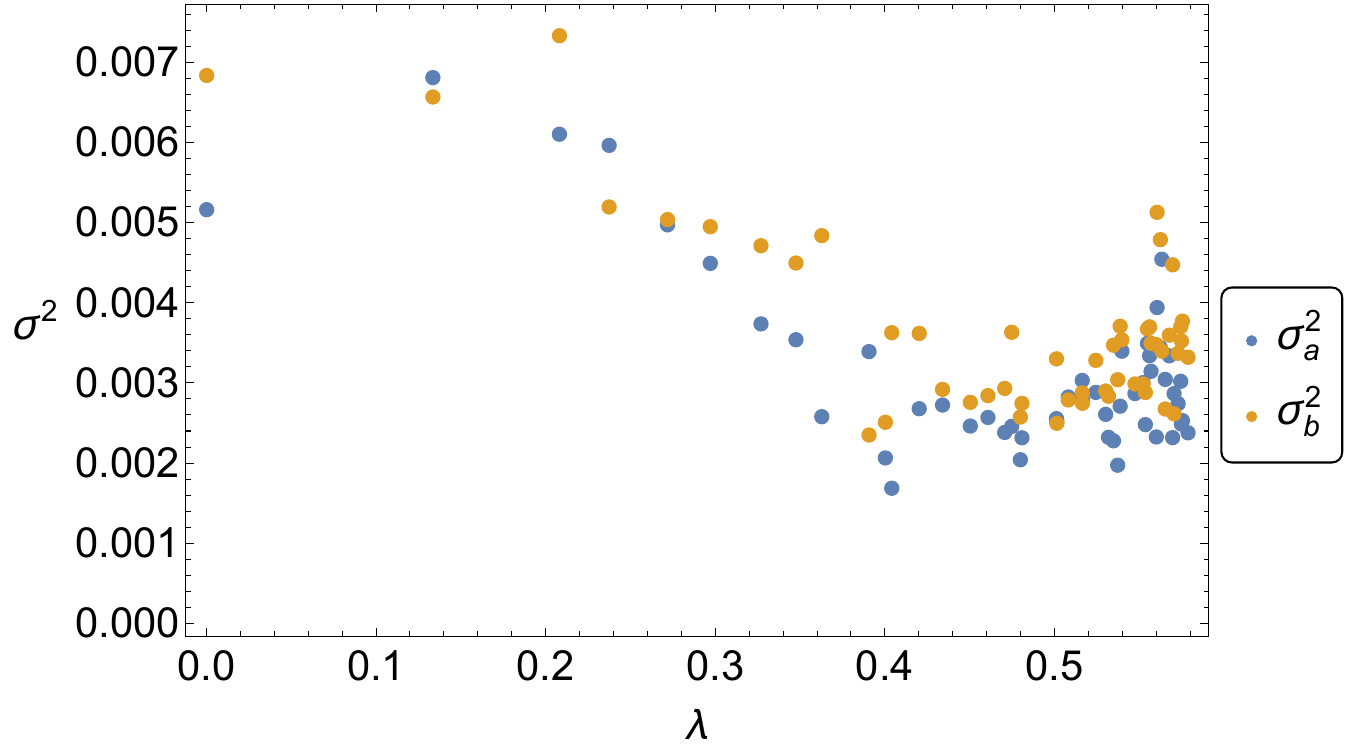}}
    \hspace{5mm}
    \subfigure[$\langle\tilde{r}\rangle$ vs $\sigma_{a,b}^2$]{\includegraphics[width=7cm]{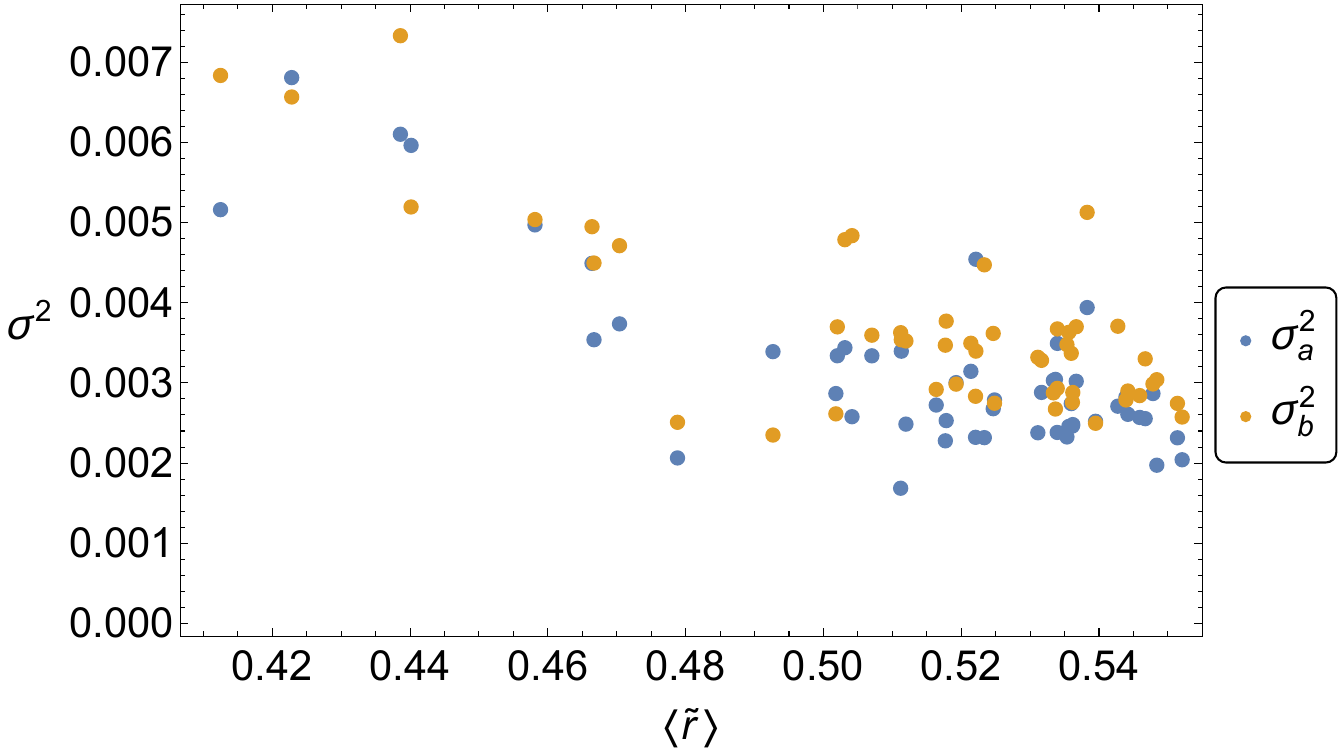}}
    \caption{The scatter plots for the wave packet initial state of (a) the Lyapunov exponents and the variances of the Krylov state complexity, and (b) the ratios and the variances.
    The points are sampled from $0\leq a/R \leq 0.75$.
    }
    \label{fig:wavepacket scatter plots}
\end{figure}

The numerical results for $N_\text{max}=500$
are shown in Figures~\ref{fig:wavepacket_ab}, \ref{fig:wavepacket_C-varab} and \ref{fig:wavepacket scatter plots}.
They are qualitatively the same as those for flat initial state in Sec.~\ref{sec:5} except for the following features:
(i) the Lanczos coefficients show linear growth for the first few indices ($a_n, b_n$ for $n\lesssim 4$, see Fig.~\ref{fig:wavepacket_ab}); (ii) fluctuation of the Krylov state complexity around the late time saturation value is relatively large, and the peak structure at early time evolution is less obvious compared to that for the flat initial state (see Fig.~\ref{fig:KSC wavepacket}).

In Table~\ref{table:wavepacket correlations}, we show the correlation coefficients for the indicators of the chaos for the wave packet initial state with $N_\text{max} = 500$. Although these values are slightly smaller compared to those for the flat initial state, they are still far from zero and indicate that the correlations between $\lambda, \sigma_{a,b}^2$ and $\langle \tilde r \rangle$ are strong.
For comparison, we also show the correlation coefficients for $N_\text{max} = 250$.
From these values we can observe some correlations, though they are slightly weaker than those $N_\text{max} = 500$. This result support that the variances of the Lanczos coefficients are robust probe of the chaos once the trancation order $N_\text{max}$ is moderately large.

\begin{table}[t]
\centering
\subtable[$N_\text{max}=500$]{
\centering
  \begin{tabular}{|c|l|}
  \hline
    $\lambda$ vs $\sigma_a^2$ & -0.682982\\
    $\lambda$ vs  $\sigma_b^2$ & -0.702761\\
    $\langle \tilde{r}\rangle$ vs $\sigma_a^2$ & -0.790287\\
    $\langle \tilde{r}\rangle$ vs $\sigma_b^2$ & -0.770899\\
    \hline
  \end{tabular}
      \label{table:wavepacket correlations 1}}
\hspace{1cm}
\subtable[$N_\text{max}=250$]{
\centering
  \begin{tabular}{|c|l|}
  \hline
    $\lambda$ vs $\sigma_a^2$ & -0.223911\\
    $\lambda$ vs  $\sigma_b^2$ & -0.546823\\
    $\langle \tilde{r}\rangle$ vs $\sigma_a^2$ & -0.410688\\
    $\langle \tilde{r}\rangle$ vs $\sigma_b^2$ & -0.599216\\
    \hline
  \end{tabular}
      \label{table:wavepacket correlations 2}
}
 \caption{Correlations between $\lambda$, $\langle \tilde{r}\rangle$, $\sigma_{a,b}^2$ for the state complexity of stadium billiard for $N_\text{max}=500$ (Table (a)) and $N_\text{max}=250$ (Table (b)).}
  \label{table:wavepacket correlations}
\end{table}


\end{document}